\documentclass[11pt,a4paper]{article}
\usepackage{amsmath,amssymb,bm}
\usepackage{physics}
\usepackage{graphicx}
\usepackage{xcolor}
\usepackage{booktabs}
\usepackage{soul}
\newif\ifhaveslashed
\IfFileExists{slashed.sty}{\haveslashedtrue}{\haveslashedfalse}
\ifhaveslashed\usepackage{slashed}\else
  \newcommand{\slashed}[1]{\ensuremath{#1\!\!\!/}}\fi
\providecommand{\half}{\tfrac{1}{2}}
\providecommand{\GeV}{\,\mathrm{GeV}}
\providecommand{\fm}{\,\mathrm{fm}}

\providecommand{\nfock}{n}
\providecommand{\Phat}{\widehat{P}_{3}}

\newif\ifjhep
\IfFileExists{jheppub.sty}{\jheptrue}{\jhepfalse}

\ifjhep
  \usepackage[numbers,sort&compress]{natbib}
  \usepackage{jheppub}
  \makeatletter
  \renewcommand{\abstract}[1]{\gdef\@abstract{#1}}
  \makeatother
\else
  \usepackage[margin=2.4cm]{geometry}
  \usepackage[numbers,sort&compress]{natbib}
  \usepackage{hyperref}
  \hypersetup{colorlinks=true,linkcolor=blue,citecolor=blue,urlcolor=blue}
  \makeatletter
  \let\@jhepauthors\@empty
  \let\@jhepaffils\@empty
  \let\@jhepemails\@empty
  \let\@jhepabstract\@empty
  \let\@jhepkeys\@empty
  \renewcommand{\author}[2][]{%
    \g@addto@macro\@jhepauthors{#2\,$^{\mathrm{#1}}$\quad}}
  \newcommand{\affiliation}[2][]{%
    \g@addto@macro\@jhepaffils{\par\small$^{\mathrm{#1}}$\,\itshape #2}}
  \newcommand{\collaboration}[2][]{%
    \g@addto@macro\@jhepaffils{\par\small #2}}
  
  \newcommand{\emailAdd}[1]{%
    \ifx\@jhepemails\@empty
      \g@addto@macro\@jhepemails{\par\small\textit{E-mail:} \texttt{#1}}%
    \else
      \g@addto@macro\@jhepemails{, \texttt{#1}}%
    \fi}
  \renewcommand{\abstract}[1]{\gdef\@jhepabstract{#1}}
  \newcommand{\keywords}[1]{\gdef\@jhepkeys{#1}}
  \renewcommand{\maketitle}{%
    \begin{center}
      {\LARGE\bfseries\@title\par}\medskip
      {\large\@jhepauthors\par}
      \@jhepaffils\par
      \@jhepemails\par
    \end{center}
    \medskip
    \begin{quote}\small\textbf{Abstract.} \@jhepabstract\end{quote}
    \ifx\@jhepkeys\@empty\else
      \par\noindent{\small\textbf{Keywords:} \@jhepkeys}\par
    \fi
    \medskip}
  \makeatother
  \providecommand{\flushbottom}{}
  
\fi

\newcommand{\kperp}{k_{\perp}}
\newcommand{\Dperp}{\Delta_{\perp}}
\newcommand{\qqbar}{|q\bar q\rangle}
\newcommand{\qqbarg}{|q\bar q g\rangle}
\newcommand{\Mz}{\mathcal{M}_0}
\newcommand{\Mt}{\widetilde{M}}

\title{\boldmath Three Dimensional Structure of Pion with One Dynamical Gluon in the Light-Cone Quark Model}
\author[a]{Satyajit Puhan}
\author[b]{Shubham Sharma}
\author[c]{Narinder Kumar}
\affiliation[a]{Institute of Physics,\\Academia Sinica, Taipei, 11529, Taiwan}
\affiliation[b]{Moscow Institute of Physics and Technology,\\Dolgoprudny
141700, Russia}
\affiliation[c]{Computational Theoretical High Energy Physics Lab,\\Department
of Physics, Doaba College, Jalandhar 144004, India}
\emailAdd{puhansatyajit@gmail.com}

\abstract{
We study the three-dimensional structure of the pion beyond the leading
Fock state by including a dynamical gluon,
$|\pi\rangle=|q\bar q\rangle+|q\bar q g\rangle$. Within the light-cone
quark model, we calculate the pion GPDs, TMDs, and PDFs from light-front
wave function overlaps, using a modified Brodsky--Huang--Lepage wave
function with the Melosh--Wigner rotation for the valence sector and the
light-front quark--gluon vertex for the $q\bar q g$ sector. The quark-number
and momentum sum rules are satisfied without additional constraints. 
We further obtain the electromagnetic form factor and the mass decomposition. After NLO evolution, we compare the PDFs with
pion-induced Drell--Yan, $J/\psi$, $\Upsilon$, and HERA leading-neutron
$F_2^\pi$ data. The $q\bar q g$ sector contributes little to Drell--Yan but
becomes important for gluon-sensitive observables such as $J/\psi$
production and the pion structure function at small $x$.
}

\begin{document}
\maketitle
\flushbottom

\section{Introduction}
\label{sec:intro}

Understanding how quarks and gluons build up the observed properties of
hadrons is one of the central problems of quantum chromodynamics
(QCD)~\cite{Gross:1973id,Politzer:1973fx,Gross:2022hyw}. Due to color
confinement, quarks and gluons are never observed as free particles, and the
information about their distribution inside a hadron has to be extracted from
measured cross-sections. This is possible through the QCD factorization
theorems, which separate the short-distance perturbative part of a hard
process from the long-distance non-perturbative part that belongs to the
hadron~\cite{Collins:1989gx}. The non-perturbative part is described by a set
of multidimensional distribution functions. The one-dimensional parton
distribution functions (PDFs) give the probability of finding a quark or a
gluon carrying a longitudinal momentum fraction $x$ of the parent hadron, and
they are extracted from deep inelastic scattering (DIS) and Drell--Yan (DY)
processes~\cite{Collins:1981uw,Holt:2010vj}. The three-dimensional generalized
parton distributions
(GPDs)~\cite{Ji:1996ek,Radyushkin:1997ki,Diehl:2003ny,Belitsky:2005qn}
and transverse-momentum-dependent parton distributions
(TMDs)~\cite{Meissner:2008ay,Angeles-Martinez:2015sea,Boussarie:2023izj} add
the transverse direction to this picture, in position space and in momentum space, respectively. All these distributions can be written as overlaps of the
same light-front wave functions (LFWFs)~\cite{Brodsky:1997de,Brodsky:2000xy}.
Therefore, a model that describes the hadron through its LFWFs should be able
to give all of them in a consistent way.

The GPDs depend on $x$, the skewness $\xi$ and the momentum transfer $t$, and
they can be accessed in hard exclusive processes such as deeply virtual
Compton scattering (DVCS)~\cite{Ji:1996nm} and deeply virtual meson production
(DVMP). At zero skewness, the two-dimensional Fourier transform of the GPDs
with respect to the transverse momentum transfer gives the density of partons
at a transverse distance $b_\perp$ from the center of momentum of the
hadron~\cite{Burkardt:2000za,Burkardt:2002hr}. The first Mellin moment of the
GPDs gives the electromagnetic form factor and the second Mellin moment gives the gravitational form factors $A$ and $B$ from the unpolarized spin non-flip and spin flip GPDs. 
%of The QCD energy-momentum tensor
%(EMT)~\cite{Kobzarev:1962wt,Pagels:1966zza,Ji:1996ek} allows one to decompose the hadron mass into the quark and gluon
%kinetic energies, the quark mass term and the trace anomaly
%contribution~\cite{Ji:1994av,Ji:1995sv,Lorce:2017xzd,Metz:2020vxd}. The trace
%anomaly arises from the breaking of scale invariance by quantum
%corrections~\cite{Collins:1976yq,Nielsen:1977sy}.
The TMDs, on the other hand,
depend on $x$ and the transverse momentum $k_\perp$ of the parton, and they can
be extracted from semi-inclusive DIS and DY
processes~\cite{Angeles-Martinez:2015sea,Boussarie:2023izj}.

Among all the hadrons, the pion has a special place. It is the lightest hadron
and the pseudo-Goldstone boson of the spontaneously broken chiral symmetry of
QCD~\cite{Nambu:1961tp,Goldstone:1961eq}. At the same time, it is a bound
state of a quark and an antiquark, and it mediates the long-range part of the
nucleon-nucleon interaction~\cite{Yukawa:1935xg}. Because of this dual nature,
the pion is an important system for understanding the emergence of hadron mass
in QCD~\cite{Horn:2016rip,Roberts:2021nhw}. Experimentally, the valence quark
distribution of the pion is mainly known from pion-induced DY measurements by
the NA3~\cite{NA3:1983ejh}, NA10~\cite{NA10:1985ibr} and
E615~\cite{E615:1989bda} collaborations. The pion electromagnetic form factor
has been measured through elastic pion-electron scattering~\cite{NA7:1986vav}
and pion electroproduction~\cite{JeffersonLabFpi:2000nlc,JeffersonLab:2008jve}.
The leading-neutron electroproduction data from HERA provide additional
information on the pion structure function at small $x$ through the Sullivan
process~\cite{Sullivan:1971kd,ZEUS:2002gig,H1:2010hym,Chang:2026tgb}. The gluon
distribution of the pion, however, is still poorly known. In the DY process, the gluon enters only at next-to-leading order in $\alpha_s$, and the global
analysis of the pion
PDFs~\cite{Gluck:1991ey,Barry:2018ort,Novikov:2020snp,Barry:2021osv}
give quite different values for the gluon momentum fraction. Charmonium
production in pion-nucleon collisions is dominated by gluon fusion at
fixed-target energies, and it is therefore more sensitive to the pion
gluon~\cite{Chang:2020rdy,Chang:2022pcb}. The gluon PDF and the gluon momentum
fraction of the pion have also been computed recently in lattice
QCD~\cite{Fan:2021bcr,ExtendedTwistedMass:2024kjf}. The upcoming Electron-Ion
Collider
(EIC)~\cite{Accardi:2012qut,AbdulKhalek:2021gbh,Aguilar:2019teb,Arrington:2021biu},
the Electron-Ion Collider in China (EicC)~\cite{Anderle:2021wcy} and the AMBER
experiment at CERN~\cite{Adams:2018pwt} are expected to improve our knowledge
of the quark and gluon distributions of the pion considerably. This makes a
model prediction for the pion gluon distribution timely.

The structure of the pion has been studied in many theoretical approaches,
such as lattice QCD~\cite{Shanahan:2018pib,Lin:2023gxz,Hackett:2023nkr,Ding:2024saz},
Dyson--Schwinger equations~\cite{Ding:2019lwe,Cui:2020tdf,Cui:2022bxn}, the
Nambu--Jona-Lasinio model~\cite{Zhang:2021uak}, AdS/QCD and light-front
holographic models~\cite{deTeramond:2008ht,Kaur:2018ewq}, basis light-front
quantization (BLFQ)~\cite{Vary:2009gt,Lan:2021wok} and light-cone
quark models
(LCQM)~\cite{Puhan:2023ekt,Choi:2025rto,Puhan:2025kzz,Puhan:2024jaw}. In the
LCQM, the pion is described by a Brodsky--Huang--Lepage (BHL)-type momentum-space wave function~\cite{Brodsky:1981jv,Huang:1994dy,Dwibedi:2025vhr,Dwibedi:2026ozl,Puhan:2023ekt,Puhan:2025pfs}, and its spin structure is
obtained through the Melosh--Wigner
rotation~\cite{Melosh:1974cu,Terentev:1976jk,Jaus:1989au,Choi:1997iq}. In our
previous works, we have used this framework to study the TMDs, GPDs,
generalized transverse-momentum-dependent distributions (GTMDs), form factors and PDFs
of the pion and other
mesons~\cite{Puhan:2023ekt,Puhan:2024jaw,Acharyya:2024enp,Puhan:2025kzz,Puhan:2025pfs,P:2026crg,Dwibedi:2026ozl}.
The gluon
contribution to the pion mass has been discussed in Ref.~\cite{Han:2024yzj}.

Most of these model calculations, including our earlier works, describe the
pion only through its leading $|q\bar q\rangle$ Fock state. In that case, the gluon distribution vanishes at the model scale and is generated completely by
QCD evolution. This puts a strong assumption on the model scale and makes it
difficult to describe gluon-sensitive observables. The pion state, however,
also contains higher Fock components. In the BLFQ approach, the $|q\bar
qg\rangle$ sector is coupled to the valence sector through the light-front QCD
Hamiltonian, and it carries a large part of the pion
normalization~\cite{Lan:2021wok,Zhu:2023lst,Lan:2024ais,Lan:2025qio}. The role
of higher Fock states has also been shown in a phenomenological analysis where
the $q\bar q$, $q\bar qq\bar q$, $q\bar qg$ and $q\bar qgg$ components are
fitted to the pion PDF and form factor data~\cite{Pasquini:2023aaf}. The pion
gluon PDF has also been obtained from the gluon dressing of the
quarks~\cite{Freese:2021zne}, and in a light-front model where the active
gluon recoils against a spectator system~\cite{Kaur:2025gyr}. A light-cone
quark model with an explicit dynamical gluon, which gives the quark and gluon
distributions together with the form factor and the mass decomposition from the same
wave function, is still missing.

In this work, we study the three-dimensional structure of the pion in the LCQM
by including one dynamical gluon in the pion state. We truncate the Fock
expansion as $|\pi\rangle=|q\bar q\rangle+|q\bar q g\rangle$. The valence
sector is described by a modified BHL wave function with the Melosh--Wigner
rotation, while the $|q\bar q g\rangle$ wave function is generated from the
valence one through the light-front quark--gluon vertex in the light-cone
gauge~\cite{Zhang:1994ti,Harindranath:1996hq}. The strong coupling is fixed by
$\alpha_s$ at the model scale, so the three-body sector does not introduce an
independent strength. Using this wave function, we calculate the unpolarized
quark and gluon TMDs, the zero-skewness GPDs with their impact-parameter dependent parton
distributions, and the PDFs. From the GPDs, we also obtain the electromagnetic form
factor. The quark and gluon PDFs are evolved to higher scales with the next-to-leading order DGLAP
equations~\cite{Dokshitzer:1977sg,Gribov:1972ri,Altarelli:1977zs} using
HOPPET~\cite{Salam:2008qg,Karlberg:2025hxk}. The evolved PDFs are then used to
calculate the cross sections of pion-induced DY, $J/\psi$ and $\Upsilon$
production, as well as the pion structure function $F_2^\pi$ from leading-neutron production at HERA, and we compare them with the available data.

The paper is organized as follows. In Sec.~\ref{sec:lfqm}, we present the
light-cone quark model with the $|q\bar q\rangle$ and $|q\bar q g\rangle$ Fock
sectors. The unpolarized TMDs are discussed in Sec.~\ref{sec:tmd}. In
Sec.~\ref{sec:gpd}, we present the GPDs and the electromagnetic form factor of
the pion. The PDFs and their
evolution are discussed in Sec.~\ref{sec:pdf}. The pion-induced processes are presented
in Sec.~\ref{sec:pheno}. Finally, we summarize our results in
Sec.~\ref{sec:conclusion}.

\section{Light-Cone Quark Model}\label{sec:lfqm}
In light-front quantization, \footnote{The light-front conventions, the definition of the intrinsic transverse momentum and the notation
used throughout are given in Appendix~\ref{app:conv}.} the meson eigenstate $|M\rangle$ is expanded on the null plane $x^+=0$ in a Fock-state series that includes quarks, gluons, and sea quarks as~\cite{Brodsky:1997de,Lepage:1980fj}
\begin{equation}
|M\rangle = \sum \psi_{q\bar q}|q\bar q\rangle + \sum \psi_{q\bar q g}|q\bar q g\rangle + \sum \psi_{q\bar q g g}|q\bar q g g\rangle + \sum \psi_{q\bar q q^\prime \bar q^\prime}|q\bar q q^\prime \bar q^\prime\rangle+\cdots,
\label{eq:fockseries}
\end{equation}
where each Fock component is described by a corresponding light-front wave function (LFWF) $\psi$, which is invariant under kinematic light-front boosts and therefore provides a frame-independent description of hadron structure.
The meson Fock-state expansion can be expressed explicitly in terms of the kinematic variables of its individual constituents. For a meson with four-momentum $P=(P^+,P^-,\bm{P}_\perp)$, the expansion reads
\begin{equation}
 \begin{aligned}
 |M(P,J_z)\rangle ={}& \sum_{n,\lambda_i}
 \int \prod_{i=1}^{n}
 \frac{\mathrm{d}x_i\, \mathrm{d}^2 \bm{k}_{\perp i}}{{16\pi^3}\sqrt{x_i}}
 \,16\pi^3\,\delta\!\Big(1-\sum_{i=1}^{{n}} x_i\Big)\,
 \delta^{(2)}\!\Big(\sum_{i=1}^{{n}}\bm{k}_{\perp i}\Big)
 \\
 &\times\Psi_{n}(x_i,\bm{k}_{\perp i},\lambda_i)\,
 |{n};k_i^+,\bm{k}_{\perp i},\lambda_i\rangle ,
 \end{aligned}
 \label{eq:fock}
\end{equation}
where, $x_i=k_i^+/P^+$ and $\bm{k}_{\perp i}$ are the longitudinal momentum fraction and intrinsic transverse momentum of the $i$-th constituent, respectively. The constituent four-momentum is given by $k_i=(k_i^+,k_i^-,\bm{k}_{\perp i})$, and $\lambda_i$ denotes its light-front helicity. $\Psi_{n}(x_i,\bm{k}_{\perp i},\lambda_i)$ is the $n$-particle light-front wave function, which encodes the radial and spin-orbit properties of the state, subject to the momentum-conservation constraints $\sum_i x_i=1$ and $\sum_i \bm{k}_{\perp i}=0$. In this work, $\Psi_2\equiv\Psi^{\pi}_{q\bar q}$ and $\Psi_3\equiv\Psi_{q\bar q g}$. Here, $J_z$ is the spin projection of the hadron, which is zero for the pseudoscalar meson.

 Most LCQM calculations of the pion truncate Eq.~\eqref{eq:fockseries} after the first term~\cite{Puhan:2023ekt,Puhan:2025kzz,Puhan:2024jaw}. A pure $\qqbar$ state has no gluon and no sea at the model scale, so all of the gluon and sea content of the pion has to be generated by QCD evolution. In this work, we keep the second term of Eq.~\eqref{eq:fockseries} for the pion ($u\bar{d}$), and the truncated expansion reads
\begin{equation}
 |\pi^+(P,J_z=0)\rangle = {\sqrt{n}} \Big[ \psi_{q\bar q}|q\bar q\rangle + \psi_{q\bar q g}|q\bar q g\rangle \Big] ,
 \label{eq:truncation}
\end{equation}
where $\sqrt{n}$ is the normalization constant. The strong coupling of the emission vertex is not a free parameter. It is tied to the coupling that drives the evolution through Eq.~\eqref{eq:gslock} below.

\subsection{The valence sector}
\label{sec:valence}
The minimal Fock state of a pseudoscalar meson is given by \cite{Pasquini:2023aaf,Ji:2003yj,Kaur:2018ewq,Puhan:2025kzz}
\begin{equation}
 |\pi^+(P,J_z=0)\rangle_{q\bar q} =
 \sum_{s_q,s_{\bar q}}
 \int \frac{\mathrm{d}x\,\mathrm{d}^2\bm{\kperp}}{16\pi^3\sqrt{x(1-x)}}\,
 \Psi_{q \bar q}^{\pi}(x,\bm{\kperp},s_q,s_{\bar q})\,
 |xP^+,\bm{\kperp},s_q,s_{\bar q}\rangle ,
 \label{eq:valstate}
\end{equation}
where the quark and antiquark four-momenta are defined as
\begin{equation}
 k_q \equiv \Big(xP^+, \frac{\bm{\kperp}^2+m_q^2}{xP^+}, \bm{\kperp}\Big),
 \qquad
 k_{\bar q} \equiv \Big((1-x)P^+, \frac{\bm{\kperp}^2+m_{\bar q}^2}{(1-x)P^+}, -\bm{\kperp}\Big),
 \label{eq:kqkqb}
\end{equation}
and the free invariant mass of the pair is
\begin{equation}
 \Mz^2(x,\bm{\kperp}) = \frac{\bm{\kperp}^2+m_q^2}{x} + \frac{\bm{\kperp}^2+m_{\bar q}^2}{1-x} .
 \label{eq:M0}
\end{equation}
The LFWF $\Psi^{\pi}_{q \bar q}(x,\bm{\kperp},s_q,s_{\bar q})$ splits into a spin-orbit wave function $J(x,\bm{\kperp},s_q,s_{\bar q})$ and a radial wave function $\phi^{\pi}(x,\bm{\kperp})$ as
\begin{equation}
 \Psi_{q \bar q}^{\pi}(x,\bm{\kperp},s_q,s_{\bar q}) = J(x,\bm{\kperp},s_q,s_{\bar q})\phi^{\pi}(x,\bm{\kperp}) .
 \label{eq:factorise}
\end{equation}
For the radial wave function, we use the Brodsky--Huang--Lepage (BHL) ansatz~\cite{Brodsky:1981jv}, which expresses the amplitude as a Gaussian in the free invariant mass of the Fock state as
\begin{equation}
 \phi^{\pi}(x,\bm{\kperp}) = \frac{A_\pi}{\sqrt{2x(1-x)}} 
 \exp\!\left[ -\frac{\Mz^2(x,\bm{\kperp})}{8\beta^2} + \frac{m_q^2+m_{\bar q}^2}{4\beta^2} \right] ,
 \label{eq:bhl}
\end{equation}
where $\beta$ is the harmonic scale and $A_\pi$ is fixed by the normalization. Both are given in Table~\ref{tab:params}. The prefactor $[2x(1-x)]^{-1/2}$ is the Terent'ev--Karmanov static Jacobian of the BHL wave function, and it gives the proper endpoint behavior of the pion. Other choices, such as using the exact light-front Jacobian $\sqrt{\mathrm{\partial}k_z/\mathrm{\partial}x} = \sqrt{\Mz/(4x(1-x))}$ or omitting the Jacobian entirely, differ from Eq.~\eqref{eq:bhl} only in this $x$-dependent prefactor and leave the Gaussian unchanged~\cite{Puhan:2025ujg}.

The spin-orbit wave function defined in Eq.~\eqref{eq:factorise} can also be obtained from the quark--antiquark vertex~\cite{Choi:1996mq,Qian:2008px,Dwibedi:2025vhr,Puhan:2025ujg,Puhan:2025kzz}
\begin{eqnarray}
 J(x,\bm{\kperp},s_q,s_{\bar q}) = \frac{\bar{u}(k_q,{s_q}) \gamma_5 v(k_{\bar q},{s_{\bar q}})}{\sqrt{2}\big[\sqrt{\Mz^2 - (m_q - m_{\bar q})^2}\big]} .
 \label{eq:vertex}
\end{eqnarray}
Equivalently, the $J_z=0$ instant-form singlet is transformed to the light-front form via the Melosh--Wigner rotation. Defining $k^{R,L}=k^1\pm i k^2$, $\Mt=(1-x)m_q+x m_{\bar q}$, and 
\begin{eqnarray}
 \omega &=& (\Mz+m_q+m_{\bar q}) \sqrt{x(1-x)\big[\Mz^2-(m_q-m_{\bar q})^2\big]} ,
 \nonumber\\
 \mathcal{N} &=& \frac{\Mz+m_q+m_{\bar q}}{\sqrt{2}\omega} ,
 \label{eq:NN}
\end{eqnarray}
the four helicity amplitudes $J(x,\bm{\kperp},{s_q},{s_{\bar q}})$ are
\begin{equation}
 \begin{aligned}
 J(\uparrow,\uparrow) &= -\mathcal{N} k^L, (L_z=-1) &\qquad
 J(\uparrow,\downarrow) &= +\mathcal{N} \Mt, (L_z=0) \\
 J(\downarrow,\uparrow) &= -\mathcal{N} \Mt, (L_z=0) &\qquad
 J(\downarrow,\downarrow) &= -\mathcal{N} k^R, (L_z=+1).
 \end{aligned}
 \label{eq:melosh}
\end{equation}
With the normalization of Eq.~\eqref{eq:NN}, the spin-orbit wave function is a unit vector,
\begin{equation}
 \sum_{s_q,s_{\bar q}} \big|J(s_q,s_{\bar q})\big|^2
 = \frac{(\bm{\kperp}^2+\Mt^2)(\Mz+m_q+m_{\bar q})^2}{\omega^2}
 = 1 .
 \label{eq:meloshnorm}
\end{equation}
The amplitudes of Eq.~\eqref{eq:melosh} contain two independent radial functions,
\begin{equation}
 U(x,\kperp) = \kperp\mathcal{N}\phi^\pi ,
 \qquad
 S(x,\kperp) = \Mt\mathcal{N}\phi^\pi ,
 \label{eq:US}
\end{equation}
and the total probability density is $\sum_{s_q,s_{\bar q}}|\Psi^\pi_{q\bar q}|^2 = 2(U^2+S^2) = (\phi^\pi)^2$. For a pion with $J_z=0$, angular momentum conservation requires $s_q+s_{\bar q}+L_z=0$, with $s_q,s_{\bar q}=\pm\tfrac12$ written as $\uparrow,\downarrow$ or $\pm$ in the Table \ref{tab:Lz_configurations}.

\subsection{\texorpdfstring{The $\qqbarg$ sector}{The $|q qbar g \rangle $sector}}
\label{sec:qqg}
To have a gluon already at the model scale, we keep the next Fock sector and calculate the dynamical gluon explicitly. Following Brodsky and collaborators~\cite{Brodsky:1981jv,Brodsky:1997de}, and the later BLFQ studies~\cite{Lan:2025qio,Lan:2024ais}, we describe the pion as an eigenstate of the light-front invariant-mass operator,
\begin{equation}
 H_{\mathrm{LC}} |\pi(P)\rangle = M_\pi^2 |\pi(P)\rangle ,
 \qquad
 H_{\mathrm{LC}} = P^{+}P^{-} - \bm{P}_\perp^2 ,
 \label{eq:lceig}
\end{equation}
where $P^{-}$ is the light-front QCD Hamiltonian in the light-cone gauge ($A^{+}=0$). We split this Hamiltonian into a free part $H_0$ and an interaction part $H_I$, $H_{\mathrm{LC}} = H_0 + H_I$. The free part is diagonal in the Fock basis and gives the free invariant mass of the state,
\begin{equation}
 H_0 |{\pi}\rangle = {\mathcal{M}_{0,n}^2} |{\pi}\rangle ,
 \qquad
 {\mathcal{M}_{0,n}^2} = \sum_{j=1}^{{n}} \frac{\bm{k}_{\perp j}^2+m_j^2}{x_j} ,
 \label{eq:h0}
\end{equation}
while $H_I$ changes the particle content. Among its interaction terms, we keep only the quark--gluon vertex ($q \to qg$). The instantaneous four-point terms first contribute at the next order in $g_s$, and the $g \to gg$ vertex first contributes in the next Fock sector.

Writing the expansion of Eq.~\eqref{eq:fock} as a column vector $(\psi_{\qqbar},\psi_{\qqbarg},\dots)$ and projecting Eq.~\eqref{eq:lceig} onto each sector turns the eigenvalue problem into an infinite coupled system,
\begin{equation}
 \begin{aligned}
% \big(M_\pi^2 - {\mathcal{M}_{0,2}^2}\big) \psi_{\qqbar} 
 %&= \mathcal{V}_{22}\psi_{\qqbar} + \mathcal{V}_{23}\psi_{\qqbarg} + \mathcal{V}_{24}\psi_{4} + \cdots , \\
 \big(M_\pi^2 - {\mathcal{M}_{0,3}^2}\big) \psi_{\qqbarg} 
 &= \mathcal{V}_{32}\psi_{\qqbar} + \mathcal{V}_{33}\psi_{\qqbarg} + \mathcal{V}_{34}\psi_{4} + \cdots ,
 \end{aligned}
 \label{eq:coupled}
\end{equation}
where $\mathcal{V}_{{NN'}} = \langle {N}|H_I|{N'}\rangle$ and ${\mathcal{M}_{0,2}^2} \equiv \Mz^2$ from Eq.~\eqref{eq:M0}. This system is exact but infinite. To obtain the $|q\bar{q}\rangle + |q\bar{q}g\rangle$ state, we make two approximations.

The first is the Fock-space truncation itself. The tower is cut after the $\qqbarg$ sector, so $\psi_4$ and everything above it are dropped. The second is an ordering in the strong coupling $g_s$. The transition matrix elements $\mathcal{V}_{32}$ and $\mathcal{V}_{23}$ carry one power of $g_s$, so $\psi_{\qqbarg}$ is of order $g_s\psi_{\qqbar}$. The diagonal element $\mathcal{V}_{33}$ comes from the instantaneous interactions and is of order $g_s^2$. The term $\mathcal{V}_{33}\psi_{\qqbarg}$ is therefore of order $g_s^3\psi_{\qqbar}$, two powers of $g_s$ below the term $\mathcal{V}_{32}\psi_{\qqbar}$, and is dropped at leading order. The second line of Eq.~\eqref{eq:coupled} then becomes
\begin{equation}
 \big(M_\pi^2 - {\mathcal{M}_{0,3}^2}\big) \psi_{\qqbarg} = \mathcal{V}_{32}\psi_{\qqbar} .
 \label{eq:secondline}
\end{equation}
Since ${\mathcal{M}_{0,3}^2} > M_\pi^2$ for any configuration of three massive constituents, the factor in parentheses is negative and never vanishes, and we can divide by it. In agreement with Ref.~\cite{Lan:2024ais}, the $\qqbarg$ wave function is
\begin{equation}
 \psi^{(s_q,s_{\bar q},\lambda)}_{q\bar q g} = \frac{1}{M_\pi^2-{\mathcal{M}_{0,3}^2}} \big[ \mathcal{V} \psi^{(s_q,s_{\bar q})}_{q\bar q} \big] ,
 \qquad
 {\mathcal{M}_{0,3}^2} = \sum_{j=q,\bar q,g} \frac{\bm{k}_{\perp j}^2+m_j^2}{x_j} ,
 \label{eq:qqg}
\end{equation}
where $m_g$ is the dressed gluon mass. 
The BHL ansatz of Eq.~\eqref{eq:bhl} takes its place, and the harmonic scale $\beta$ absorbs the effect of the interaction in the bracket. Equation~\eqref{eq:qqg} is then used to generate the three-body sector that was integrated out. The vertex that builds the three-body sector is the one whose exchange binds the two-body sector, so $g_s$ cannot be chosen independently of $\beta$.
In the light-cone gauge, the interaction $\mathcal{V}$ that couples the two sectors gives, after the sum over the colors of a color-singlet $q\bar q$ pair,
\begin{equation}
 \begin{aligned}
 \Psi_3^{(s_q s_{\bar q}\lambda)}(x_q,\bm{k}_{q\perp};x_g,\bm{k}_{g\perp})
 =\;&
 \;
 \frac{g_s \ \sqrt{2 C_F}}{M_\pi^2-{\mathcal{M}_{0,3}^2}}
 \\ &\times
 \Bigg\{
 \frac{1}{\sqrt{x_q x_g}\,\sqrt{x_1}}
 \sum_{s_1} W^{(s_q,s_1)}_{\lambda}(x_q,x_g,\bm{\kappa})\,
 {\Psi^{\pi}_{q\bar q}(x_1,\bm{k}_{1\perp},s_1,s_{\bar q})}
 \\ &\qquad -\;
 \frac{1}{\sqrt{x_{\bar q} x_g}\,\sqrt{\bar x_1}}
 \sum_{s_1} \overline W^{(s_1,s_{\bar q})}_{\lambda}
 (x_{\bar q},x_g,\bar{\bm{\kappa}})\,
 {\Psi^{\pi}_{q\bar q}(x_q,\bm{k}_{q\perp},s_q,s_1)}
 \Bigg\} ,
 \end{aligned}
 \label{eq:psi3}
\end{equation}
with
\begin{equation}
 \begin{gathered}
 x_1 = x_q+x_g, \qquad \bm{k}_{1\perp}= \bm{k}_{q\perp}+\bm{k}_{g\perp}
 = -\bm{k}_{\bar q\perp}, \qquad
 \bm{\kappa} = x_q \bm{k}_{g\perp} - x_g \bm{k}_{q\perp}, \\
 \bar x_1 = x_{\bar q}+x_g, \qquad
 \bar{\bm{k}}_{1\perp} = -\bm{k}_{q\perp}, \qquad
 \bar{\bm{\kappa}} = x_{\bar q}\bm{k}_{g\perp}-x_g\bm{k}_{\bar q\perp} .
 \end{gathered}
 \label{eq:kin3}
\end{equation}
where, $\Psi_3\equiv\Psi_{q\bar q g}$ denotes this three-body LFWF, $s_q$ and $s_{\bar q}$ the quark and antiquark helicities, and $\lambda$ the gluon helicity. Equation~\eqref{eq:psi3} is obtained  using the Lepage--Brodsky (LB) normalization~\cite{Lepage:1980fj}. The
color factor $C_F=4/3$ is common to the direct and interference terms for a
color-singlet $q\bar q$ pair. The arguments of the valence wave function are
those of the \emph{spectator} in each emission.

\paragraph{The quark--gluon vertex.}
For the quark--gluon vertex, we use the LB spinors~\cite{Lepage:1980fj,Brodsky:1997de},
\begin{equation}
 u_{{s}}(p) = \frac{p^+ + \beta_{\mathrm{LB}} m + \bm{\alpha}_\perp \cdot \bm{p}_\perp}{\sqrt{p^+}} X_{{s}} ,
 \qquad
 v_{{s}}(p) = \frac{p^+ - \beta_{\mathrm{LB}} m + \bm{\alpha}_\perp \cdot \bm{p}_\perp}{\sqrt{p^+}} X_{{-s}} ,
 \label{eq:spinors}
\end{equation}
where $\beta_{\mathrm{LB}} = \gamma^0$, written this way to distinguish it from the harmonic scale $\beta$. Here, $\alpha^i = \gamma^0\gamma^i$, and the basis spinors in the Dirac representation are $X_\uparrow = \frac{1}{\sqrt{2}}(1,0,1,0)^{T}$ and $X_\downarrow = \frac{1}{\sqrt{2}}(0,1,0,-1)^{T}$. The polarization vectors in the light-cone gauge ($A^+=0$) are given by
\begin{equation}
 \epsilon^\mu_\lambda(k) = \left( \epsilon^+=0, \epsilon^-=\frac{2\bm{\epsilon}_{\perp\lambda} \cdot \bm{k}_\perp}{k^+}, \bm{\epsilon}_{\perp\lambda} \right) ,
 \qquad
 \bm{\epsilon}_{\perp\pm} = \mp\frac{1}{\sqrt{2}}(1,\pm i) ,
 \label{eq:pol}
\end{equation}
which yields $\slashed{\epsilon}^{\,*} = \frac{1}{2}\epsilon^{-*}\gamma^+ - \bm{\epsilon}^{\,*}_\perp \cdot \bm{\gamma}_\perp$. The quark--gluon vertex functions are then defined as
\begin{align}
 W^{(s_q,s_1)}_\lambda(p_q,p_g) &= \frac{1}{\sqrt{2}} \bar{u}_{s_q}(p_q) \slashed{\epsilon}^{\,*}_\lambda(p_g) u_{s_1}(p_1) , 
 \label{eq:Wdef} \\
 \overline{W}^{(s_1,s_{\bar q})}_\lambda(p_{\bar q},p_g) &= \frac{1}{\sqrt{2}} \bar{v}_{s_1}(p_1) \slashed{\epsilon}^{\,*}_\lambda(p_g) v_{s_{\bar q}}(p_{\bar q}) , 
 \label{eq:Wbardef}
\end{align}
where the parent momentum is $p_1=p_q+p_g$ (or $p_1=p_{\bar q}+p_g$ for the antiquark vertex), with the longitudinal fractions $x_1=x_q+x_g$ ($\bar x_1=x_{\bar q}+x_g$) and transverse momenta $\bm{k}_{1\perp} = \bm{k}_{q\perp} + \bm{k}_{g\perp}$, as in Eq.~\eqref{eq:kin3}.
\begin{table*}[t]
\centering
\renewcommand{\arraystretch}{1.6}
\begin{tabular}{c|l|l}
\hline
$L_z$ 
& \multicolumn{1}{c|}{$q\bar q$ configurations:
$|J_z=0\rangle\rightarrow |{s_q}\,{s_{\bar q}}\rangle$}
& \multicolumn{1}{c}{$q\bar q g$ configurations:
$|J_z=0\rangle\rightarrow |{s_q}\,{s_{\bar q}}\,{\lambda}\rangle$}
\\
\hline

$-2$
&
&
$|0\rangle\rightarrow
\left|+\frac{1}{2}\,+\frac{1}{2}\,+1\right\rangle$
\\[6pt]

$-1$
&
$|0\rangle\rightarrow
\left|+\frac{1}{2}\,+\frac{1}{2}\right\rangle$
&
$|0\rangle\rightarrow
\left|+\frac{1}{2}\,-\frac{1}{2}\,+1\right\rangle$,
\quad
$|0\rangle\rightarrow
\left|-\frac{1}{2}\,+\frac{1}{2}\,+1\right\rangle$
\\[6pt]

$0$
&
$|0\rangle\rightarrow
\left|+\frac{1}{2}\,-\frac{1}{2}\right\rangle$,
\quad
$|0\rangle\rightarrow
\left|-\frac{1}{2}\,+\frac{1}{2}\right\rangle$
&
$|0\rangle\rightarrow
\left|+\frac{1}{2}\,+\frac{1}{2}\,-1\right\rangle$,
\quad
$|0\rangle\rightarrow
\left|-\frac{1}{2}\,-\frac{1}{2}\,+1\right\rangle$
\\[6pt]

$+1$
&
$|0\rangle\rightarrow
\left|-\frac{1}{2}\,-\frac{1}{2}\right\rangle$
&
$|0\rangle\rightarrow
\left|+\frac{1}{2}\,-\frac{1}{2}\,-1\right\rangle$,
\quad
$|0\rangle\rightarrow
\left|-\frac{1}{2}\,+\frac{1}{2}\,-1\right\rangle$
\\[6pt]

$+2$
&
&
$|0\rangle\rightarrow
\left|-\frac{1}{2}\,-\frac{1}{2}\,-1\right\rangle$
\\
\hline
\end{tabular}
\caption{Allowed $L_z$ of the $q\bar q$ and $q\bar q g$ helicity configurations, fixed by the rule $L_z+s_q+s_{\bar q}(+\lambda)=0$.}
\label{tab:Lz_configurations}
\end{table*}

The Dirac algebra is carried out in terms of the relative transverse momentum of the splitting of Eq.~\eqref{eq:kin3} and its complex components,
\begin{equation}
{\bm{\kappa}}=x_q{\bm{k}_{g\perp}}-x_g{\bm{k}_{q\perp}},\qquad
\kappa^{R,L}=\kappa^1\pm i\kappa^2 .
\label{eq:kappa}
\end{equation}
The LFWFs satisfy angular-momentum conservation along the $z$-axis, i.e., $J_z=\sum_{i=1}^{{N}}s_{iz}+\sum_{j=1}^{{N}-1}L_{jz}$, where ${N}=2$ for the $q\bar q$ Fock sector and ${N}=3$ for the $q\bar q g$ sector. The intrinsic spin contribution is given by the sum of the constituent helicities, while $L_z$ denotes the total relative orbital angular momentum for a given LFWF configuration. For the $q\bar q$ sector, $s_{1z}$ and $s_{2z}$ correspond to the quark and antiquark helicities, ${s_q}$ and ${s_{\bar q}}$, respectively, such that $L_z=J_z-({s_q}+{s_{\bar q}})$. For the $q\bar q g$ sector, the gluon helicity ${\lambda}$ also contributes to the intrinsic spin, giving $L_z=J_z-({s_q}+{s_{\bar q}}+{\lambda})$. For the pion, $J_z=0$, and hence the allowed $L_z$ values are determined by the corresponding helicity configurations. All possible configurations of the $q\bar q$ and $q\bar q g$ Fock sectors, together with their corresponding $L_z$ values, are presented in Table~\ref{tab:Lz_configurations}. By explicitly evaluating Eqs.~\eqref{eq:Wdef} and \eqref{eq:Wbardef} over all combinations of gluon polarizations and quark--antiquark helicities, we find that the non-vanishing entries take the form
\begin{equation}
 \begin{aligned}
 W^{(+,+)}_{+1} &= -\frac{\sqrt{x_1}\kappa^L}{x_g\sqrt{x_q}} , &\qquad
 W^{(+,+)}_{-1} &= +\frac{\sqrt{x_q}\kappa^R}{x_g\sqrt{x_1}} , \\
 W^{(-,-)}_{+1} &= -\frac{\sqrt{x_q}\kappa^L}{x_g\sqrt{x_1}} , &\qquad
 W^{(-,-)}_{-1} &= +\frac{\sqrt{x_1}\kappa^R}{x_g\sqrt{x_q}} , \\
 W^{(-,+)}_{+1} &= W^{(+,-)}_{-1} = \frac{m_f x_g}{\sqrt{x_qx_1}} , & &
 \end{aligned}
 \label{eq:Wtable}
\end{equation}
and, with the relative transverse momentum $\bar{\bm{\kappa}} = x_{\bar q}\bm{k}_{g\perp} - x_g\bm{k}_{\bar q\perp}$ and the longitudinal fraction $\bar x_1 = x_{\bar q} + x_g$ of Eq.~\eqref{eq:kin3}, the antiquark vertex elements evaluate to
\begin{equation}
 \begin{aligned}
 \overline{W}^{(+,+)}_{+1} &= -\frac{\sqrt{\bar x_1}\bar\kappa^L}{x_g\sqrt{x_{\bar q}}} , &\qquad
 \overline{W}^{(+,+)}_{-1} &= +\frac{\sqrt{x_{\bar q}}\bar\kappa^R}{x_g\sqrt{\bar x_1}} , \\
 \overline{W}^{(-,-)}_{+1} &= -\frac{\sqrt{x_{\bar q}}\bar\kappa^L}{x_g\sqrt{\bar x_1}} , &\qquad
 \overline{W}^{(-,-)}_{-1} &= +\frac{\sqrt{\bar x_1}\bar\kappa^R}{x_g\sqrt{x_{\bar q}}} , \\
 \overline{W}^{(+,-)}_{+1} &= \overline{W}^{(-,+)}_{-1} = \frac{m_f x_g}{\sqrt{x_{\bar q}\bar x_1}} . & &
 \end{aligned}
 \label{eq:Wbartable}
\end{equation}
Here, $m_f$ is the vertex mass, which replaces the constituent quark mass $m_q$ in the helicity-flip entries. It is a separate parameter because the helicity-flip amplitude is the only part of the quark--gluon vertex that is not fixed by the kinematics of the constituents. $m_f$ thus describes the non-perturbative enhancement of the spin-flip matrix element, which is responsible for the $\pi$--$\rho$ mass splitting~\cite{Lan:2021wok,Lan:2024ais}. The vertex functions of Eqs.~\eqref{eq:Wtable} and \eqref{eq:Wbartable} respect parity,
\begin{equation}
 \big|W^{(s,s_1)}_\lambda\big| = \big|W^{(-s,-s_1)}_{-\lambda}\big| ,
 \label{eq:parity}
\end{equation}
and the antiquark vertex $\overline{W}$ has the same structure under $x_q\to x_{\bar q}$ and $\bm{\kappa}\to\bar{\bm{\kappa}}$, with its helicity-flip entries at $\overline{W}^{(+,-)}_{+1}=\overline{W}^{(-,+)}_{-1}$.

\begin{figure}[t]
 \centering
 \includegraphics[width=0.99\textwidth]{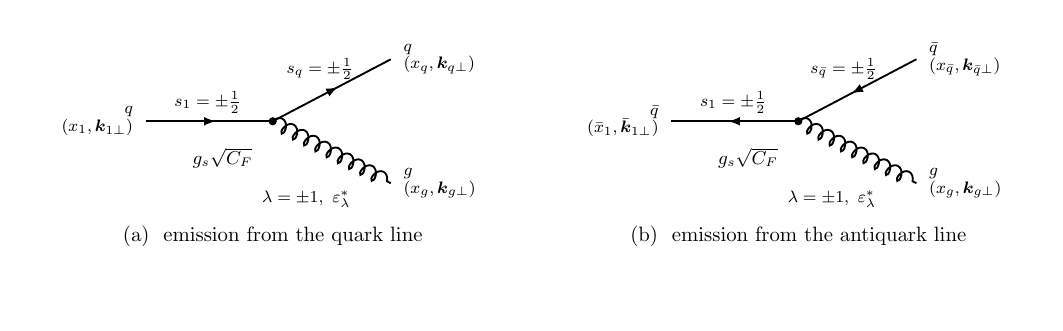}
 \caption{Light-front quark--gluon vertex for gluon emission from (a) the quark and (b) the antiquark line. The helicity-conserving terms scale as $\kappa^{R,L}/x_g$ and vanish in the collinear limit, while the helicity-flip terms are proportional to the vertex mass $m_f$.}
 \label{fig:vertexfeynman}
\end{figure}

The two emission processes are illustrated in Fig.~\ref{fig:vertexfeynman}. The relative transverse momentum $\bm{\kappa}$ vanishes when the gluon is collinear with its parent, so the helicity-conserving entries switch off in that limit and only the $m_f$ term survives. This is why the vertex mass controls the size of the $\qqbarg$ sector, and it links that size to the parameter behind the $\pi$--$\rho$ splitting.

Two properties of Eq.~\eqref{eq:Wtable} should be noted. First, because of the parity relation of Eq.~\eqref{eq:parity}, the amplitudes are kept for both gluon helicities. Reference~\cite{Lan:2024ais} gives a vanishing spin-flip element $\overline{W}^{(-,+)}_{\lambda}=0$. Our explicit evaluation shows that this entry vanishes only for $\lambda=+1$. The $\lambda=-1$ entry is finite, and we keep it. Otherwise, our expressions agree with Ref.~\cite{Lan:2024ais}.

Second, we checked the vertex expressions in two independent ways. For $m_f\to0$, $x_1\to1$ and $\bm{k}_{1\perp}=0$, the squared amplitudes summed over all helicities are proportional to the Altarelli--Parisi splitting function
\begin{equation}
 P_{gq}(y) = C_F \frac{1+(1-y)^2}{y} ,
 \label{eq:AP}
\end{equation}
where $y = x_g/x_1$. We also compared the entries of Eq.~\eqref{eq:Wtable} with a numerical evaluation of the spinor contractions of Eq.~\eqref{eq:Wdef}. The color weight is the same for the two direct terms and the interference
term. With the valence singlet $N_c^{-1/2}\sum_i|i\bar i\rangle$, emission
from the quark carries $T^{a}_{i_qi_{\bar q}}/\sqrt{N_c}$ and emission from the
antiquark $-T^{a}_{i_qi_{\bar q}}/\sqrt{N_c}$, the antiquark generator being
$-(T^a)^{T}$, so that summing the square over $a$, $i_q$ and $i_{\bar q}$,
\begin{equation}
 \frac{1}{N_c}\sum_a\Tr\big(T^aT^a\big)=\frac{N_c^2-1}{2N_c}=C_F=\frac43
 \label{eq:colour}
\end{equation}
in all three cases. $C_F$ therefore factors out of Eq.~\eqref{eq:psi3}, and
only the relative minus sign remains, which reflects the color coherence of a
neutral dipole. The coupling of the vertex is fixed by the strong coupling at
the model scale,
\begin{equation}
 g_s = \sqrt{4\pi\alpha_s(\mu_0)} .
 \label{eq:gslock}
\end{equation}
The model scale $\mu_0$ is determined below from the E615 data \cite{E615:1989bda}. The quark mass $m_q$ and the harmonic scale $\beta$ are determined from the decay constant
and the charge radius of the pion. The decay constant is defined by the matrix element of the
axial-vector current between the vacuum and the pion as
\begin{equation}
 \big\langle 0\big|\bar\psi_d(0)\gamma^+\gamma_5\psi_u(0)\big|\pi^+(P)\big\rangle
 = i f_\pi P^+ .
 \label{eq:fpidef}
\end{equation}
With the color factor $\sqrt{N_c}$ and the normalization $n=1/(1+g_s^2\widehat P_3)$ of the
$\qqbar$ sector, we obtain the LFWF overlap
\begin{equation}
\begin{aligned}
 f_\pi
 &=2\sqrt{n\,N_c}\int\frac{\mathrm{d}x\,\mathrm{d}^2\bm k_\perp}{16\pi^3}
 \Big[\Psi^{\pi}_{q\bar q}(x,\bm k_\perp,+,-)-\Psi^{\pi}_{q\bar q}(x,\bm k_\perp,-,+)\Big]
 \\
 &=2\sqrt{2\,n\,N_c}\int\frac{\mathrm{d}x\,\mathrm{d}^2\bm k_\perp}{16\pi^3}\,
 \phi^{\pi}(x,\bm k_\perp)\,\frac{\Mt}{\sqrt{\bm k_\perp^2+\Mt^2}} .
\end{aligned}
 \label{eq:fpilfwf}
\end{equation}
The charge form factor
receives contributions from both sectors,
$F(Q^2)=n\big[F^{\qqbar}(Q^2)+g_s^2F^{\qqbarg}(Q^2)\big]$, and the charge radius follows from
its slope at $Q^2=0$. We fit $m_q$ and $\beta$, with the three-body scale $\beta_3=\beta$ of Eq.~\eqref{eq:damp}, to the PDG values
$f_\pi=130.2(1.2)$ MeV and $\langle r^2\rangle^{1/2}=0.659(4)$ fm~\cite{ParticleDataGroup:2024cfk}.
The fit gives $m_q=0.304(2)$ GeV and $\beta=0.463(16)$ GeV. We then keep $m_q$ fixed. With these values, we obtain
$f_\pi=129.3$ MeV and $\langle r^2\rangle^{1/2}=0.663$ fm.
\begin{table*}[htbp]
\centering
\caption{Model parameters used in this work. $m_q$ and $\beta$ are determined from the decay constant and the charge radius of the full state, self-consistently with $\mu_0$. $m_g$ and $m_f$ are taken from the scan of Appendix~\ref{app:mgmf} and checked against the BLFQ values, and $M_\pi$ is the PDG value~\cite{ParticleDataGroup:2024cfk}. Only $\mu_0$ is fitted to the E615 data~\cite{E615:1989bda}, which fixes $\alpha_s(\mu_0)$ and $g_s$ through Eq.~\eqref{eq:gslock}. Masses and $\mu_0$ are in GeV, $A_\pi$ in GeV$^{-1}$, and $\alpha_s$ and $g_s$ are dimensionless.}
\label{tab:params}
\begin{tabular}{ccccccccc}
\hline\hline
 $m_q=m_{\bar q}$ & $\beta$ & $A_\pi$ & $m_g$ & $m_f$ & $M_\pi$ & $\mu_0$ & $\alpha_s(\mu_0)$ & $g_s$ \\ 
\hline
$0.304$ & $0.463$ & $22.50$ & $0.55$ & $7.00$ & $0.1396$ & $0.593$ & $0.80$ & $3.17$ \\
\hline\hline
\end{tabular}
\end{table*}

\paragraph{Regularization of the soft-gluon region.}
The wave function in Eq.~\eqref{eq:psi3} is not normalizable in its
present form. Each term in Eq.~\eqref{eq:psi3} contains the two-body
Gaussian factor
$\exp[-{\mathcal{M}_{0,2}^2}/(8\beta^2)]$ through the valence wave function
$\Psi^{\rm val}$, while no corresponding suppression factor is
introduced for the gluon. As $x_g\to0$ at fixed $\bm{k}_{g\perp}$, the
vertex grows as $1/x_g$, the energy denominator behaves as $1/x_g$, and
the prefactor contributes a factor $x_g^{-1/2}$. In the soft-gluon limit, the
three-body wave function therefore behaves as
\begin{equation}
\Psi_3
\xrightarrow{x_g\to0}
\frac{{\kappa^L}}
{\sqrt{x_g}\left(k_{g\perp}^2+m_g^2\right)}
\left[
\Psi_2\left(x_q,\bm{k}_{q\perp}+\bm{k}_{g\perp}\right)
-
\Psi_2\left(x_q,\bm{k}_{q\perp}\right)
\right].
\label{eq:soft_limit}
\end{equation}
Thus,
\begin{equation}
|\Psi_3|^2\sim\frac{1}{x_g},
\qquad
\int\mathrm{d}x_g\,|\Psi_3|^2
\sim
\int\frac{\mathrm{d}x_g}{x_g},
\end{equation}
which gives a logarithmic divergence as $x_g\to0$. The introduction of
a finite gluon mass $m_g$ does not remove this divergence, since it
regulates the energy denominator but does not eliminate the
$1/x_g$ behavior of the squared amplitude. In the BLFQ approach~\cite{Lan:2024ais}, this
divergence is regulated by the longitudinal discretization,
$x_g\geq1/K$. In a continuum formulation, the soft-gluon region has to be
suppressed explicitly to obtain a normalizable three-body wave function. We
follow the BHL ansatz, in which the wave function of an
$N$-body Fock state has a Gaussian dependence on the free
invariant mass of that state. We therefore multiply each
term in Eq.~\eqref{eq:psi3} by the factor
\begin{equation}
 \mathcal{D}
 =
 \exp\!\left[
 -\frac{{\mathcal{M}_{0,3}^2}-{\mathcal{M}_{0,2}^2}}{8\beta_3^2}
 \right],
 \label{eq:damp}
\end{equation}
which turns the two-body Gaussian into the three-body one. This factor
suppresses the soft-gluon region, so the three-body wave function can be
normalized without a cutoff parameter. $\beta_3$ is the BHL scale of the
three-body Fock state, and we use the same harmonic scale for both sectors,
$\beta_3=\beta$.

Since Eq.~\eqref{eq:psi3} is constructed from the valence wave function
rather than introduced as an independent state, the relative weight of the
two Fock sectors is fixed by a single integral and is not an
additional free parameter.
The $|q\bar q\rangle$ Fock state is first normalized to unity, which fixes
the normalization constant $A_\pi$ in Eq.~\eqref{eq:bhl},
\begin{equation}
 \int \frac{\mathrm{d}x\,\mathrm{d}^2\bm{k}_{\perp}}{16\pi^3}
 \sum_{{s_qs_{\bar q}}}
 \left|\Psi^\pi\right|^2 = 1 .
 \label{eq:valnorm}
\end{equation}
For the three-body $|q\bar q g\rangle$ Fock state, we define
\begin{equation}
 \widehat{P}_3 =
 {\frac{1}{g_s^2}}\int \frac{\mathrm{d}x_q\,\mathrm{d}^2\bm{k}_{q\perp}\,
 \mathrm{d}x_g\,\mathrm{d}^2\bm{k}_{g\perp}}{(16\pi^3)^2}
 \sum_{s_qs_{\bar q}\lambda}
 \left|\Psi_3\right|^2 .
 \label{eq:P3}
\end{equation}
$\widehat{P}_3$ is the norm of the three-body component relative to the
unit-normalized valence state, without the explicit factor $g_s^2$. The
truncated pion state of Eq.~\eqref{eq:truncation} is then normalized with the
common factor
\begin{equation}
 n\equiv\frac{1}{1+g_s^2\widehat{P}_3},
 \qquad
 P_{q\bar q}=n,
 \qquad
 P_{q\bar q g}=g_s^2\widehat{P}_3\,n,
 \qquad
 P_{q\bar q}+P_{q\bar q g}=1 .
 \label{eq:probs}
\end{equation}
The relative probability of the $q\bar q g$ sector is set by
$g_s^2\widehat{P}_3$, and the overall normalization by the single factor
$n$. Every distribution below is calculated from the unit-normalized valence
wave function and multiplied by the same factor $n$, with no separate
normalization for the individual distributions.

\begin{figure*}[t]
 \centering

 \begin{minipage}[t]{0.48\textwidth}
 \centering
 \includegraphics[width=\textwidth]{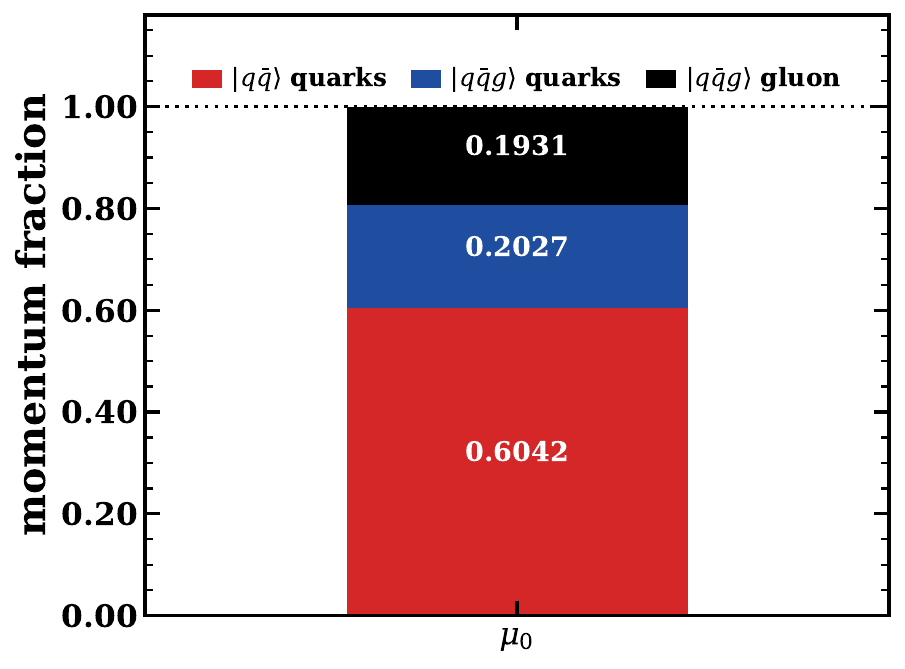}
 \end{minipage}
 \hfill
 \begin{minipage}[t]{0.48\textwidth}
 \centering
 \includegraphics[width=\textwidth]{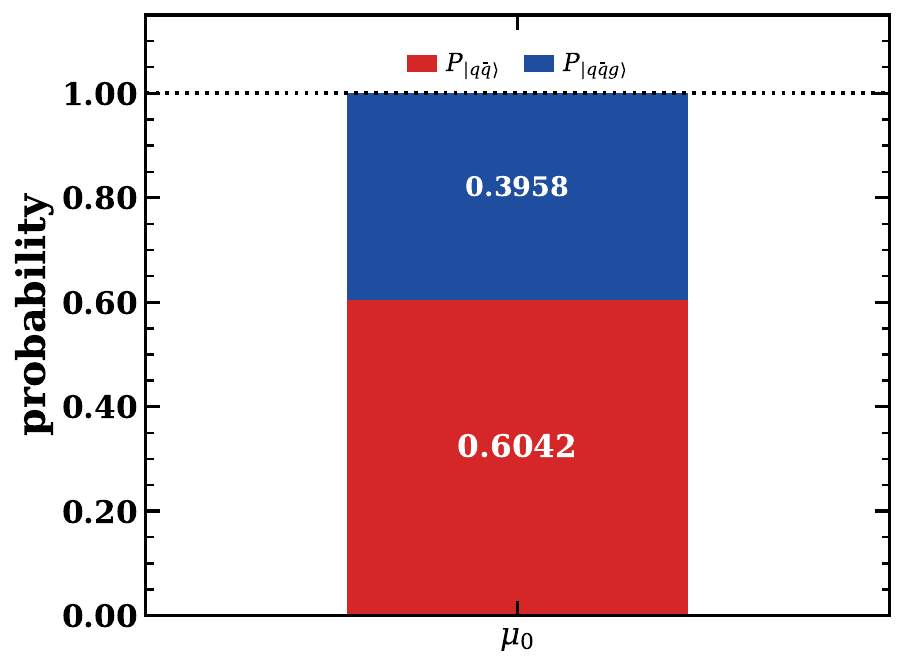}
 \end{minipage}

 \caption{(Color online) Momentum fractions (left) and Fock-sector probabilities (right) of the pion at the model scale.}
 \label{fig:mdecomp_fock}
\end{figure*}

\begin{figure}[t]
 \centering
 \includegraphics[width=0.99\textwidth]{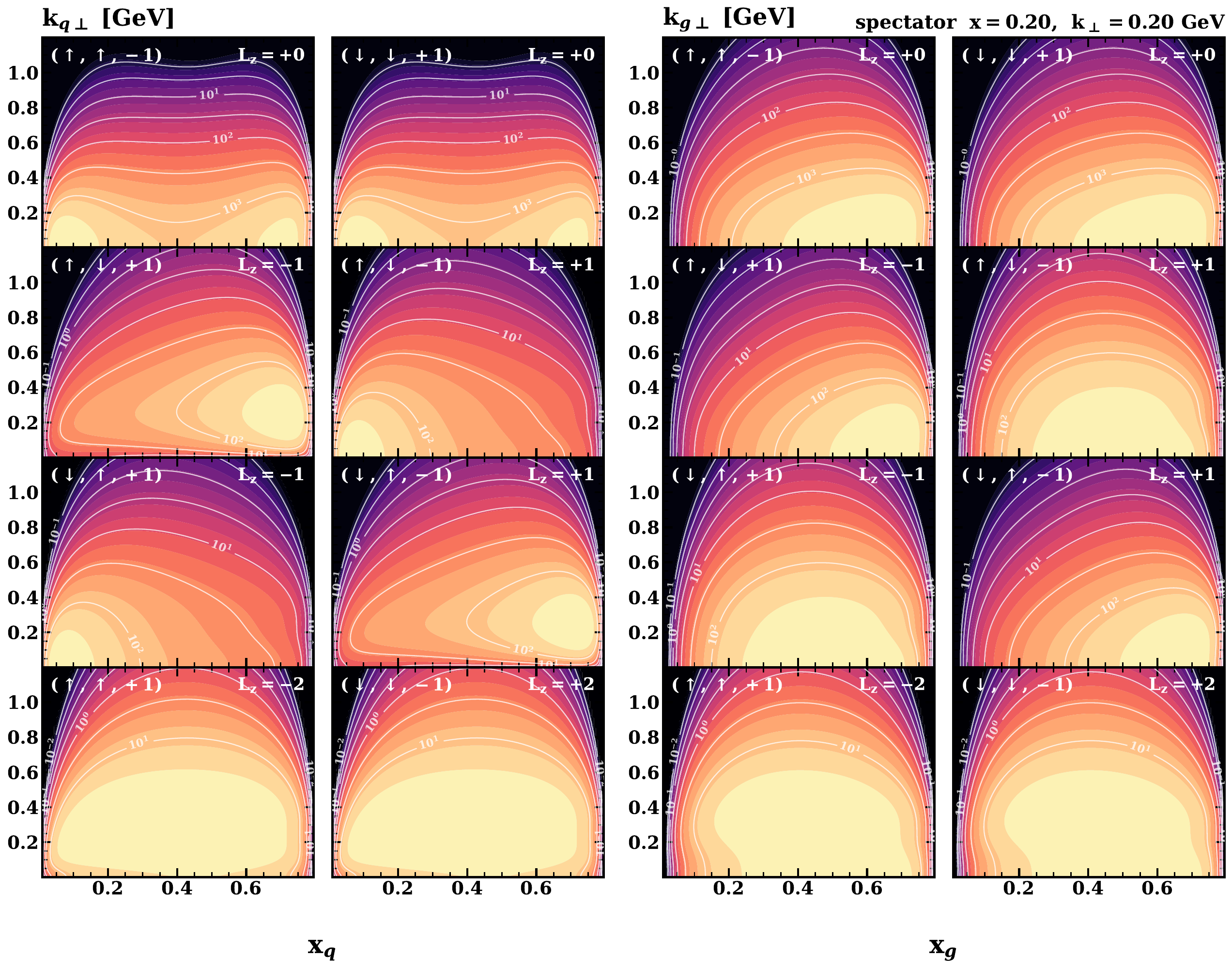}
 \caption{(Color online) Contour plots of the $q\bar q g$ LFWF for the eight helicity configurations of Table~\ref{tab:Lz_configurations}, against the quark variables $(x_q,k_{q\perp})$ (left) and the gluon variables $(x_g,k_{g\perp})$ (right), with the spectator fixed at $x=0.20$ and $k_\perp=0.20$ GeV.}
 \label{fig:momentumdecompositionmodelscale}
\end{figure}

The $q\bar q g$ sector accounts for $40\%$ of the pion probability, and
the constituent gluon carries $19\%$ of the pion momentum at the model scale.
The valence sector carries $60\%$ of the momentum, and the quark and
antiquark of the three-body sector carry the remaining $21\%$. In
Fig.~\ref{fig:mdecomp_fock}, we present the momentum decomposition and the
Fock-sector probabilities at the model scale. With the normalization
$n=(1+{g_s^2}\widehat{P}_3)^{-1}$, the quark-number and
momentum sum rules read
\begin{equation}
 \int_0^1\!\mathrm{d}x\,
 \big[q(x)+q^{(3)}(x)\big] = 1 ,
 \qquad
 \int_0^1\!\mathrm{d}x\,x\,
 \big[2q(x)+2q^{(3)}(x)+g(x)\big] = 1 ,
 \label{eq:sumrules}
\end{equation}
where $q^{(3)}(x)$ denotes the quark distribution from the
$q\bar q g$ sector and $g(x)$ is the gluon distribution. At the model
scale, the sea-quark distribution is absent because the present
Fock-space truncation does not include higher Fock components such as
$|q\bar q q\bar q\rangle$. The sea-quark contribution is therefore
generated by QCD evolution. In Fig.~\ref{fig:momentumdecompositionmodelscale},
we present the LFWFs of the $q\bar q g$ sector for the $L_z$ configurations of
Table~\ref{tab:Lz_configurations}, as functions of the momentum fraction and
transverse momentum of the active constituent, with the spectator fixed at
$x=0.20$ and $k_{\perp}=0.20~\mathrm{GeV}$.

\begin{figure}[t]
 \centering
 \includegraphics[width=0.99\textwidth]{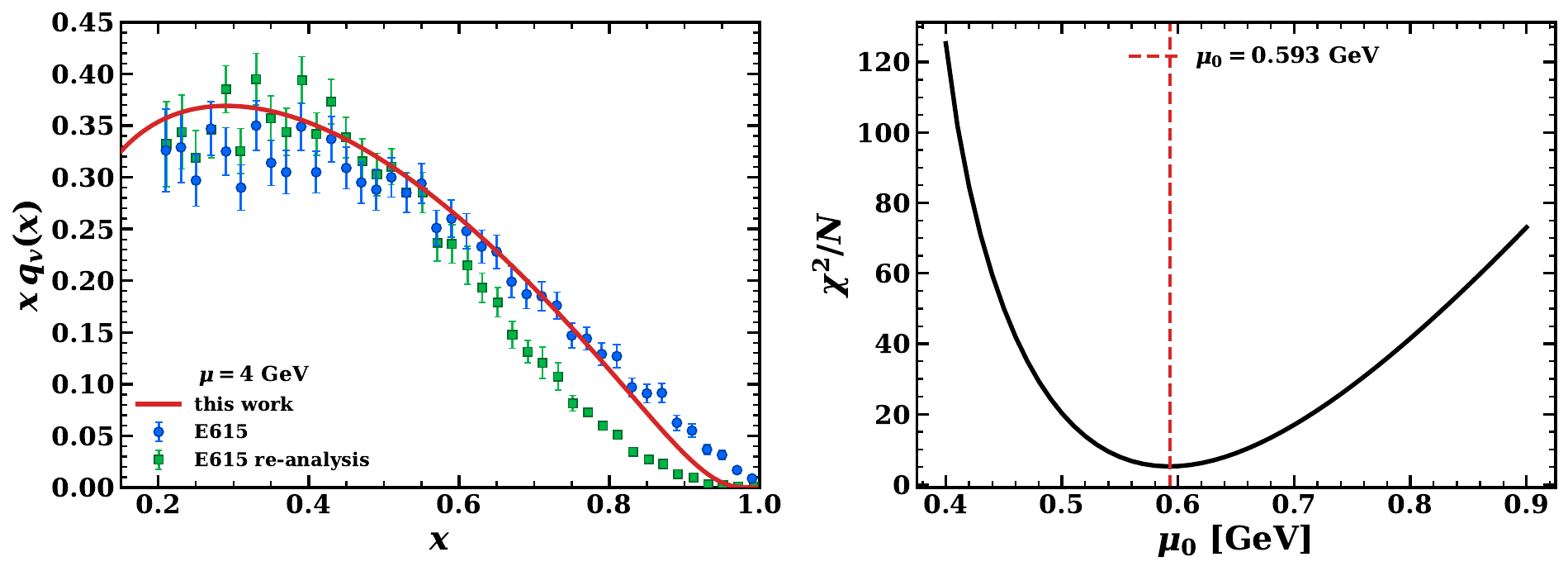}
 \caption{Valence PDF at $\mu=4$ GeV compared with the E615 data~\cite{E615:1989bda} and their re-analysis~\cite{Aicher:2010cb} (left), and $\chi^2/N$ against $\mu_0$ (right). The rise at small $\mu_0$ comes from the Landau pole of the two-loop coupling.}
 \label{fig:e615fit}
\end{figure}

All parameters of the present work are summarized in
Table~\ref{tab:params}. For the $q\bar q$ sector, the
distributions depend on two parameters, the quark mass and the harmonic
scale. The $q\bar q g$ sector adds two more, the gluon mass $m_g$ and the
vertex mass $m_f$. In the BLFQ analysis
\cite{Lan:2021wok,Lan:2024ais,Lan:2025qio}, these are $0.60$ and $5.69$ GeV, respectively. Those values were
fixed within a different Fock-space truncation and a different treatment of the
coupling, so we do not carry them over. Instead, we take $m_g=0.55$ GeV and $m_f=7.00$ GeV from the scan of the E615 fit and the JAM21 gluon~\cite{Barry:2021osv}, with the BLFQ values as a check, as discussed in
Appendix~\ref{app:mgmf}. The strong coupling $g_s$ is not a fit
parameter. It follows from the model scale $\mu_0$ through
Eq.~\eqref{eq:gslock}. To determine $\mu_0$, we evolve the valence PDF of
the model to $\mu=4~\mathrm{GeV}$ with NLO DGLAP evolution and compare it
with the E615 data. We repeat this for a range of initial scales, and the
minimum of $\chi^2/N$ gives $\mu_0=0.593~\mathrm{GeV}$. At this scale,
$\alpha_s(\mu_0)=0.80$, which gives $g_s=\sqrt{4\pi\alpha_s(\mu_0)}=3.17$.
In Fig.~\ref{fig:e615fit}, we present the evolved valence PDF together with
the E615 data and their re-analysis, and the dependence of $\chi^2/N$ on
the initial scale. We do not assign uncertainty bands to the distributions.
An estimate of the uncertainty can be obtained by varying $\mu_0$ within its
allowed range and propagating the variation through the evolution.
\section{Transverse Momentum Dependent Parton Distributions}
\label{sec:tmd}
\begin{figure}[t]
 \centering
 \includegraphics[width=0.99\textwidth]{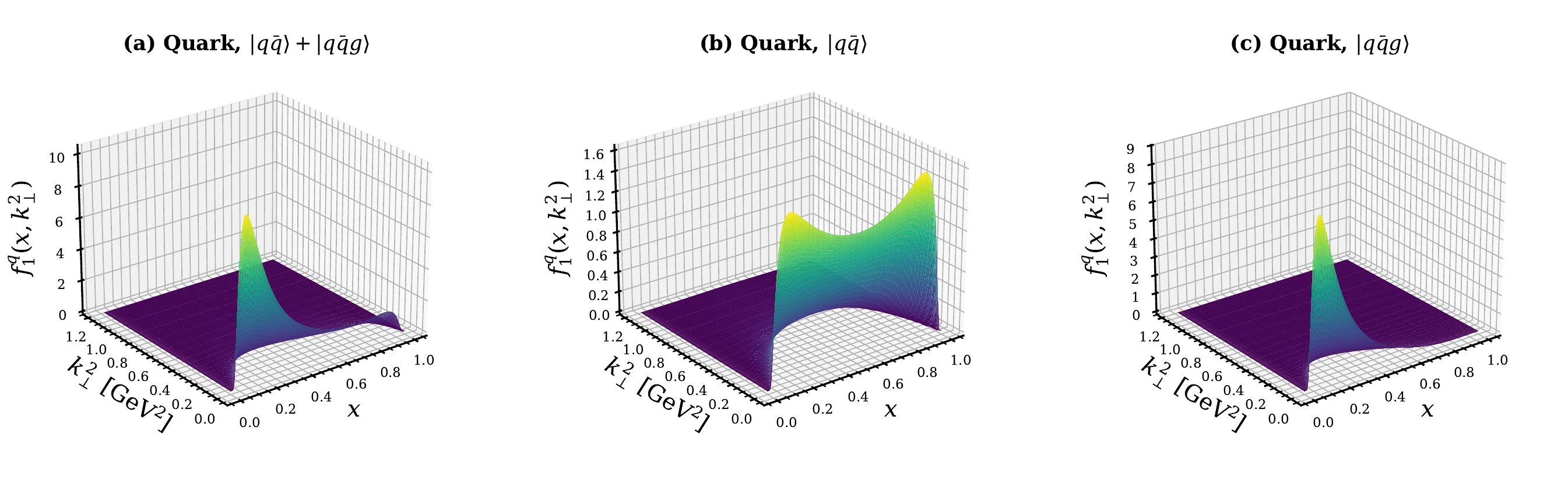}
 \caption{Unpolarized quark TMD $f_1^q(x,k_\perp^2)$ at the model scale. (a) Both Fock sectors, (b) $\qqbar$, (c) $\qqbarg$.}
 \label{fig:tmdpanel}
\end{figure}
\begin{figure}[t]
 \centering
 \includegraphics[width=0.70\textwidth]{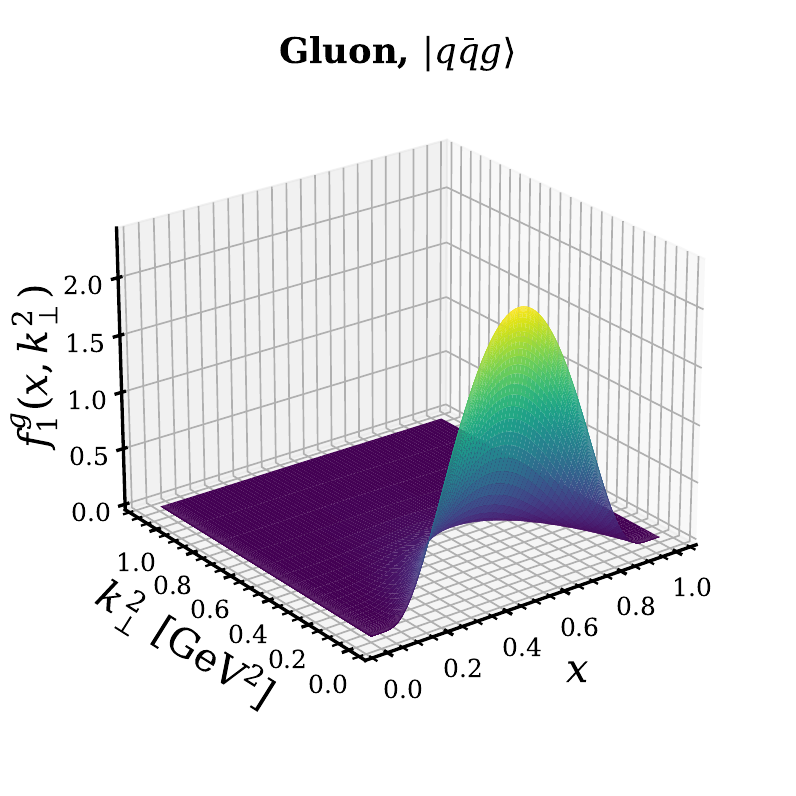}
 \caption{Unpolarized gluon TMD $f_1^g(x,k_\perp^2)$ at the model scale.}
 \label{fig:tmdglu}
\end{figure}
\begin{figure}[t]
 \centering
 \includegraphics[width=0.99\textwidth]{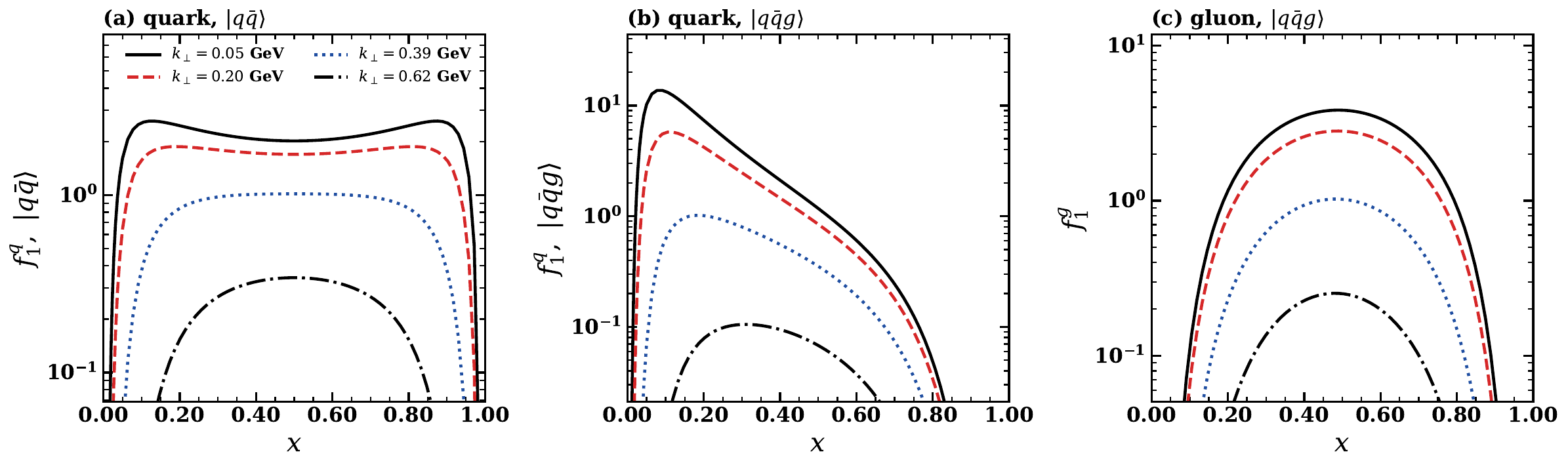}
 \caption{(Color online) Model-scale TMDs against $x$ at fixed $k_\perp$. (a) Quark, $\qqbar$, (b) quark, $\qqbarg$, (c) gluon, $\qqbarg$.}
 \label{fig:tmd_x}
\end{figure}
\begin{figure}[t]
 \centering
 \includegraphics[width=0.99\textwidth]{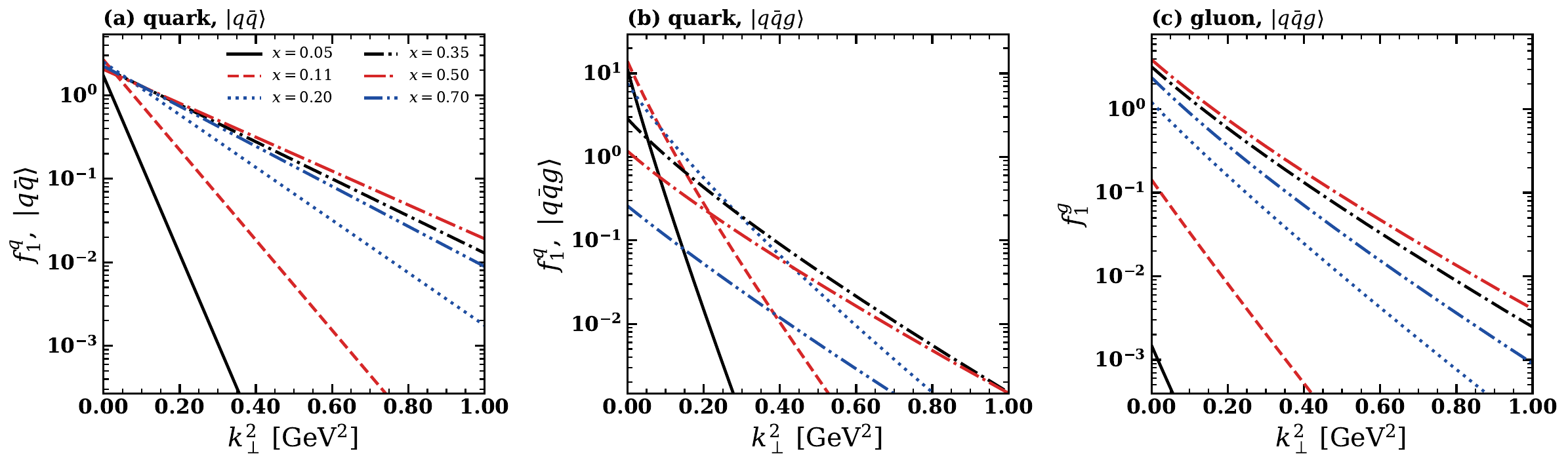}
 \caption{(Color online) Model-scale TMDs against $k_\perp^2$ at fixed $x$, with the panels of Fig.~\ref{fig:tmd_x}.}
 \label{fig:tmd_kperp}
\end{figure}

\begin{figure}[t]
 \centering
 \includegraphics[width=0.99\textwidth]{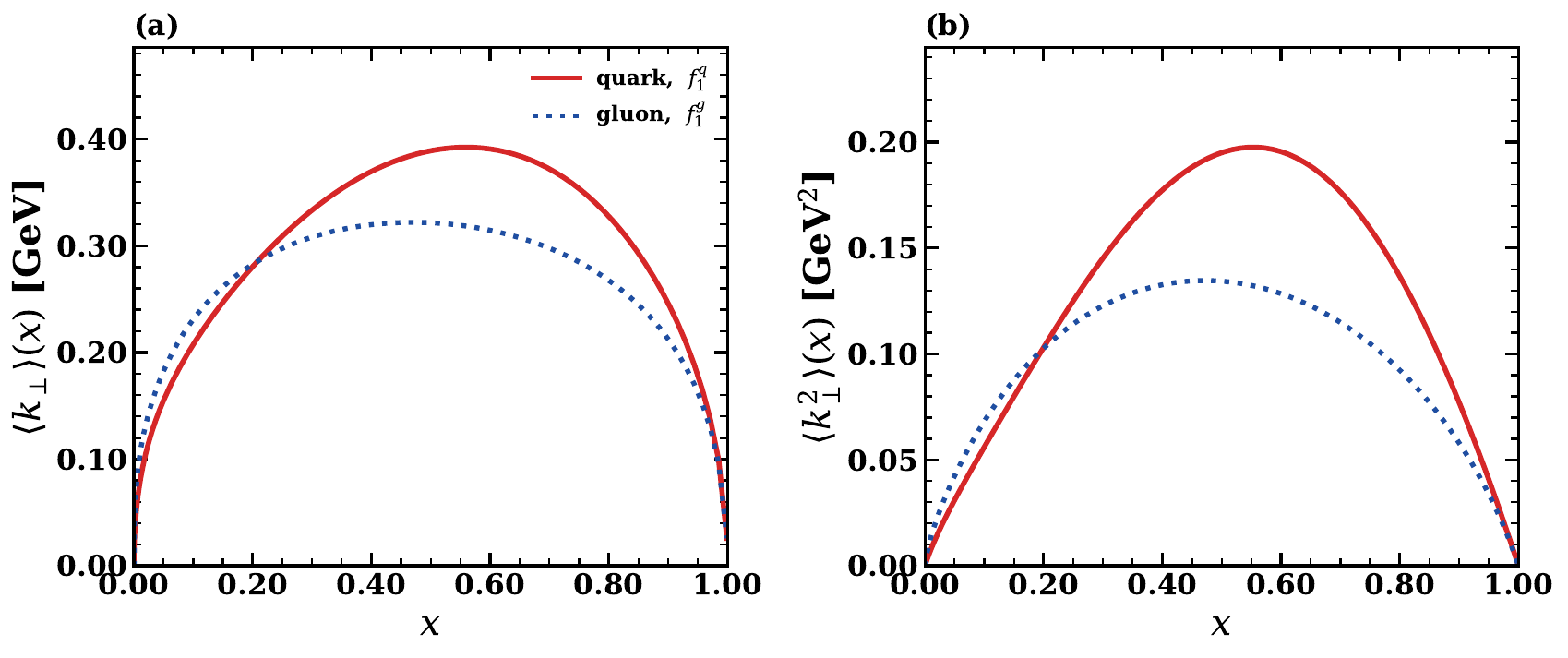}
 \caption{$\langle k_\perp\rangle(x)$ (a) and $\langle k_\perp^2\rangle(x)$ (b) for the total quark and the gluon TMD at the model scale.}
 \label{fig:tmdxs}
\end{figure}

\begin{figure}[t]
 \centering
 \includegraphics[width=0.99\textwidth]{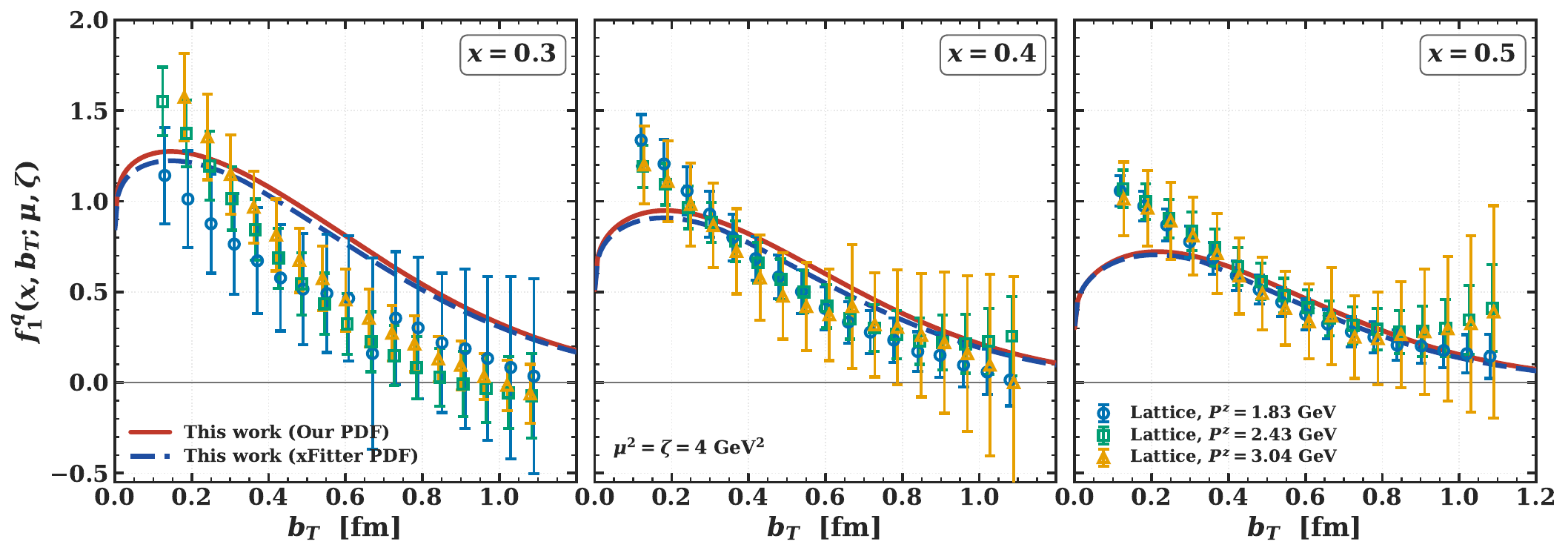}
 \caption{Quark TMD after CSS evolution to $\mu=2$ GeV, with our PDF and the xFitter PDF~\cite{Novikov:2020snp} as collinear input, compared with the lattice QCD results of Ref.~\cite{Bollweg:2025iol}.}
 \label{fig:tmdevolvedmu2}
\end{figure}

\begin{figure}[t]
 \centering
 \includegraphics[width=0.99\textwidth]{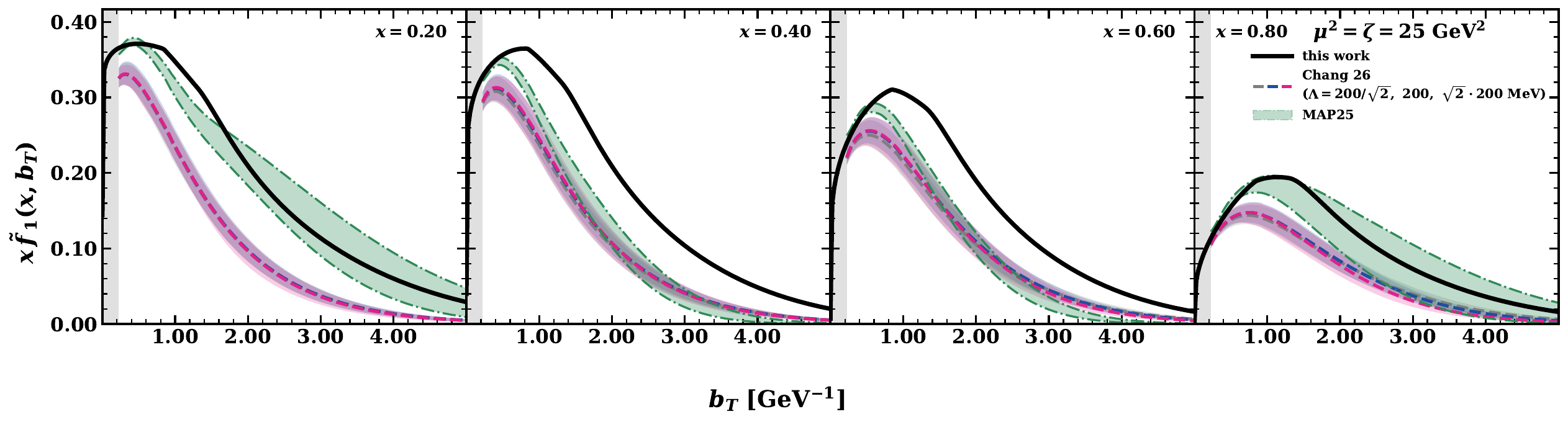}
 \caption{Quark TMD after CSS evolution to $\mu^2=25$ GeV$^2$, compared with the Drell--Yan extraction results of Refs.~\cite{Chang:2026rft,Rossi:2025pwh}.}
 \label{fig:tmdevolvedmu25}
\end{figure}
TMDs describe the three-dimensional structure of hadrons in momentum
space, through both the longitudinal momentum fraction and the transverse
momentum of the constituents.
For a spin-$1/2$ target, there are eight leading-twist TMDs, of which
six are $T$-even and two are $T$-odd \cite{Meissner:2009ww}. For a spin-1 target, there are
eighteen leading-twist TMDs, including nine $T$-even distributions \cite{Bacchetta:2000jk}. The
pion is a spin-zero meson and has only two leading-twist quark TMDs, the
$T$-even unpolarized distribution $f_1(x,k_\perp^2)$ and the $T$-odd
Boer--Mulders function $h_1^{\perp}(x,k_\perp^2)$. The Boer--Mulders
function describes the correlation between the transverse polarization of
a quark and its transverse momentum in an unpolarized hadron, and it arises
from the gauge link. With the gauge link set to unity, it vanishes, and
$f_1(x,k_\perp^2)$ is the only nonvanishing leading-twist TMD in the
present framework. In this section, we therefore study $f_1$ for the quark
and the gluon, with emphasis on the role of the dynamical gluon.
\par The unpolarized quark and gluon TMDs are defined through the correlation functions~\cite{Lan:2025qio}
\begin{align}
f^{q}_{1,\pi}(x,k_\perp^2)&=\half\!\int\!\frac{dz^-d^2{\bm{z}_\perp}}{2(2\pi)^3}
e^{ik\cdot z}\;\big\langle\pi(P)\big|
\bar\psi\big(-\tfrac z2\big)\,\gamma^+\,\psi\big(\tfrac z2\big)
\big|\pi(P)\big\rangle\Big|_{z^+=0},
\label{eq:tmdq}\\
f^{g}_{1,\pi}(x,k_\perp^2)&=\frac{1}{xP^+}\!\int\!
\frac{dz^-d^2{\bm{z}_\perp}}{2(2\pi)^3}e^{ik\cdot z}\;
\big\langle\pi(P)\big|G^{+i}\big(-\tfrac z2\big)G^{+i}\big(\tfrac z2\big)
\big|\pi(P)\big\rangle\Big|_{z^+=0},
\label{eq:tmdg}
\end{align}
with the gauge links set to unity. In terms of LFWF overlaps, they found to be
\begin{align}
f_1^{q}(x,k_\perp^2)&=\sum_{N,\{\lambda_i\}}\int\![N]\;
\big|\Psi_N\big|^2\,\delta(x-x_q)\,\delta^{(2)}({\bm{k}_\perp-\bm{k}_{q\perp}}),
\label{eq:tmdqov}\\
f_1^{g}(x,k_\perp^2)&=\sum_{N,\{\lambda_i\}}\int\![N]\;
\big|\Psi_N\big|^2\,\delta(x-x_g)\,\delta^{(2)}({\bm{k}_\perp-\bm{k}_{g\perp}}).
\label{eq:tmdgov}
\end{align}
Here, the $N$-particle integration measure is defined as
\begin{equation}
[N]\equiv
\prod_{i=1}^{N}
\frac{\mathrm{d}x_i\,\mathrm{d}^2{\bm{k}_{\perp i}}}{16\pi^3}
\,
16\pi^3\,
\delta\!\left(1-\sum_{j=1}^{N}x_j\right)
\delta^{(2)}\!\left(\sum_{j=1}^{N}{\bm{k}_{\perp j}}\right),
\label{eq:Nmeasure}
\end{equation}
where $N$ is the number of constituents in the Fock state, $\{\lambda_i\}$
are their helicities, and $x_i$ and ${\bm{k}_{\perp i}}$ are their
longitudinal momentum fractions and transverse momenta. The quark TMD
receives contributions from both sectors and the gluon TMD only from the
$\qqbarg$ sector,
\begin{align}
f_1^{q}(x,k_\perp^2)&=n\left[{\sum_{s_qs_{\bar q}}}\frac{|{\Psi^{\pi}_{q\bar q}}(x,{\bm{k}_\perp},{s_q,s_{\bar q}})|^2}{16\pi^3}
+\frac{1}{(16\pi^3)^2}\!\int\!\! dx_g\,d^2{\bm{k}_{g\perp}}
\!\!\sum_{s_q s_{\bar q}\lambda}\!\!
\big|{\Psi_3}\big|^2\right],
\label{eq:tmdqexpl}\\
f_1^{g}(x,k_\perp^2)&=\frac{n}{(16\pi^3)^2}\!\int\!\! dx_q\,d^2{\bm{k}_{q\perp}}
\!\!\sum_{s_q s_{\bar q}\lambda}\!\!\big|{\Psi_3}\big|^2 ,
\label{eq:tmdgexpl}
\end{align}
where the struck parton is fixed at $x_q=x$, $\bm{k}_{q\perp}=\bm{k}_\perp$ in Eq.~\eqref{eq:tmdqexpl} and at $x_g=x$, $\bm{k}_{g\perp}=\bm{k}_\perp$ in Eq.~\eqref{eq:tmdgexpl}.
In Fig.~\ref{fig:tmdpanel}, we present the three-dimensional profiles
of the unpolarized quark TMD $f_1^q(x,k_\perp^2)$ as functions of the
longitudinal momentum fraction $x$ and the squared transverse momentum
$k_\perp^2$. Panel (a) shows the total quark TMD, and panels (b) and (c)
show the contributions of the $q\bar q$ and $q\bar q g$ sectors. In the
$q\bar q$ sector, the quark TMD is symmetric under $x\leftrightarrow1-x$
and is enhanced toward small and large $x$. This differs from our previous
BHL results~\cite{Puhan:2024jaw} and comes mainly from the modified BHL wave
function of Eq.~\eqref{eq:bhl}. The quark TMD of the $q\bar q g$ sector is
larger than the $q\bar q$ one at small $x$. This enhancement comes from the
gluon emission, which shifts the quark toward smaller $x$, and it is also
seen in the BLFQ calculation of Ref.~\cite{Lan:2025qio}.

In Fig.~\ref{fig:tmdglu}, we present the unpolarized gluon TMD
$f_1^g(x,k_\perp^2)$ in the pion as a function of the longitudinal
momentum fraction $x$ and the squared transverse momentum $k_\perp^2$.
The valence sector contains no gluon, so the gluon TMD comes entirely
from the $q\bar q g$ sector. It peaks near $x\simeq0.5$ and falls toward
both endpoints. In $k_\perp^2$, it is concentrated below
$k_\perp^2\simeq0.4~\mathrm{GeV}^2$. At the model scale, the dynamical gluon
therefore sits at intermediate momentum fractions and low transverse
momenta.

In Figs.~\ref{fig:tmd_x} and \ref{fig:tmd_kperp}, we present the quark and
gluon TMDs of the different Fock sectors as functions of $x$ at fixed
$k_\perp^2$ and as functions of $k_\perp^2$ at fixed $x$. The valence
quark, the quark of the $q\bar q g$ sector, and the gluon are shown in
separate panels.

\subsection{Transverse dependence of TMDs}
The $x$-dependent mean transverse momentum (${j}=1$) and mean
squared transverse momentum (${j}=2$) are defined as \cite{Puhan:2023ekt}
\begin{equation}
\langle k_\perp^{{j}}\rangle_i(x)=
\frac{\displaystyle\int d^2\kperp\;k_\perp^{{j}}\,
f_1^i(x,k_\perp^2)}
{\displaystyle\int d^2\kperp\;
f_1^i(x,k_\perp^2)} .
\label{eq:kmom}
\end{equation}
In Fig.~\ref{fig:tmdxs}, we present $\langle k_\perp\rangle_i(x)$ and
$\langle k_\perp^2\rangle_i(x)$ for the total quark and the gluon TMDs. The
$x$-integrated moments are listed in Table~\ref{tab:kmom}. The valence
sector gives the largest moments and the quark of the $q\bar q g$ sector the
smallest, and the gluon lies in between. The quark moments of the
$q\bar q g$ sector agree with the BLFQ quark values~\cite{Lan:2025qio}, while
our gluon moments are about $25\%$ and $45\%$ larger than the BLFQ ones.

\begin{table}[htbp]
\centering
\caption{$x$-integrated transverse-momentum moments of Eq.~\eqref{eq:kmom} at the model scale, compared with BLFQ~\cite{Lan:2025qio}.}
\label{tab:kmom}
\begin{tabular}{lcccc}
\hline\hline
 & \multicolumn{2}{c}{quark} & \multicolumn{2}{c}{gluon} \\
 & $\langle k_\perp\rangle$ [GeV]
 & $\langle k_\perp^2\rangle$ [GeV$^2$]
 & $\langle k_\perp\rangle$ [GeV]
 & $\langle k_\perp^2\rangle$ [GeV$^2$]\\
\hline
This work, $\qqbar$ & 0.36 & 0.17 & --- & ---\\
This work, $\qqbarg$ & 0.26 & 0.09 & 0.31 & 0.13\\
This work, total & 0.32 & 0.14 & 0.31 & 0.13\\
\hline
BLFQ~\cite{Lan:2025qio} & 0.26 & 0.09 & 0.25 & 0.09\\
\hline\hline
\end{tabular}
\end{table}
\subsection{Evolution of the pion TMD}
The TMDs discussed above are the model-scale distributions, defined at
the initial scale $\mu_0=0.593~\GeV$ with the gauge link set to unity.
To compare them with TMD extractions and lattice QCD at higher scales, we
evolve them with the Collins--Soper--Sterman (CSS)
formalism~\cite{Collins:1984kg}. The evolution is formulated in the
transverse-separation space conjugate to the parton transverse momentum,
\begin{equation}
 \tilde f_1(x,b_T;\mu)
 =
 \int\!\mathrm{d}^2\kperp\,
 e^{-i\kperp\cdot\bm{b}_T}
 f_1(x,k_\perp^2;\mu)
 =
 2\pi\!\int_0^\infty\!\!\mathrm{d}k_\perp\,k_\perp\,
 J_0(k_\perp b_T)\,
 f_1(x,k_\perp^2;\mu),
 \label{eq:bspace}
\end{equation}
where $b_T$ is the transverse separation conjugate to the parton
transverse momentum $k_\perp$. In the CSS framework, the evolution is
multiplicative in $b_T$ space. For the present calculation, the model
TMD is separated into its collinear part and a non-perturbative
transverse-momentum-dependent factor. The evolved $b_T$-space TMD is
then constructed from the collinear distribution, the perturbative
Sudakov factor, the non-perturbative input, and the non-perturbative
evolution kernel. Schematically, it takes the form
\begin{equation}
 \tilde f_1(x,b_T;\mu)
 =
 f_{\rm coll}(x,\mu_b)\,
 e^{-S_{\rm pert}(b_*,\mu)}
 f_{\rm NP}(x,b_T)\,
 e^{-S_{\rm NP}(b_T,\mu)} .
 \label{eq:css}
\end{equation}
Here, the collinear PDF $f_{\rm coll}(x,\mu_b)$ is obtained from the
pion PDF of the present work, given in Sec.~\ref{sec:pdf}, evolved from
the model scale $\mu_0=0.593~\GeV$. For comparison, we also use the xFitter
pion PDF~\cite{Novikov:2020snp} from the LHAPDF library~\cite{Buckley:2014ana}
as an alternative collinear input. The non-perturbative transverse
factor $f_{\rm NP}(x,b_T)$ is obtained directly from the model TMD
through the normalized Hankel transform as
\begin{equation}
 f_{\rm NP}(x,b_T)
 =
 \frac{
 \displaystyle\int_0^\infty\!\mathrm{d}k_\perp\,k_\perp\,
 J_0(k_\perp b_T)\,
 f_1(x,k_\perp^2;\mu_0)}
 {\displaystyle\int_0^\infty\!\mathrm{d}k_\perp\,k_\perp\,
 f_1(x,k_\perp^2;\mu_0)} ,
 \label{eq:fnp1}
\end{equation}
which satisfies $f_{\rm NP}(x,0)=1$. The model TMD contains a nontrivial
transverse-momentum prefactor from the light-front Jacobian, so the Hankel
transform of Eq.~\eqref{eq:fnp1} is evaluated numerically. The normalized
factor $f_{\rm NP}$ does not depend on the overall normalization of the
input TMD, and the Fock-sector factor $\nfock$ cancels between the numerator
and the denominator. It carries only the shape of the model TMD in
$k_\perp$.

The matching scale is determined using the $b_*$ prescription,
\begin{equation}
 b_*=
 \frac{b_T}{\sqrt{1+b_T^2/b_{\rm max}^2}},
 \qquad
 \mu_b=\frac{2e^{-\gamma_E}}{b_*},
 \label{eq:bstar}
\end{equation}
which prevents the perturbative scale from entering the Landau-pole
region. The perturbative Sudakov factor is evaluated using the NNLL
coefficients implemented in the evolution, while the collinear part is
matched at NLO for the pion. The complete expressions and details of
the evolution procedure used in the present work are given in
Appendix~\ref{app:sec:tmdevol}. The non-perturbative evolution is
implemented through the parameter $g_2=0.248$, adopted from the MAP Collaboration analysis~\cite{Bacchetta:2022awv}, with its CSS reference
scale $Q_0=1~\GeV$. We keep $g_2$ together with this reference scale and do
not identify $Q_0$ with the model scale $\mu_0=0.593~\GeV$, at which the
input TMD is constructed. A determination of the non-perturbative evolution
parameters at the lower scale $\mu_0$ is left for future work.

In Fig.~\ref{fig:tmdevolvedmu2}, we present the quark TMD evolved to
$\mu^2=4~\GeV^2$ as a function of $b_T$ for $x=0.3$, $0.4$ and $0.5$,
together with the lattice QCD results at three values of the pion momentum
$P^z$~\cite{Bollweg:2025iol}. Using the xFitter PDF as collinear input
instead of ours changes the result only slightly. At $x=0.3$, our result
agrees with the lattice data at small $b_T$. At $x=0.4$, it agrees at larger
$b_T$ and lies below the lattice data for $b_T\lesssim0.2$~fm. At $x=0.5$, it
is consistent with the lattice data for $b_T\gtrsim0.3$~fm. The $b_T$
dependence of our model is therefore broadly consistent with lattice QCD.

In Fig.~\ref{fig:tmdevolvedmu25}, we compare our TMD at $\mu^2=25~\mathrm{GeV}^2$ with the recent extraction of the pion TMD from Drell--Yan data~\cite{Chang:2026rft} and with the MAP25 results~\cite{Rossi:2025pwh}. Our distributions have a similar $b_T$ dependence, but they are larger than those of Ref.~\cite{Chang:2026rft} at all four values of $x$, by up to a factor of about $1.5$ at $b_T\simeq1$--$2\GeV^{-1}$, and they fall more slowly at large $b_T$. Our results agree with MAP25 in size and shape at $x=0.2$ and $x=0.8$ and lie above it at $x=0.4$ and $0.6$. The differences in normalization show that the collinear input and the non-perturbative evolution parameters have a large effect on the size of the evolved TMD. The Sudakov factors suppress the large-$b_T$ region, and since $b_T$ is conjugate to $k_\perp$, the evolved distribution becomes broader in $k_\perp$.

\section{Generalized Parton Distributions}
\label{sec:gpd}
TMDs describe the transverse momentum of the constituents of a hadron,
while GPDs describe its spatial structure and the correlation between the
longitudinal momentum and the transverse position of the
constituents~\cite{Diehl:2003ny}. Together, they give a more complete picture
of hadron structure. GPDs depend on the longitudinal momentum fraction $x$,
the squared momentum transfer $t=-\Delta_\perp^2$, and the skewness
\begin{equation}
\xi=-\frac{\Delta^+}{{P'^{+}}+P^+},
\end{equation}
which describes the longitudinal momentum transfer between the initial and final hadrons. A spin-0 hadron has two leading-twist quark GPDs, the
chiral-even unpolarized GPD $H(x,\xi,t)$ and the chiral-odd GPD
$E_T(x,\xi,t)$. The first moment of $H$ gives the electromagnetic form factor.

The unpolarized quark GPDs of pseudoscalar mesons have been studied in
many works~\cite{Chavez:2021llq,Zhang:2021mtn}, while the gluon GPDs have
received much less attention~\cite{Lan:2025qio}, even for the nucleon. Here,
we calculate the unpolarized quark and gluon GPDs of the pion at zero
skewness, $\xi=0$, and extend them to finite skewness with the
double-distribution (DD) formalism and the Radyushkin profile
function~\cite{Radyushkin:1998es}.

The unpolarized quark and gluon GPDs of the pion are defined through the correlation functions~\cite{Diehl:2003ny}
\begin{align}
H^{q}_\pi(x,0,t)&=\!\int\!\frac{dz^-}{2(4\pi)}e^{ixP^+z^-/2}
\big\langle\pi(P')\big|{\bar\psi}\big(-\tfrac z2\big)\gamma^+{\psi}\big(\tfrac z2\big)
\big|\pi(P)\big\rangle\Big|_{z^+={\bm{z}_\perp}=0},
\label{eq:gpdq}\\
H^{g}_\pi(x,0,t)&=\frac{1}{xP^+}\!\int\!\frac{dz^-}{4\pi}e^{ixP^+z^-/2}
\big\langle\pi(P')\big|G^{+i}\big(-\tfrac z2\big)G^{+i}\big(\tfrac z2\big)
\big|\pi(P)\big\rangle\Big|_{z^+={\bm{z}_\perp}=0},
\label{eq:gpdg}
\end{align}
with $\Delta=P'-P$ and $t=-\Delta_\perp^2$ at zero skewness. Here, $P$ and $P'$ are the four-momenta of the initial and final pions. Our gluon GPD
is normalized so that $H^g(x,0,0)=g(x)$, which differs by a factor $x$ from the convention $H^g(x,0,0)=xg(x)$ of Ref.~\cite{Diehl:2003ny}.

In the symmetric frame, ${\bm{P}_\perp}=-\Dperp/2$ and
${\bm{P}'_\perp}=+\Dperp/2$, the intrinsic momenta shift as
\begin{equation}
{\bm{k}_{\perp i}}\;\longrightarrow\;
\begin{cases}
{\bm{k}_{\perp i}}\mp\tfrac12(1-x_i)\Dperp & \text{struck parton},\\[3pt]
{\bm{k}_{\perp i}}\pm\tfrac12 x_i\Dperp & \text{spectators},
\end{cases}
\label{eq:shifts}
\end{equation}
which preserves $\sum_i {\bm{k}_{\perp i}}=0$, while the physical momenta of the
struck parton shift by the full $\Dperp$,
\begin{equation}
{\bm{p}_\perp}={\bm{k}_\perp}-\tfrac12\Dperp,\qquad
{\bm{p}'_\perp}={\bm{k}_\perp}+\tfrac12\Dperp .
\label{eq:physshift}
\end{equation}
The LFWF overlap representations of the quark and gluon GPDs are
\begin{align}
H^{q}(x,0,t)&=\!\sum_{N,\{\lambda_i\}}\!\int\![N]\;
\Psi_N^{*}\{{\bm{k}''_{\perp i}}\}\,\Psi_N\{{\bm{k}'_{\perp i}}\}\;
\delta(x-x_q),
\label{eq:gpdqov}\\
H^{g}(x,0,t)&=\!\sum_{N,\{\lambda_i\}}\!\int\![N]\;
\Psi_N^{*}\{{\bm{k}''_{\perp i}}\}\,\Psi_N\{{\bm{k}'_{\perp i}}\}\;
\delta(x-x_g),
\label{eq:gpdgov}
\end{align}
with ${\bm{k}''_{\perp i}}$ (${\bm{k}'_{\perp i}}$) the final (initial) set of Eq.~\eqref{eq:shifts}, i.e.\ the lower (upper) signs.

\begin{figure}[t]
 \centering
 \includegraphics[width=0.99\textwidth]{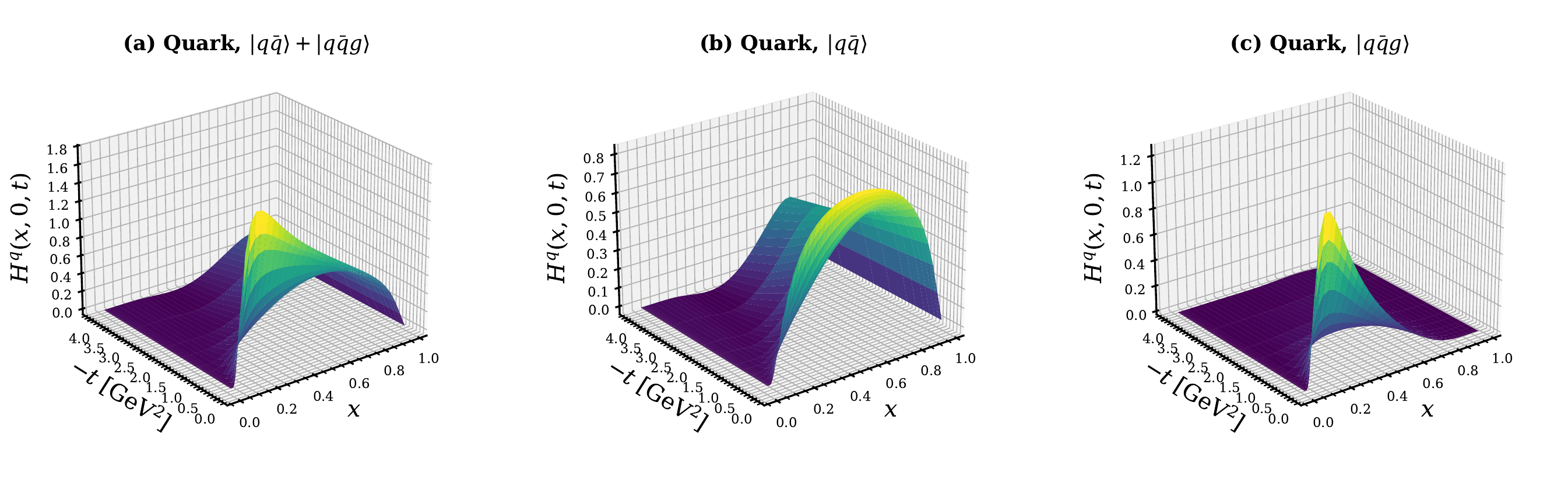}
 \caption{(Color online) Quark GPD $H^q(x,0,t)$. (a) Both Fock sectors, (b) $\qqbar$, (c) $\qqbarg$.}
 \label{fig:gpd}
\end{figure}

As for the TMDs, the quark GPD receives contributions from both Fock
sectors and the gluon GPD only from the $q\bar q g$ sector. In
Fig.~\ref{fig:gpd}, we present the quark GPDs of the two sectors at zero
skewness as functions of $x$ and $-t$. They are largest at small $x$ and
near $t=0$, especially in the $q\bar q g$ sector, and decrease with
increasing momentum transfer.

\begin{figure}[t]
 \centering
 \includegraphics[width=0.99\textwidth]{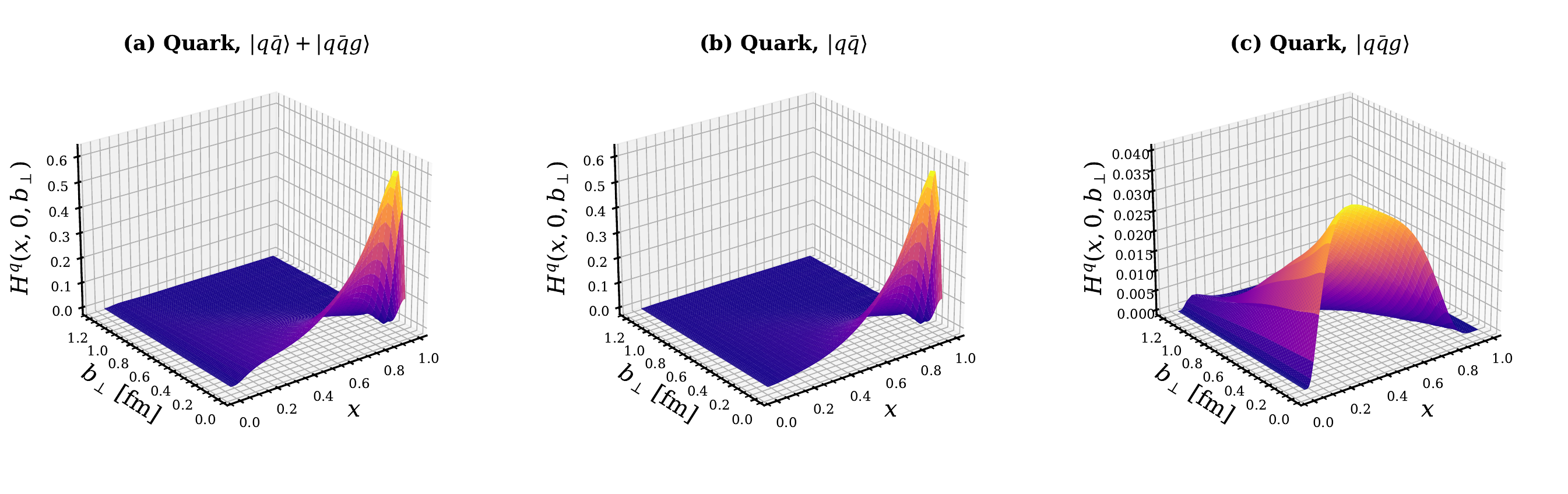}
 \caption{(Color online) Quark GPD in impact-parameter space, $H^q(x,0,b_\perp)$, with the panels of Fig.~\ref{fig:gpd}.}
 \label{fig:gpd_impact}
\end{figure}

The GPDs in impact-parameter space are obtained from the Fourier transform~\cite{Diehl:2002he}
\begin{equation}
\begin{split}
H^{i}(x,0,\boldsymbol{b}_{\perp})
&=
\int\!\frac{d^2\boldsymbol{\Delta}_{\perp}}{(2\pi)^2}\;
e^{-i\boldsymbol{\Delta}_{\perp}\cdot\boldsymbol{b}_{\perp}}\,
H^{i}(x,0,-\Delta_\perp^2)\\
&=
\frac{1}{2\pi}\!\int_0^\infty\!\! d\Delta_\perp\;\Delta_\perp\,
J_0(\Delta_\perp b_\perp)\,
H^i(x,0,-\Delta_\perp^2).
\end{split}
\label{eq:impact}
\end{equation}
In Fig.~\ref{fig:gpd_impact}, we present the quark GPDs in
impact-parameter space. The valence sector dominates, and the $q\bar q g$
sector is small, with a maximum of about $0.04$. The BLFQ calculation of
Ref.~\cite{Lan:2025qio} shows the same pattern with a larger peak value.

\begin{figure}[t]
 \centering
 \begin{minipage}[t]{0.48\textwidth}
 \centering
 \includegraphics[width=\textwidth]{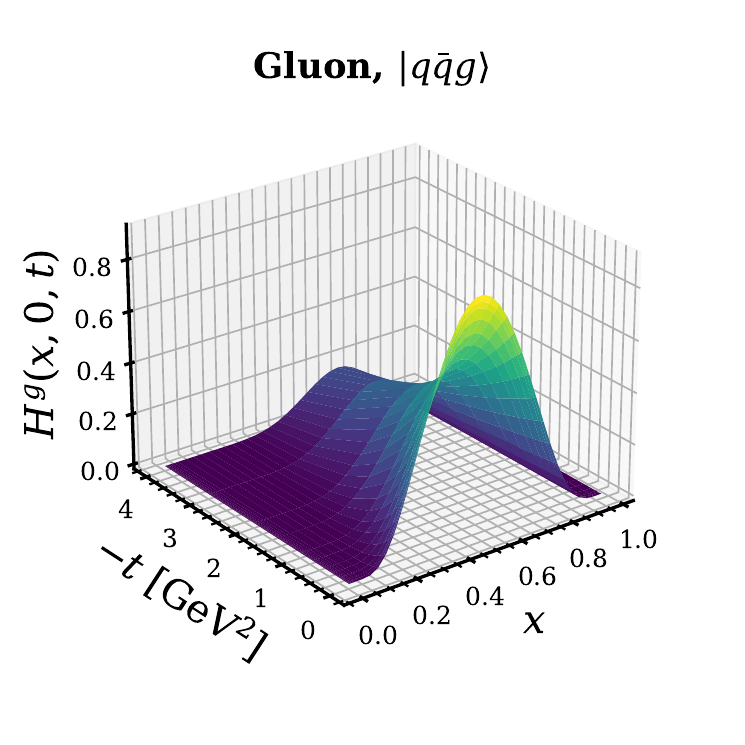}
 \end{minipage}
 \hfill
 \begin{minipage}[t]{0.48\textwidth}
 \centering
 \includegraphics[width=\textwidth]{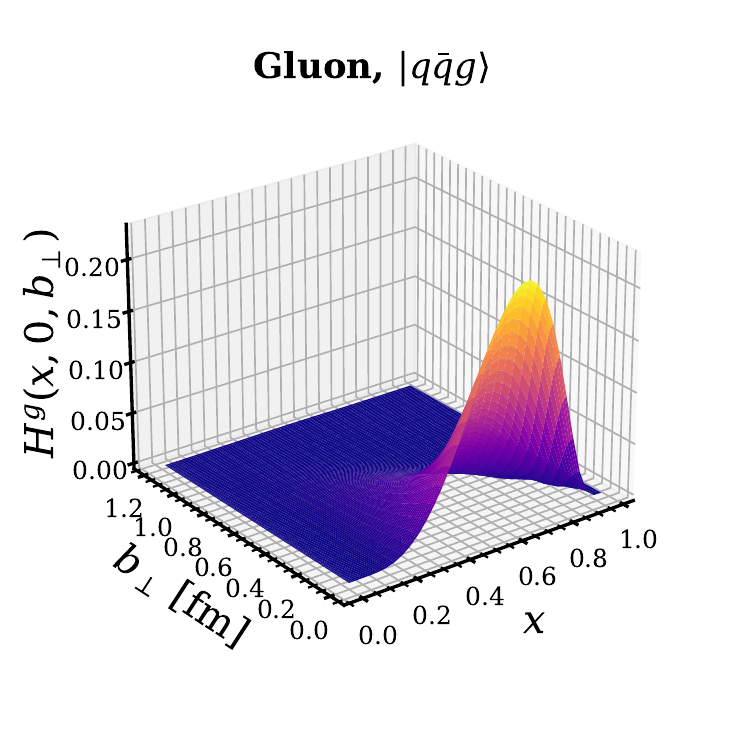}
 \end{minipage}

 \caption{(Color online) Gluon GPD $H^g(x,0,t)$ (left) and $H^g(x,0,b_\perp)$ (right).}
 \label{fig:gpd_gluon}
\end{figure}

\begin{figure}[t]
 \centering
 \includegraphics[width=0.99\textwidth]{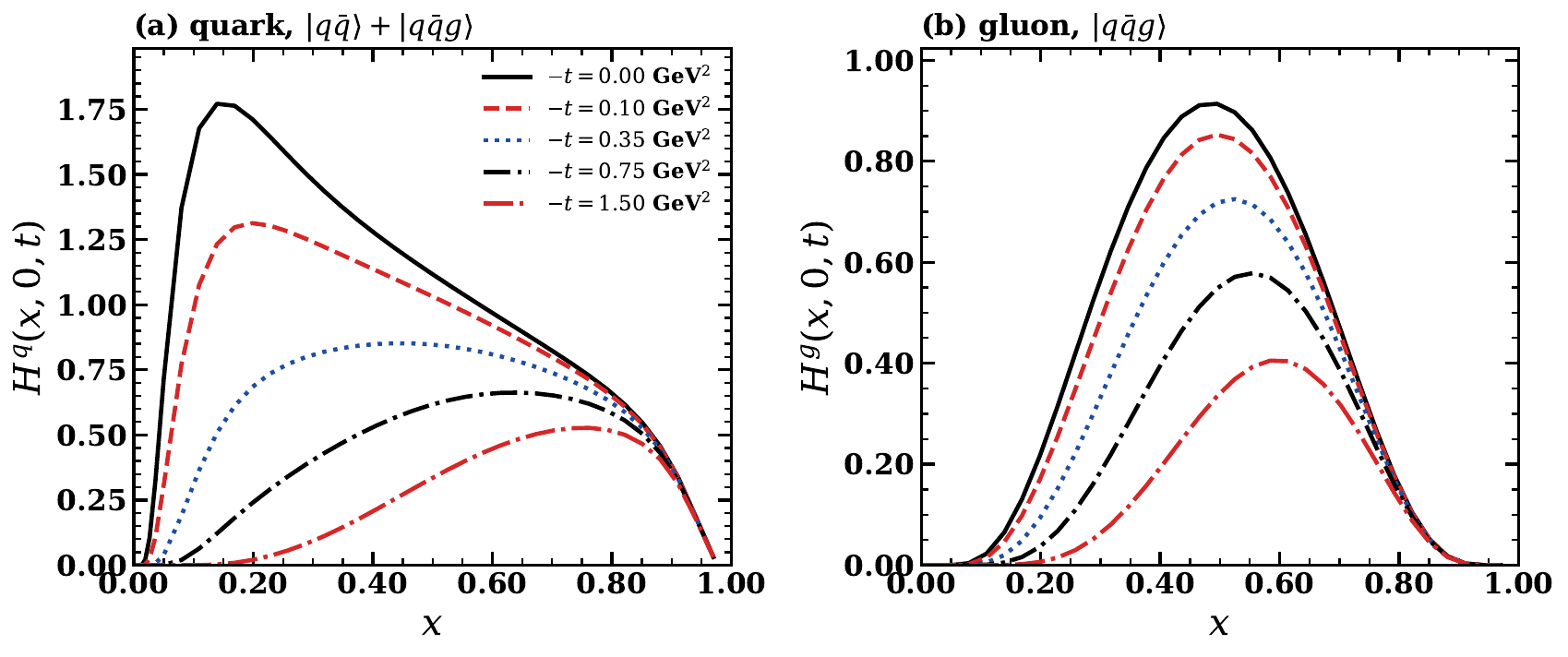}
 \caption{Quark (a) and gluon (b) GPDs at fixed $-t$. Larger $-t$ moves the distributions to larger $x$.}
 \label{fig:gpdslice}
\end{figure}

In Fig.~\ref{fig:gpd_gluon}, we present the gluon GPD $H^g(x,0,t)$. In
momentum-transfer space, it moves toward intermediate and large $x$ as $-t$
grows. In impact-parameter space, it is largest at small $b_\perp$ and
large $x$, as in the BLFQ results~\cite{Lan:2025qio}. In
Fig.~\ref{fig:gpdslice}, we present the quark and gluon GPDs as functions of
$x$ at fixed $t$.

\begin{figure}[t]
 \centering
 \includegraphics[width=0.99\textwidth]{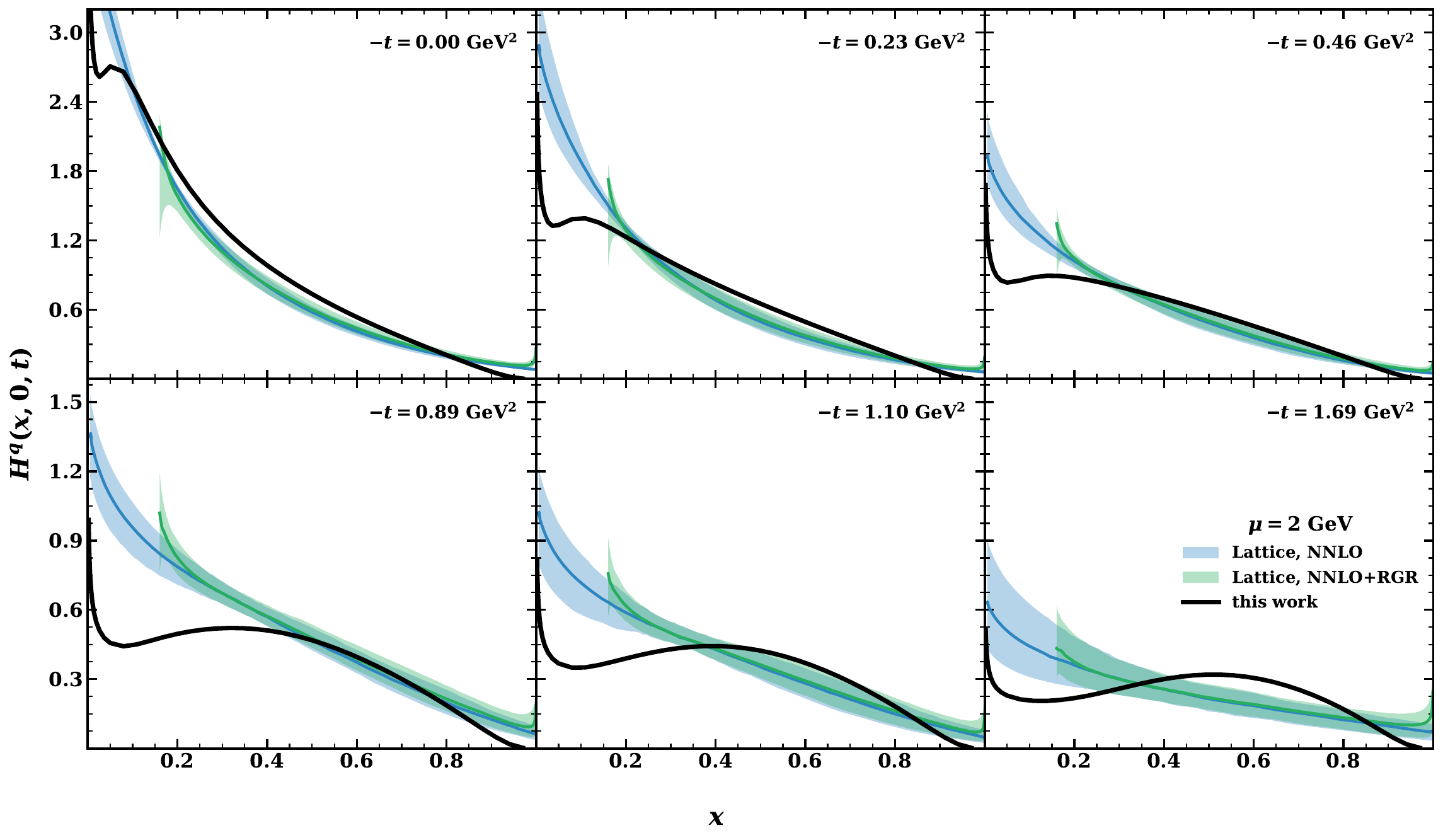}
 \caption{(Color online) Quark GPD evolved to $\mu=2$ GeV compared with the lattice QCD results of Ref.~\cite{Ding:2024saz} at six values of $-t$.}
 \label{fig:gpdlat}
\end{figure}

\begin{figure}[t]
 \centering
 \includegraphics[width=0.99\textwidth]{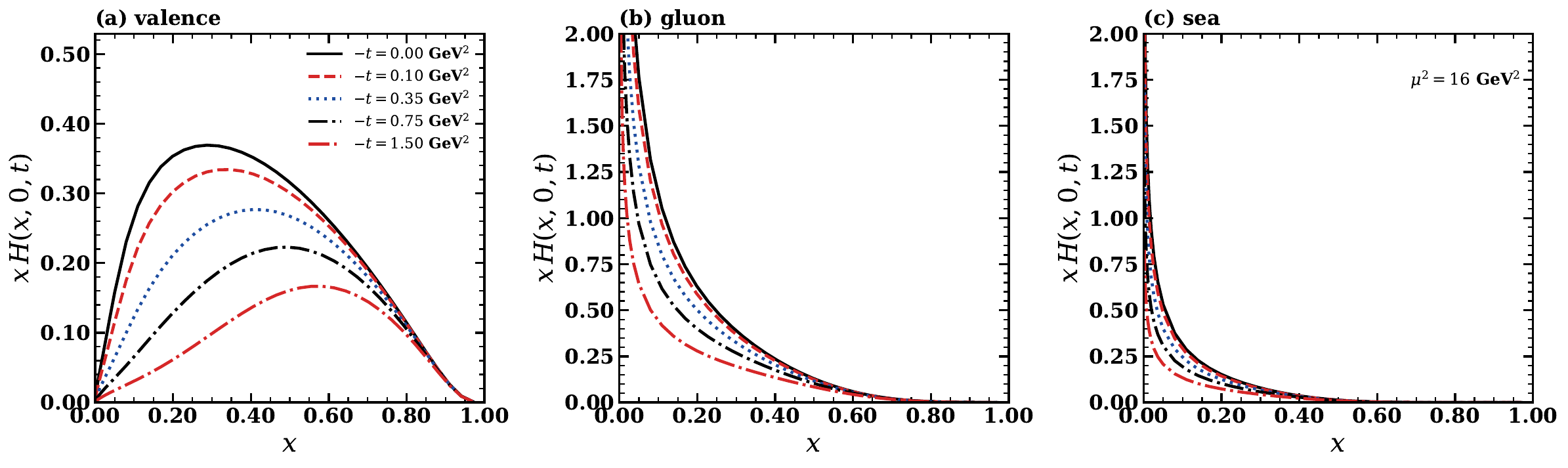}
 \caption{(a) valence, (b) gluon and (c) sea GPDs $xH(x,0,t)$ after NLO evolution to $\mu^2=16$ GeV$^2$ for five values of $-t$.}
 \label{fig:gpdevrow}
\end{figure}

At zero skewness, the GPD evolution kernel reduces to the DGLAP kernel, and
$t$ is a spectator variable. Each fixed-$t$ slice of $H^i(x,0,t)$ can
therefore be evolved like a PDF. We evolve the GPDs from
$\mu_0=0.593~\GeV$ with NLO DGLAP evolution in the
HOPPET~\cite{Karlberg:2025hxk} code. In Fig.~\ref{fig:gpdlat}, we compare the
quark GPD at $\mu=2~\GeV$ with the lattice QCD results of
Ref.~\cite{Ding:2024saz} at their six values of $-t$. At $-t=0$, our result
agrees with the lattice data for $x\gtrsim0.15$. At $-t=0.23$ and
$0.46~\GeV^2$, it agrees for $x\gtrsim0.25$ and lies below the lattice data
at smaller $x$. At larger $-t$, the model lies below the lattice data for
$x\lesssim0.4$ and is close to them for $0.4\lesssim x\lesssim0.8$. In
Fig.~\ref{fig:gpdevrow}, we present the valence, gluon and sea GPDs at
$\mu^2=16~\GeV^2$. The valence GPD decreases with increasing $-t$, while the
gluon and sea GPDs rise strongly toward small $x$. The sea is absent at the
model scale and is generated entirely by the evolution.

\subsection{Impact Parameter Space GPDs}
\begin{figure}[t]
 \centering
 \includegraphics[width=0.99\textwidth]{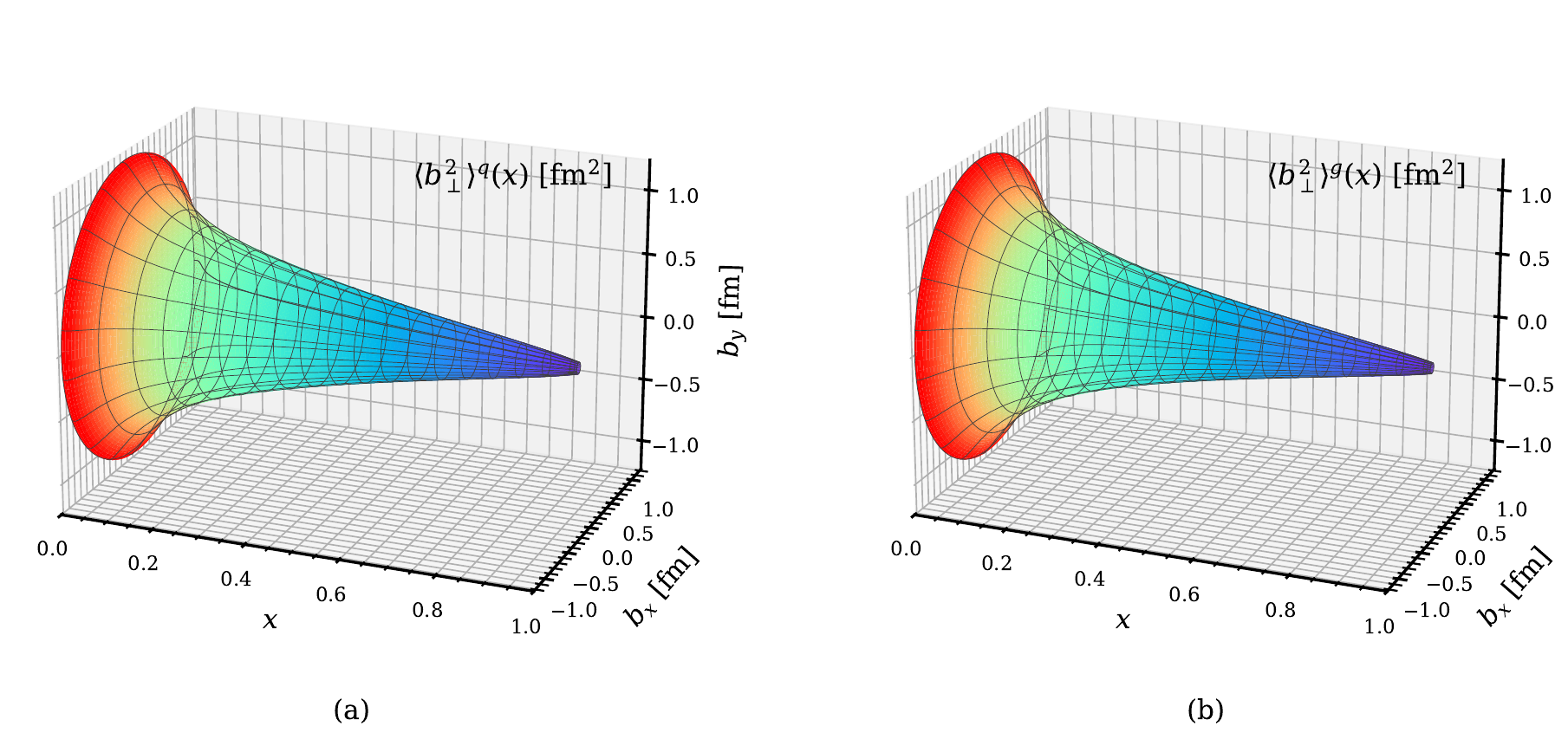}
 \caption{(Color online) Three-dimensional view of $\langle b_\perp^2\rangle(x)$ for the quark (left) and the gluon (right).}
 \label{fig:b2_trumpet}
\end{figure}

The transverse size of the pion at a given $x$ is described by the mean
squared impact parameter,
\begin{equation}
\langle b_\perp^2\rangle_i(x)=
\frac{\int d^2\boldsymbol{b}_{\perp}\;b_\perp^2\,
H^i(x,0,\boldsymbol{b}_{\perp})}
 {\int d^2\boldsymbol{b}_{\perp}\;
H^i(x,0,\boldsymbol{b}_{\perp})}
=-\frac{4}{H^i(x,0,0)}\,
\frac{\partial H^i}{\partial\Delta_\perp^2}
\bigg|_{\Delta_\perp=0}.
\label{eq:b2}
\end{equation}
In Fig.~\ref{fig:b2_trumpet}, we present $\langle b_\perp^2\rangle_i(x)$
for the quark and the gluon. Both decrease with increasing $x$, so partons
with small momentum fractions are spread over a wider transverse region,
and those with large $x$ sit close to the center. The quark and gluon
curves are close to each other over the whole range of $x$.

\begin{figure}[t]
 \centering
 \begin{minipage}[t]{0.48\textwidth}
 \centering
 \includegraphics[width=\textwidth]{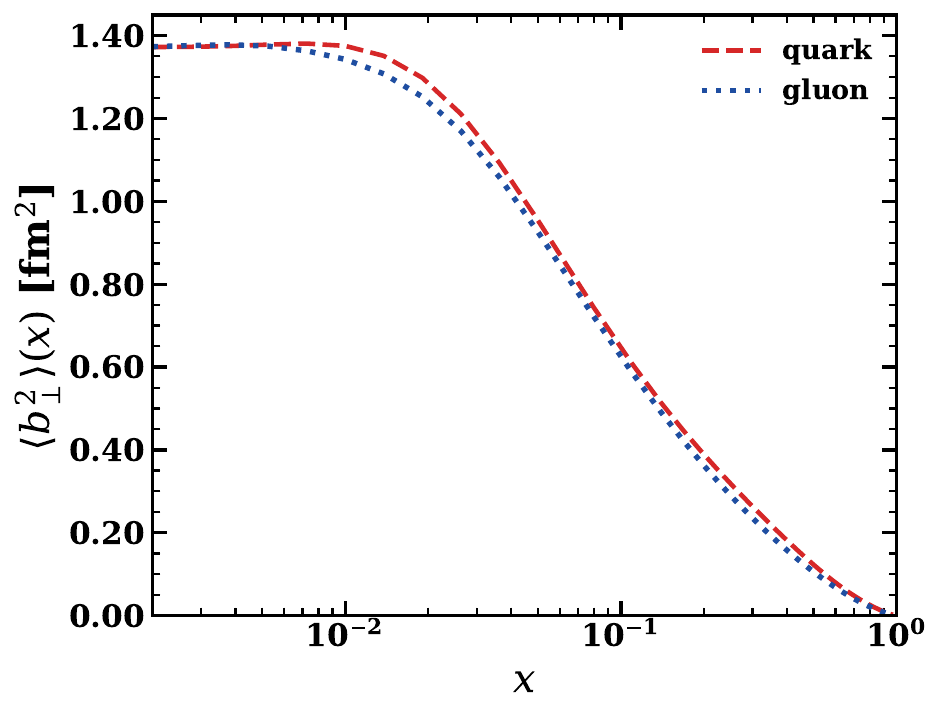}
 \end{minipage}
 \hfill
 \begin{minipage}[t]{0.48\textwidth}
 \centering
 \includegraphics[width=\textwidth]{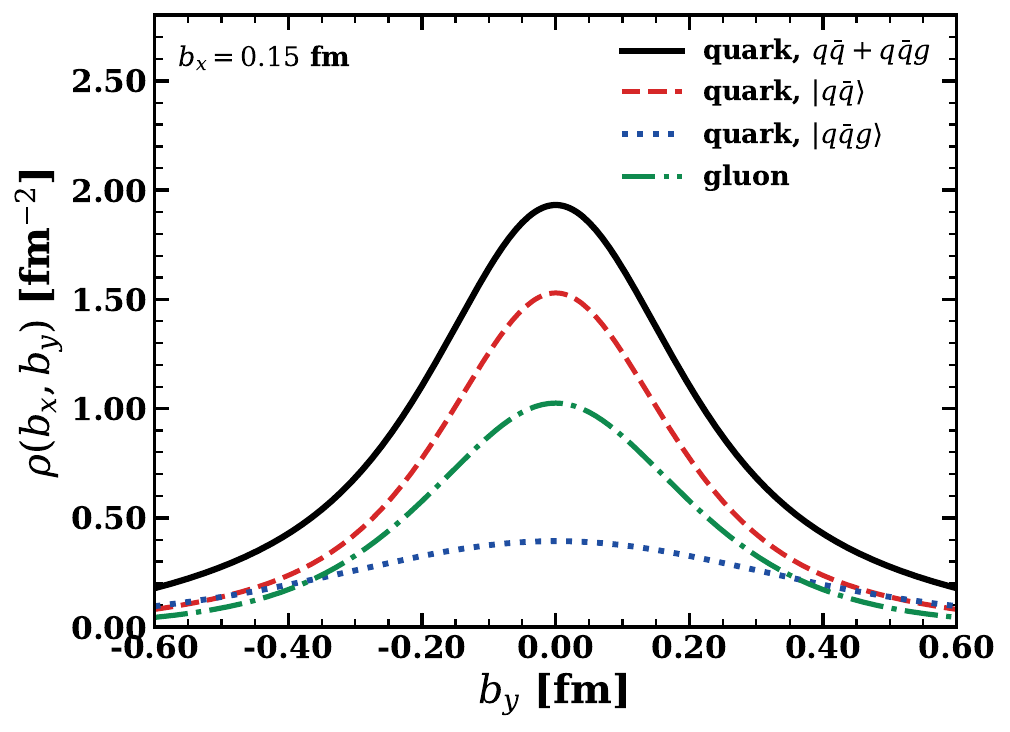}
 \end{minipage}

 \caption{(Color online) $\langle b_\perp^2\rangle(x)$ (left) and the transverse density in the impact-parameter plane (right) for the quark and the gluon.}
 \label{fig:impdens}
\end{figure}

In Fig.~\ref{fig:impdens}, we present the transverse density in the
impact-parameter plane. The total quark density is dominated by the valence
sector, and the $q\bar q g$ sector adds a small part. The gluon density is
also concentrated around the center, with a smaller magnitude than the
quark density.


\begin{figure}[t]
 \centering
 \includegraphics[width=0.99\textwidth]{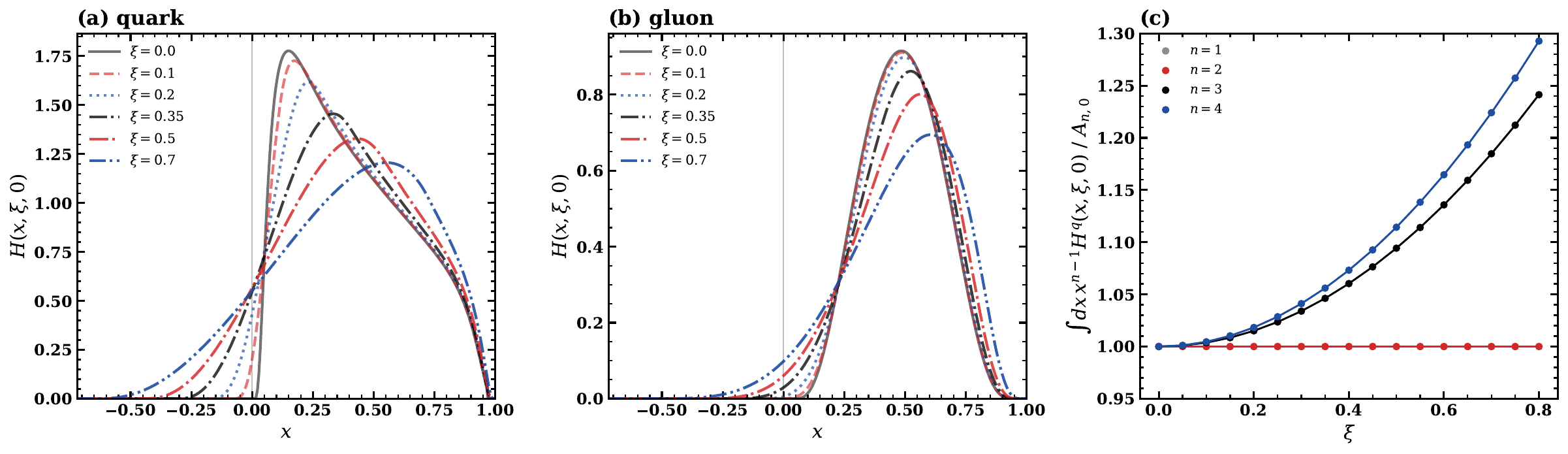}
 \caption{Skewness dependence of the quark (a) and gluon (b) GPDs from the double distribution~\cite{Radyushkin:1998es}, and the polynomiality test for $n=1$--$4$ (c).}
 \label{fig:poly}
\end{figure}

\subsection{Polynomiality}
\label{sec:poly}
The zero-skewness GPD is extended to finite skewness with the double-distribution (DD) representation and the Radyushkin
profile~\cite{Radyushkin:1998es},
\begin{equation}
 F(\beta,\alpha,t)=h(\beta,\alpha)\,H(\beta,0,t),\qquad
 h(\beta,\alpha)=\frac{3}{4}\,
 \frac{(1-|\beta|)^2-\alpha^2}{(1-|\beta|)^3},
 \label{eq:profile}
\end{equation}
where the skewness-dependent GPD is given by
\begin{equation}
 H(x,\xi,t)=\int_0^1\!\!\mathrm{d}\beta\int\!\mathrm{d}\alpha\;
 \delta(x-\beta-\xi\alpha)\,F(\beta,\alpha,t),
 \label{eq:DD}
\end{equation}
with $\int\!\mathrm{d}\alpha\,h(\beta,\alpha)=1$, so that the
zero-skewness GPD $H(x,0,t)$ is recovered at $\xi=0$. For
$\xi>0$, the allowed range of $\alpha$ at fixed $x$ is
\begin{equation}
 \max\Big[-\frac{1-x}{1+\xi},\;\frac{x-1}{\xi}\Big]
 \leq\alpha\leq
 \min\Big[\frac{1-x}{1-\xi},\;\frac{x}{\xi}\Big],
 \label{eq:alpharange}
\end{equation}
which gives the support $-\xi<x<1$. The region $-\xi<x<\xi$ is the ERBL
region, and its part at $x<0$ must be included in the Mellin moments.
Without it, polynomiality is violated, and the $n=1$ moment acquires an
unphysical $\xi$ dependence instead of staying equal to the valence number.
With the full support, the Mellin moments are
\begin{equation}
 \int_{-\xi}^{1}\!\mathrm{d}x\;x^{n-1}H(x,\xi,t)
 =
 \int\!\mathrm{d}\beta\,\mathrm{d}\alpha\;
 (\beta+\xi\alpha)^{n-1}F(\beta,\alpha,t),
 \label{eq:polyn}
\end{equation}
which is a polynomial in $\xi$ of degree $n-1$. In Fig.~\ref{fig:poly}, we
present this check for $n=1,\ldots,4$. The $n=1$ and $n=2$ moments
are independent of $\xi$. For $n=2$, the profile is symmetric under
$\alpha\rightarrow-\alpha$, which removes the term linear in $\xi$.

\subsection{The electromagnetic form factor}
\label{sec:emff}
\begin{figure}[t]
 \centering
 \includegraphics[width=0.99\textwidth]{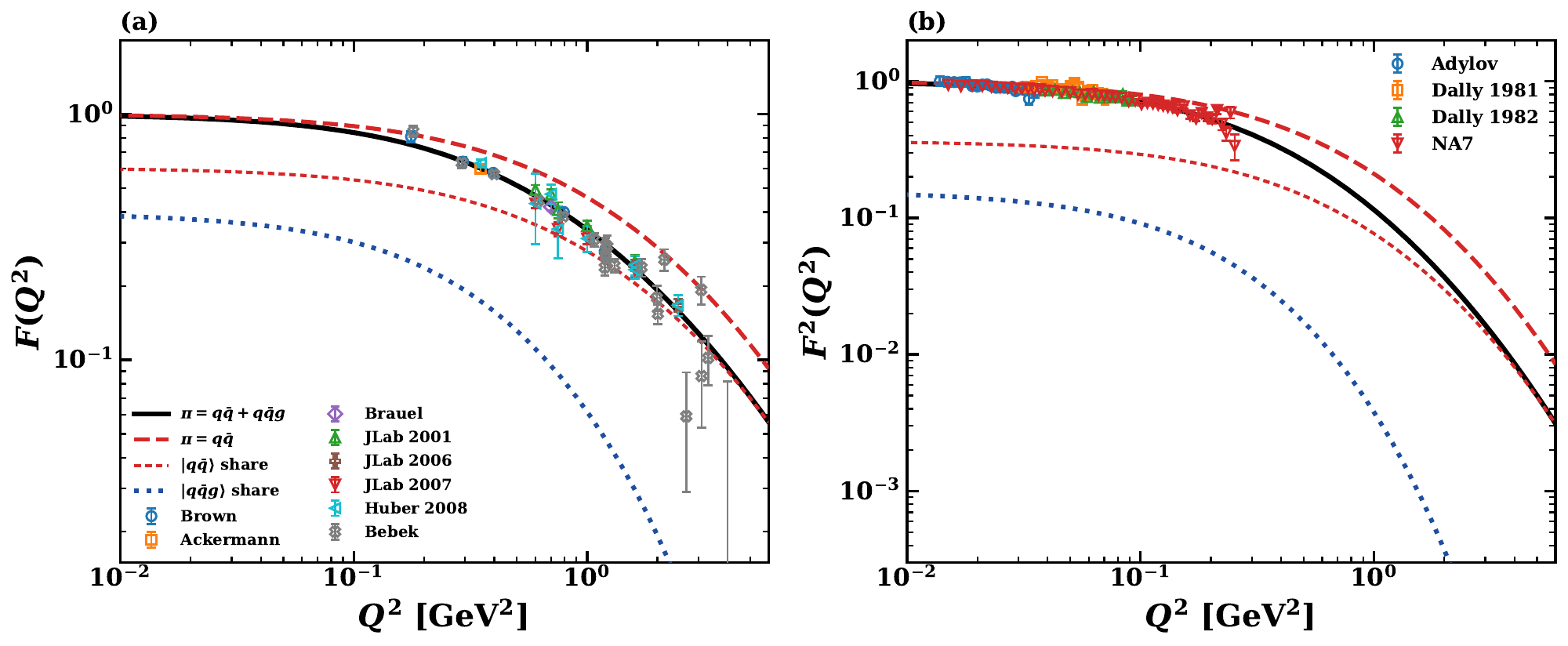}
 \caption{(Color online) $F(Q^2)$ (left) and $F^2(Q^2)$ (right) for the full state, the $\qqbar$ state with its own normalization, and the two sector contributions, compared with the data of Refs.~\cite{Brown:1973wr,Ackermann:1977rp,Brauel:1979zk,JeffersonLabFpi:2000nlc,JeffersonLabFpi-2:2006ysh,JeffersonLabFpi:2007vir,JeffersonLab:2008jve,Bebek:1977pe,Adylov:1977kj,Dally:1981ur,Dally:1982zk,NA7:1986vav}.}
 \label{fig:FQ2}
\end{figure}

\begin{figure}[t]
 \centering
 \includegraphics[width=0.99\textwidth]{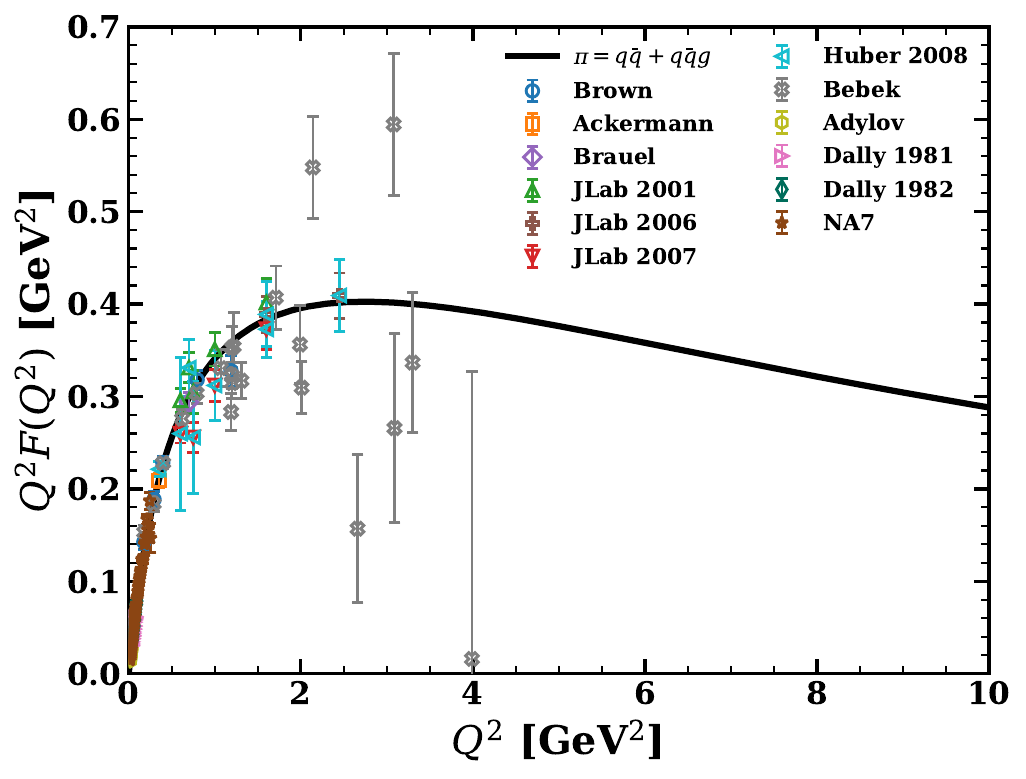}
 \caption{(Color online) $Q^2F(Q^2)$ of the full pion state compared with the data of Refs.~\cite{Brown:1973wr,Ackermann:1977rp,Brauel:1979zk,JeffersonLabFpi:2000nlc,JeffersonLabFpi-2:2006ysh,JeffersonLabFpi:2007vir,JeffersonLab:2008jve,Bebek:1977pe,Adylov:1977kj,Dally:1981ur,Dally:1982zk,NA7:1986vav}.}
 \label{fig:FQ21}
\end{figure}
The charge form factor is the first Mellin moment of the zero-skewness
quark GPD~\cite{Diehl:2003ny},
\begin{equation}
 F(Q^2)
 =
 \int_0^1\!\mathrm{d}x\;H^q(x,0,t),
 \label{eq:ffdef}
\end{equation}
with $Q^2=-t=\Delta_\perp^2 $.
The pion charge form factor is the charge-weighted sum of the quark and
antiquark contributions,
\begin{equation}
 F^{\pi}(Q^2)
 =
 e_q F^q(Q^2)
 +
 e_{\bar q}F^{\bar q}(Q^2).
 \label{eq:mesonff}
\end{equation}
In the $\qqbar$ sector, the antiquark carries the momentum fraction $1-x$, and its GPD follows from the quark one,
\begin{equation}
 H^{\bar q}_{q\bar q}(x,0,t;m_q,m_{\bar q})
 =
 H^q_{q\bar q}(1-x,0,t;m_{\bar q},m_q),
 \label{eq:cconj}
\end{equation}
so both contributions follow from the same GPD by exchanging the quark and
antiquark masses and replacing $x\rightarrow1-x$. With both Fock sectors,
the charge form factor is the sum of the two sector contributions,
\begin{equation}
 F(Q^2)
 =
 F_{q\bar q}(Q^2)
 +
 F_{q\bar q g}(Q^2),
 \label{eq:ff_fock}
\end{equation}
where $F_{q\bar q}(Q^2)$ and $F_{q\bar q g}(Q^2)$ are the first Mellin
moments of the quark GPDs of each sector,
\begin{equation}
 F_{q\bar q}(Q^2)
 =
 \int_0^1\!dx\;H^{q}_{q\bar q}(x,0,t),
 \qquad
 F_{q\bar q g}(Q^2)
 =
 \int_0^1\!dx\;H^{q}_{q\bar q g}(x,0,t).
 \label{eq:ff_fock_gpd}
\end{equation}
With both Fock sectors, the charge is correctly normalized,
\begin{equation}
 F(0)
 =
 \nfock\big(1+g_s^2\Phat\big)
 =1 .
 \label{eq:Fv0}
\end{equation}
Numerically, we obtain $F(0)=1.00$, with the individual contributions
$F_{q \bar q}(0)=0.60$ and $F_{q \bar q g}(0)=0.40$.

In Fig.~\ref{fig:FQ2}, we present the charge form factor $F(Q^2)$ of
the full pion state, the two sector contributions, and the $q\bar q$
state with its own normalization, together with the experimental
data~\cite{Brown:1973wr,Ackermann:1977rp,Brauel:1979zk,JeffersonLabFpi:2000nlc,JeffersonLabFpi-2:2006ysh,JeffersonLabFpi:2007vir,JeffersonLab:2008jve,Bebek:1977pe,Adylov:1977kj,Dally:1981ur,Dally:1982zk,NA7:1986vav}.
At small $Q^2$, the form factor of the full state falls faster than the
$q\bar q$-only result, because the $q\bar q g$ sector has a broader
transverse charge distribution. In Fig.~\ref{fig:FQ21}, we present
$Q^2F(Q^2)$. It rises at low $Q^2$, reaches a maximum near
$Q^2\simeq 2.5~\GeV^2$, and then decreases, in good agreement with the data
over the measured range. The charge radius follows from the slope of the
form factor at $Q^2=0$,
\begin{equation}
 \langle r^2\rangle
 =
 -6\left.\frac{\mathrm{d}F(Q^2)}{\mathrm{d}Q^2}\right|_{Q^2=0}.
 \label{eq:chargeradius}
\end{equation}
We obtain $r=0.663~\fm$ for the full state and
$r=0.530~\fm$ for the valence $q\bar q$ sector alone, compared with the
experimental value $r=0.659(4)~\fm$~\cite{ParticleDataGroup:2024cfk}.
The full-state radius reproduces the experimental value, to which $\beta$ is fitted, while the radius of the $q\bar q$ state alone is too small.
\section{Parton Distribution Functions}
\label{sec:pdf}
GPDs and TMDs give three-dimensional information on the pion, but they
are hard to measure because there is no pion target. Most experimental
information on pion structure comes from fixed-target experiments with
pion beams, through cross sections measured as functions of the Feynman
variable $x_F$ and the dilepton invariant mass, as discussed in
Sec.~\ref{sec:pheno}. These data constrain the pion PDFs. On the
theoretical side, only a few studies have addressed the dynamical gluon of
the pion, in the BLFQ approach~\cite{Lan:2021wok,Lan:2024ais} and in
Refs.~\cite{Han:2024yzj,Freese:2021zne}.

The quark and gluon PDFs are defined by the light-cone correlation
functions obtained from Eqs.~\eqref{eq:tmdq}--\eqref{eq:tmdg} at
$\bm{z}_{\perp}=0$,
\begin{align}
q(x)&=\half\!\int\!\frac{dz^-}{2\pi}\,e^{ixP^+z^-/2}
\big\langle\pi(P)\big|\bar\psi\big(-\tfrac z2\big)\,\gamma^+\,
\psi\big(\tfrac z2\big)\big|\pi(P)\big\rangle
\Big|_{z^+=\bm{z}_\perp=0},
\label{eq:pdfqdirect}\\
g(x)&=\frac{1}{xP^+}\!\int\!\frac{dz^-}{4\pi}\,e^{ixP^+z^-/2}
\big\langle\pi(P)\big|G^{+i}\big(-\tfrac z2\big)G^{+i}\big(\tfrac z2\big)
\big|\pi(P)\big\rangle
\Big|_{z^+=\bm{z}_\perp=0}.
\label{eq:pdfgdirect}
\end{align}
Equivalently, they follow from the TMDs of Sec.~\ref{sec:tmd}
integrated over the transverse momentum, or from the GPDs of
Sec.~\ref{sec:gpd} at $t=0$,
\begin{align}
q(x)&=\int\!d^{2}\bm{k}_{\perp}\,f_{1}^{q}(x,k_{\perp}^{2})=H^{q}(x,0,0),
\label{eq:pdfqfromtmd}\\
g(x)&=\int\!d^{2}\bm{k}_{\perp}\,f_{1}^{g}(x,k_{\perp}^{2})=H^{g}(x,0,0),
\label{eq:pdfgfromtmd}
\end{align} 
Equation~\eqref{eq:pdfgfromtmd} gives $g(x)$ and not $xg(x)$ because of
the factor $1/x$ in the gluon operator of Eq.~\eqref{eq:gpdg}. With the
gauge links set to unity, the LFWF overlap representations are
\begin{align}
q(x)&=\sum_{N,\{\lambda_i\}}\int\![N]\;\big|\Psi_N\big|^2\,\delta(x-x_q),
\label{eq:pdfqov}\\
g(x)&=\sum_{N,\{\lambda_i\}}\int\![N]\;\big|\Psi_N\big|^2\,\delta(x-x_g),
\label{eq:pdfgov}
\end{align}
with the measure of Eq.~\eqref{eq:Nmeasure}. In terms of the two sectors,
\begin{align}
q(x)&=n\left[\sum_{s_qs_{\bar q}}\int\!\frac{d^2\bm{k}_\perp}{16\pi^3}
\big|\Psi^{\pi}_{q\bar q}(x,\bm{k}_\perp,s_q,s_{\bar q})\big|^2
+\frac{1}{(16\pi^3)^2}\!\int\! d^2\bm{k}_\perp\,dx_g\,d^2\bm{k}_{g\perp}
\!\!\sum_{s_qs_{\bar q}\lambda}\!\!\big|\Psi_3\big|^2\right],
\label{eq:pdfqexpl}\\
g(x)&=\frac{n}{(16\pi^3)^2}\!\int\! d^2\bm{k}_\perp\,dx_q\,d^2\bm{k}_{q\perp}
\!\!\sum_{s_qs_{\bar q}\lambda}\!\!\big|\Psi_3\big|^2 ,
\label{eq:pdfgexpl}
\end{align}
where, as in Eqs.~\eqref{eq:tmdqexpl}--\eqref{eq:tmdgexpl}, the struck parton
is fixed at $x_q=x$ in Eq.~\eqref{eq:pdfqexpl} and at $x_g=x$ in
Eq.~\eqref{eq:pdfgexpl}.

\par
\begin{figure}[t]
 \centering
 \includegraphics[width=0.99\textwidth]{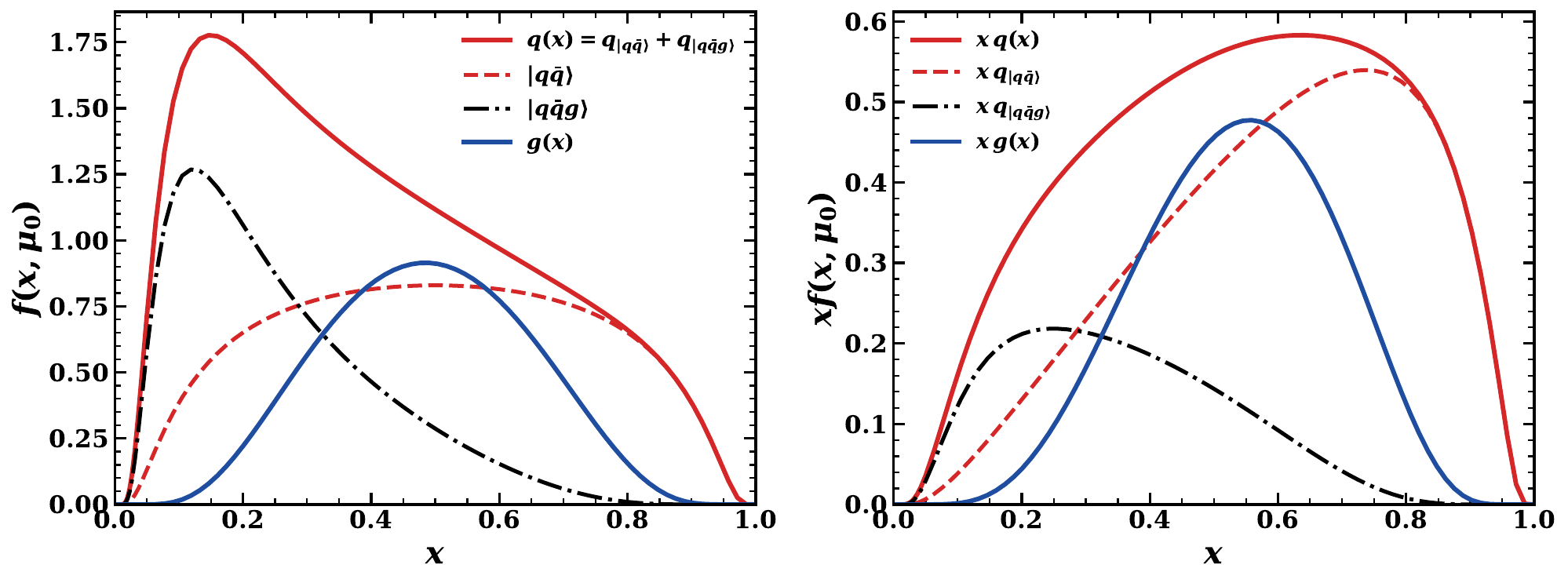}
 \caption{(Color online) Model-scale PDFs, $f(x,\mu_0)$ (left) and $xf(x,\mu_0)$ (right), for the quark, its two sector contributions and the gluon.}
 \label{fig:modelscale_pdf}
\end{figure}

 The quark distribution of the pion receives contributions from both Fock sectors,
\begin{equation}
q(x)=q_{\qqbar}(x)+q_{\qqbarg}(x).
\label{eq:qsectors}
\end{equation}
In Fig.~\ref{fig:modelscale_pdf}, we present the two contributions and
their sum for both $f(x,\mu_0)$ and $xf(x,\mu_0)$. In the valence sector,
the quark distribution is symmetric around $x=0.5$. Once the
$\qqbarg$ sector is included, the struck quark shares the pion momentum with
a gluon as well as with the antiquark, and the distribution moves toward
smaller $x$. This shift is seen directly in the quark distribution of the
$\qqbarg$ sector. The gluon comes entirely from the $\qqbarg$ sector. At
the model scale, it vanishes at both small and large $x$ and has its maximum
near $x=0.5$. The quarks of both Fock sectors carry $81\%$ of the pion momentum and the
gluon carries the remaining $19\%$, $\langle x\rangle_g=0.193$. For
comparison, the gluon carries about $6\%$ of the momentum at the initial
scale of the Bethe--Salpeter calculation of Ref.~\cite{dePaula:2023ver}. In our model, the gluon is already present in the input and is not
generated only by the evolution. The Fock space contains no
$|q\bar qq\bar q\rangle$ component, so the sea is absent at $\mu_0$ and is
produced entirely by the evolution.

\par
\begin{figure}[t]
 \centering
\includegraphics[width=0.99\textwidth]{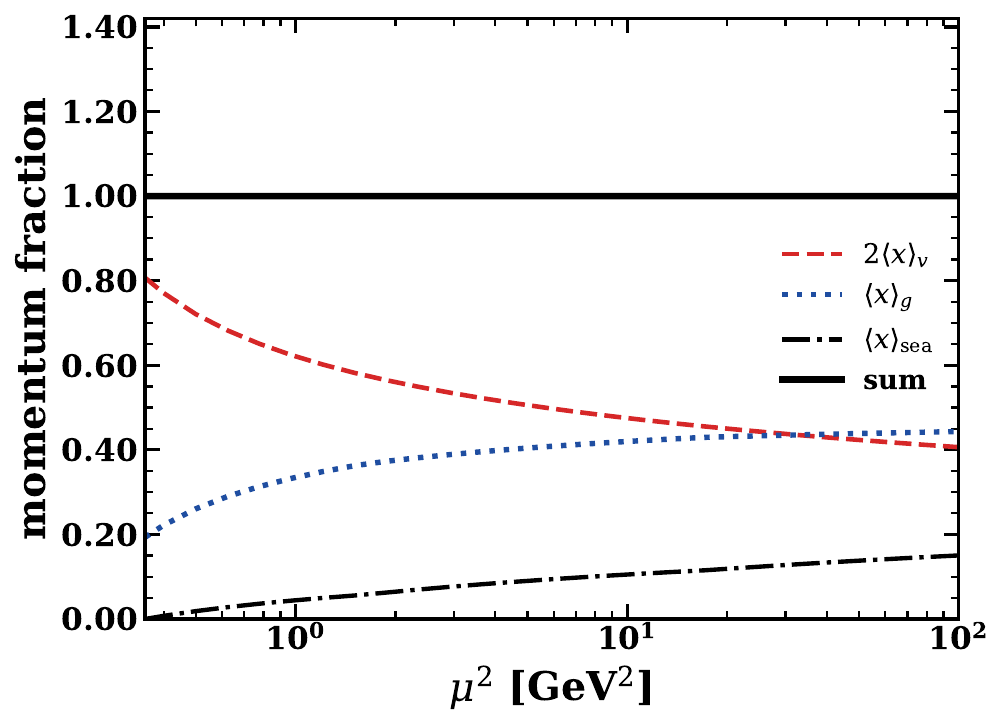}
 \caption{Momentum fractions $2\langle x\rangle_v$, $\langle x\rangle_g$, $\langle x\rangle_{\rm sea}$ and their sum, against $\mu^2$.}
 \label{fig:gpdpanelquark}
\end{figure}

 The valence normalization and the momentum sum rule are not imposed in
our model. Both follow from the normalization of the Fock expansion. The
momentum sum rule reads
\begin{equation}
\int_0^1 dx\,x\big[q_{\qqbar}(x,\mu^2)+\bar q_{\qqbar}(x,\mu^2)+q_{\qqbarg}(x,\mu^2)
+\bar q_{\qqbarg}(x,\mu^2)+g(x,\mu^2)+S(x,\mu^2)\big]=1 ,
\label{eq:momsum}
\end{equation}
where $S(x,\mu^2)$ is the total sea quark distribution, which is zero at the
model scale. The Mellin moments of the distributions are defined as
\begin{equation}
\langle x^{n}\rangle_i(\mu^2)=\int_0^1 dx\,x^{n}f_i(x,\mu^2),
\qquad n=1,2,3,
\label{eq:mellin}
\end{equation}
where $i=v,g,S$ denotes the valence quark of a single flavor, $f_v=q-\bar q$,
the gluon ($f_g=g$) and the total sea
($f_S=\sum_{q}(q+\bar q)-2f_v$). For $n=1$, the moment is the momentum fraction, and Eq.~\eqref{eq:momsum} can be written as
\begin{equation}
2\langle x\rangle_v+\langle x\rangle_g+\langle x\rangle_S=1 .
\label{eq:momsum2}
\end{equation}
At the model scale, we obtain $2\langle x\rangle_v=0.807$ and
$\langle x\rangle_g=0.193$, and their sum is one. The sum rule remains
satisfied after NLO DGLAP evolution at all $\mu^2$. In
Fig.~\ref{fig:gpdpanelquark}, we present the momentum fractions of the
valence quark, the gluon and the sea as functions of $\mu^2$, together with
their sum. The values of $\langle x^n\rangle_i$ for $n=1,2,3$ at the model
scale and at $\mu^2=4$, $5$ and $16\GeV^2$ are listed in
Table~\ref{tab:mellin}.

\par
\begin{figure}[t]
 \centering
 \includegraphics[width=0.99\textwidth]{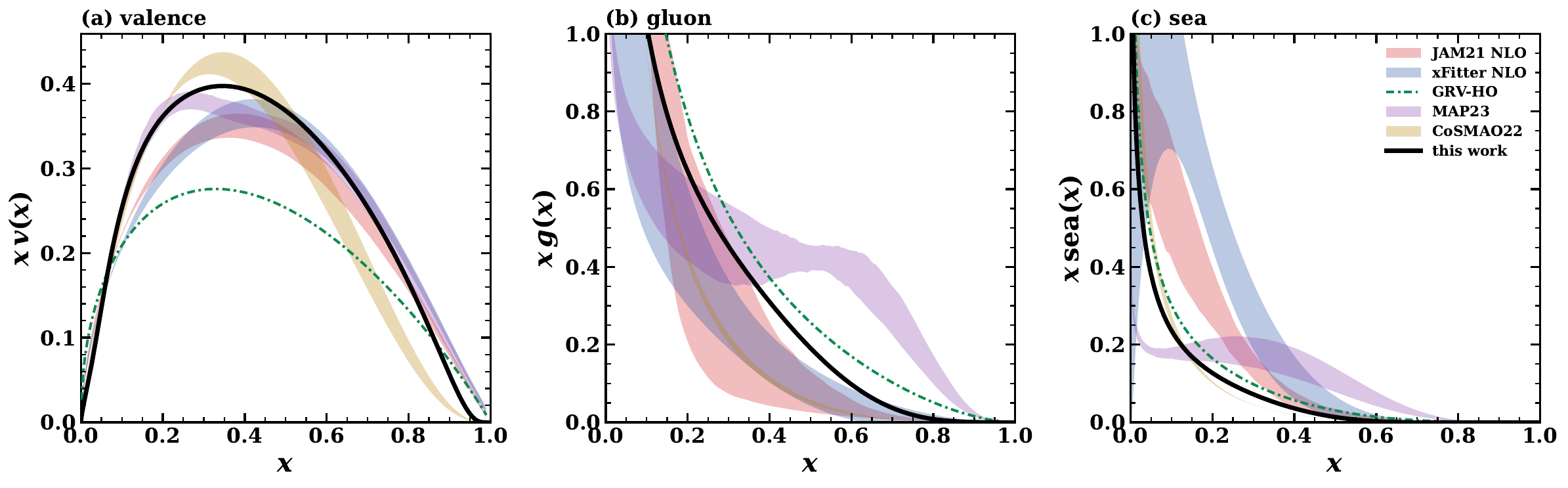}
 \caption{(Color online) Valence (a), gluon (b) and sea (c) PDFs at $\mu^2=4$ GeV$^2$ compared with JAM21~\cite{Barry:2021osv}, xFitter~\cite{Novikov:2020snp}, MAP23~\cite{Pasquini:2023aaf}, GRV~\cite{Gluck:1991ey} and CoSMAO22~\cite{Lu:2022cjx}.}
 \label{fig:pdflat}
\end{figure}

 In Fig.~\ref{fig:pdflat}, we compare our results at $\mu^2=4\GeV^2$ with CoSMAO22~\cite{Lu:2022cjx},
JAM21~\cite{Barry:2021osv}, xFitter~\cite{Novikov:2020snp},
MAP23~\cite{Pasquini:2023aaf} and GRV~\cite{Gluck:1991ey}. The gluon and sea
panels have the same vertical scale. Our valence distribution agrees with
JAM21, xFitter and MAP23 and differs from CoSMAO22. The gluon distribution
lies within the spread of the other results and becomes very small above
$x\simeq0.8$. The sea is generated only by gluon splitting in the
evolution. At low $x$, our sea distribution is smaller than JAM21 and xFitter and close to GRV. From the
model scale to $\mu^2=16\GeV^2$, $2\langle x\rangle_v$ decreases from $0.81$
to $0.46$, the gluon momentum fraction increases from $0.19$ to $0.43$ and
the sea increases from zero to $0.114$ (Table~\ref{tab:mellin}). At
$\mu^2=5\GeV^2$, our gluon momentum fraction is $0.404$, which is close to
JAM21 ($0.380$) and MAP23 ($0.389$) and larger than xFitter ($0.265$). The
xFitter set has a smaller gluon, but its sea is about three times
larger than ours. Since there is no sea at the model scale, the momentum of
the sea at higher $\mu^2$ comes from our model gluon.

\par
\begin{figure}[t]
 \centering
 \includegraphics[width=0.99\textwidth]{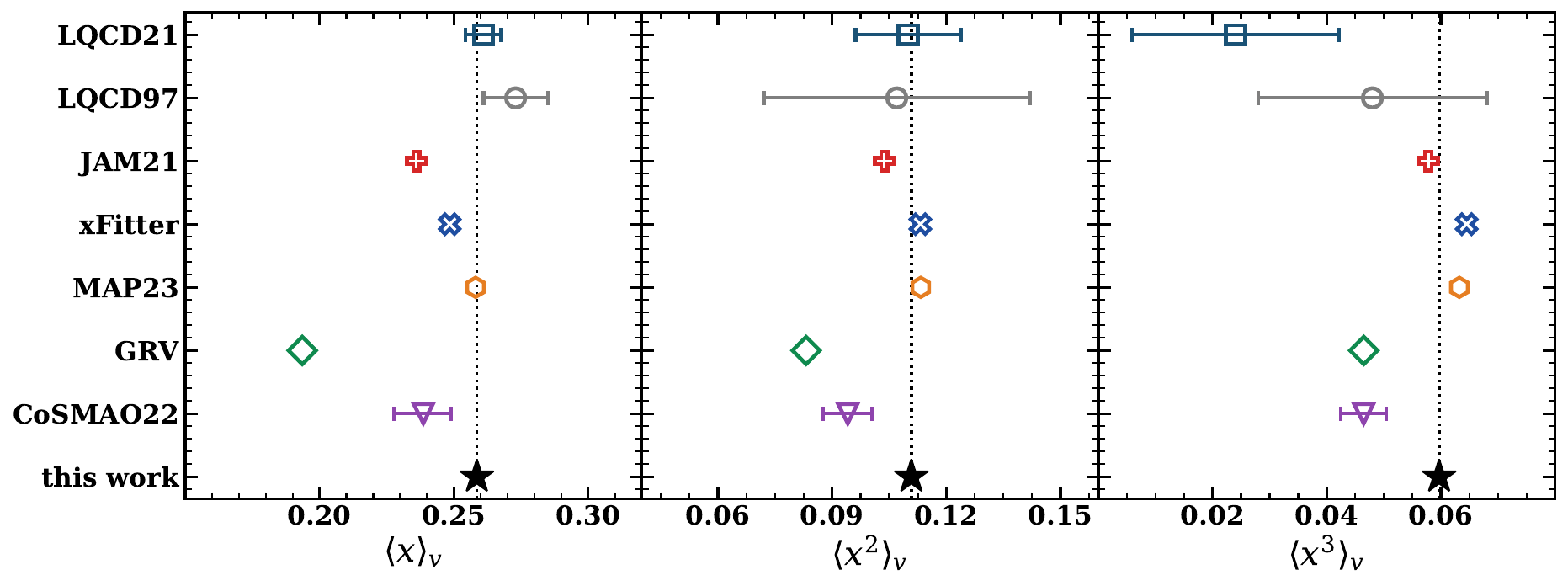}
 \caption{(Color online) Valence Mellin moments at $\mu^2=4$ GeV$^2$ compared with JAM21~\cite{Barry:2021osv}, xFitter~\cite{Novikov:2020snp}, MAP23~\cite{Pasquini:2023aaf}, GRV~\cite{Gluck:1991ey} and CoSMAO22~\cite{Lu:2022cjx} and lattice QCD, LQCD21~\cite{Alexandrou:2021mmi} and LQCD97~\cite{Best:1997qp}. The dotted line marks this work.}
 \label{fig:mellinquark}
\end{figure}

\begin{figure}[t]
 \centering
 \includegraphics[width=0.99\textwidth]{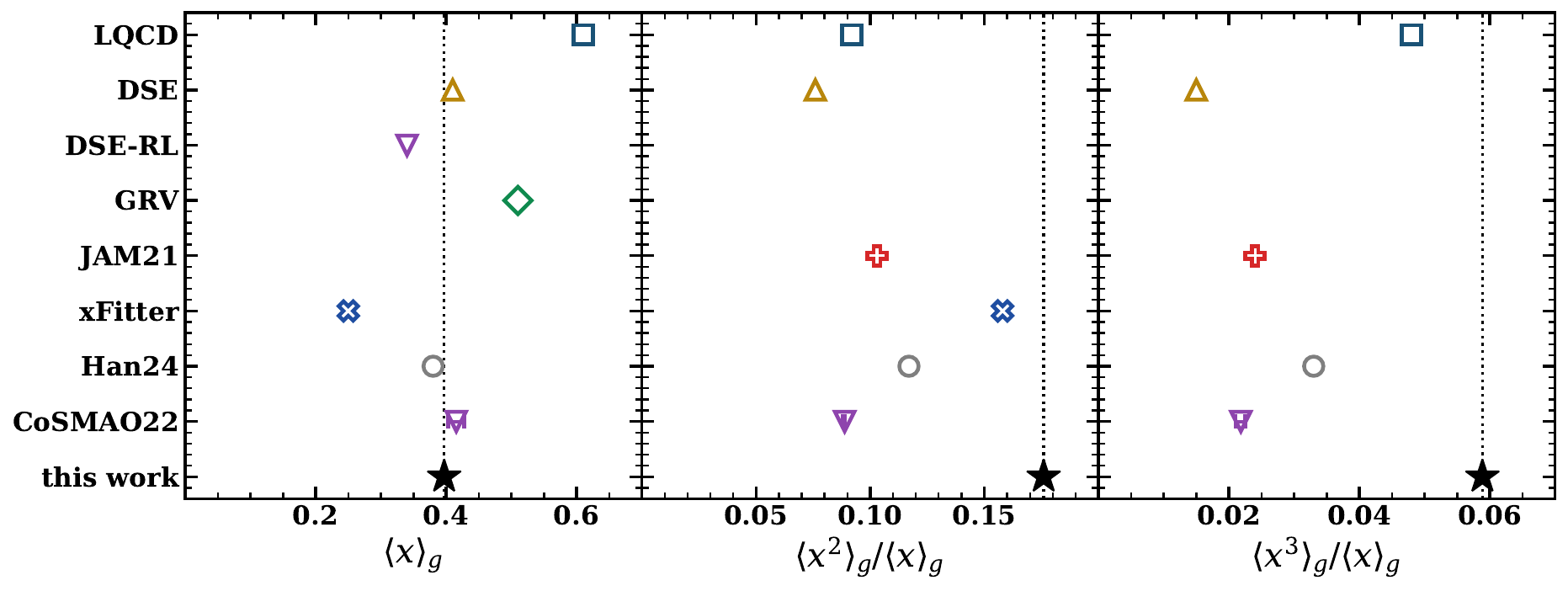}
 \caption{(Color online) $\langle x\rangle_g$ and $\langle x^n\rangle_g/\langle x\rangle_g$ at $\mu^2=4$ GeV$^2$ compared with lattice QCD~\cite{Fan:2021bcr}, DSE~\cite{Cui:2020tdf}, GRV~\cite{Gluck:1991ey}, JAM21~\cite{Barry:2021osv}, xFitter~\cite{Novikov:2020snp}, Han24~\cite{Han:2024yzj} and CoSMAO22~\cite{Lu:2022cjx}. The dotted line marks this work.}
 \label{fig:mellin}
\end{figure}

\begin{figure}[t]
 \centering
 \includegraphics[width=0.99\textwidth]{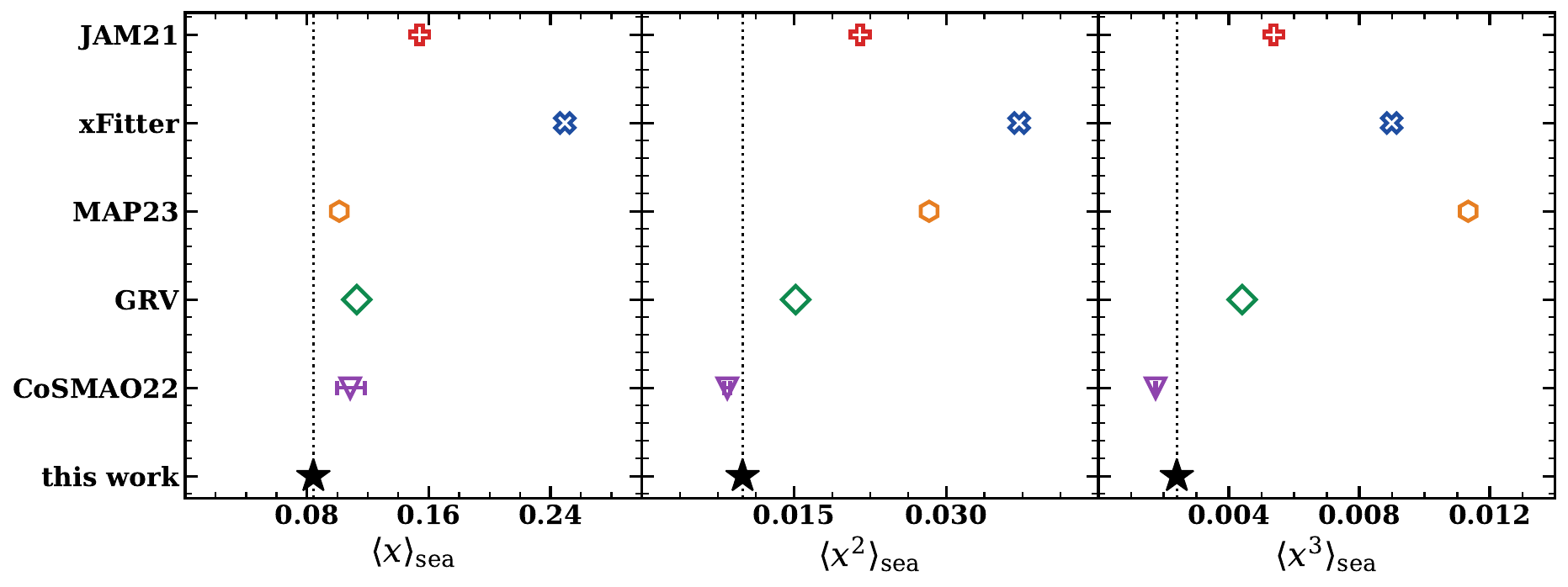}
 \caption{(Color online) Sea Mellin moments at $\mu^2=4$ GeV$^2$ compared with JAM21~\cite{Barry:2021osv}, xFitter~\cite{Novikov:2020snp}, MAP23~\cite{Pasquini:2023aaf}, GRV~\cite{Gluck:1991ey} and CoSMAO22~\cite{Lu:2022cjx}. The dotted line marks this work.}
 \label{fig:mellinsea}
\end{figure}

 The higher moments, $n=2$ and $3$, describe the shape of the
distributions. In Fig.~\ref{fig:mellinquark}, we present the valence
moments at $\mu^2=4\GeV^2$. We obtain $\langle x\rangle_v=0.259$,
$\langle x^2\rangle_v=0.111$ and $\langle x^3\rangle_v=0.060$. The first two
agree with lattice QCD, which gives $0.261(7)$ and $0.110(14)$ in
LQCD21~\cite{Alexandrou:2021mmi} and $0.273(12)$ and $0.107(35)$ in
LQCD97~\cite{Best:1997qp}. For $\langle x^3\rangle_v$, the two lattice results
differ from each other, and our value is close to LQCD97, $0.048(20)$. To compare only the shape, we divide by the first moment and find
$\langle x^2\rangle_v/\langle x\rangle_v=0.43$ and
$\langle x^3\rangle_v/\langle x\rangle_v=0.23$. The corresponding values are
$0.44$ and $0.25$ for JAM21, $0.46$ and $0.26$ for xFitter, and $0.44$ and
$0.24$ for MAP23. The shape of the valence distribution is therefore nearly
the same in all these results. CoSMAO22 gives $0.39$ and $0.19$, which means
its valence quark is softer at large $x$. In Fig.~\ref{fig:mellin}, we
present the gluon momentum fraction and the ratios
$\langle x^n\rangle_g/\langle x\rangle_g$.
At $\mu^2=4\GeV^2$, we obtain $\langle x\rangle_g=0.398$,
$\langle x^2\rangle_g/\langle x\rangle_g=0.18$ and
$\langle x^3\rangle_g/\langle x\rangle_g=0.059$. Our gluon momentum fraction
is close to the DSE result, $0.41$, and to Han24, $0.38$, and it is smaller
than the lattice value, $0.61$. Computed from the LHAPDF grids
(Table~\ref{tab:mellin}), the two ratios are $0.12$ and $0.031$ for JAM21,
$0.17$ and $0.052$ for xFitter, $0.19$ and $0.073$ for GRV, and $0.089$ and
$0.022$ for CoSMAO22. MAP23 gives much larger values, $0.33$ and $0.17$. Its
gluon momentum fraction, $0.382$, is close to ours, but its gluon
distribution is harder and has more strength at large $x$. Our gluon lies
between the softer and the harder gluon distributions of the other analysis. In Fig.~\ref{fig:mellinsea}, we present the sea moments. At $\mu^2=4\GeV^2$, our sea carries $\langle x\rangle_{\rm sea}=0.085$ of the pion
momentum. This is somewhat smaller than MAP23 ($0.101$), CoSMAO22 ($0.108$) and GRV
($0.113$), smaller than JAM21 ($0.154$) and much smaller than xFitter
($0.249$). The higher moments, $\langle x^2\rangle_{\rm sea}=0.010$ and
$\langle x^3\rangle_{\rm sea}=0.002$, are the smallest together with
CoSMAO22. In our model, the sea is generated only by the evolution from a
scale where it is zero, so it remains concentrated at small $x$.

\begin{table*}[htbp]\centering
\caption{Mellin moments of Eq.~\eqref{eq:mellin} for the single-flavor valence, gluon and sea distributions at the model scale and after NLO evolution, compared with JAM21~\cite{Barry:2021osv}, xFitter~\cite{Novikov:2020snp}, MAP23~\cite{Pasquini:2023aaf}, GRV~\cite{Gluck:1991ey} and CoSMAO22~\cite{Lu:2022cjx} from LHAPDF~\cite{Buckley:2014ana}. For JAM21 and MAP23, we use the replica mean, and for the other sets the central member.}
\label{tab:mellin}
\footnotesize
\setlength{\tabcolsep}{4pt}
\begin{tabular}{ll ccc ccc ccc}
\toprule
 & & \multicolumn{3}{c}{valence} & \multicolumn{3}{c}{gluon} & \multicolumn{3}{c}{sea}\\
\cmidrule(lr){3-5}\cmidrule(lr){6-8}\cmidrule(lr){9-11}
scale & set & $\langle x\rangle$ & $\langle x^2\rangle$ & $\langle x^3\rangle$
 & $\langle x\rangle$ & $\langle x^2\rangle$ & $\langle x^3\rangle$
 & $\langle x\rangle$ & $\langle x^2\rangle$ & $\langle x^3\rangle$\\
\midrule
$\mu_0=0.593\GeV$ & This work & 0.403 & 0.218 & 0.138 & 0.193 & 0.104 & 0.060 & 0.000 & 0.000 & 0.000\\
\midrule
$\mu=2\GeV$ & This work & 0.259 & 0.111 & 0.060 & 0.398 & 0.070 & 0.023 & 0.085 & 0.010 & 0.002\\
 & JAM21 & 0.236 & 0.104 & 0.058 & 0.374 & 0.045 & 0.011 & 0.154 & 0.022 & 0.005\\
 & xFitter & 0.248 & 0.113 & 0.065 & 0.254 & 0.042 & 0.013 & 0.249 & 0.037 & 0.009\\
 & MAP23 & 0.258 & 0.113 & 0.063 & 0.382 & 0.127 & 0.063 & 0.101 & 0.028 & 0.011\\
 & GRV & 0.194 & 0.083 & 0.046 & 0.511 & 0.098 & 0.037 & 0.113 & 0.015 & 0.004\\
 & CoSMAO22 & 0.239 & 0.094 & 0.046 & 0.415 & 0.037 & 0.009 & 0.108 & 0.008 & 0.002\\
\midrule
$\mu^2=5\GeV^2$ & This work & 0.253 & 0.107 & 0.057 & 0.404 & 0.068 & 0.022 & 0.090 & 0.010 & 0.002\\
 & JAM21 & 0.231 & 0.100 & 0.055 & 0.380 & 0.045 & 0.011 & 0.158 & 0.021 & 0.005\\
 & xFitter & 0.243 & 0.109 & 0.062 & 0.265 & 0.042 & 0.013 & 0.249 & 0.036 & 0.009\\
 & MAP23 & 0.252 & 0.109 & 0.061 & 0.389 & 0.121 & 0.059 & 0.106 & 0.029 & 0.011\\
 & GRV & 0.190 & 0.081 & 0.045 & 0.513 & 0.094 & 0.035 & 0.118 & 0.016 & 0.004\\

\midrule
$\mu^2=16\GeV^2$ & This work & 0.229 & 0.092 & 0.047 & 0.429 & 0.059 & 0.018 & 0.114 & 0.012 & 0.003\\
 & JAM21 & 0.209 & 0.086 & 0.046 & 0.408 & 0.042 & 0.010 & 0.174 & 0.020 & 0.005\\
 & xFitter & 0.220 & 0.094 & 0.051 & 0.312 & 0.041 & 0.012 & 0.249 & 0.033 & 0.008\\
 & MAP23 & 0.228 & 0.094 & 0.050 & 0.415 & 0.100 & 0.043 & 0.127 & 0.029 & 0.011\\
 & GRV & 0.173 & 0.070 & 0.038 & 0.518 & 0.078 & 0.026 & 0.141 & 0.017 & 0.005\\
\bottomrule
\end{tabular}
\end{table*}

%==============================================================
\section{Mass Decomposition}
\label{sec:mass}

\par The QCD energy--momentum tensor contains the trace anomaly, and its
forward matrix elements determine how the pion mass is shared among its
constituents. We use the decomposition of Ji~\cite{Ji:1994av,Ji:1995sv,Han:2024yzj}, in
which the hadron mass separates into four terms,
\begin{equation}
 M = M_q + M_g + M_m + M_a ,
 \label{eq:jidecomp}
\end{equation}
namely the quark and gluon kinetic energies, the quark mass term and the trace
anomaly term~\cite{Baikov:2014qja}. In units of the hadron mass, they are~\cite{Ji:1995sv}
\begin{equation}
 M_q=\frac34\left(a-\frac{b}{1+\gamma_m}\right),\quad
 M_g=\frac34\,(1-a),\quad
 M_m=\frac{(4+\gamma_m)\,b}{4(1+\gamma_m)},\quad
 M_a=\frac{1-b}{4},
 \label{eq:jiterms}
\end{equation}
where $a=2\langle x\rangle_v+\langle x\rangle_S$ is the total quark momentum
fraction, $b$ is the fraction of the mass coming from the quark scalar term
and $\gamma_m$ is the anomalous dimension of the quark mass. The momentum
fraction $a$ is taken from our evolved distributions, and at
$\mu^2=4\GeV^2$ we have $a=0.602$. For the pion, $b$ is fixed by chiral
symmetry. The Feynman--Hellmann theorem together with the
Gell-Mann--Oakes--Renner relation gives
$\langle\pi|m_q\bar\psi\psi|\pi\rangle=M_\pi^2$, that is, $b=1/2$, to leading
order in the quark mass. The same decomposition of the pion was studied in
Refs.~\cite{Ji:1995sv,Han:2024yzj}.

\par
\begin{figure}[t]
 \centering
 \includegraphics[width=0.99\textwidth]{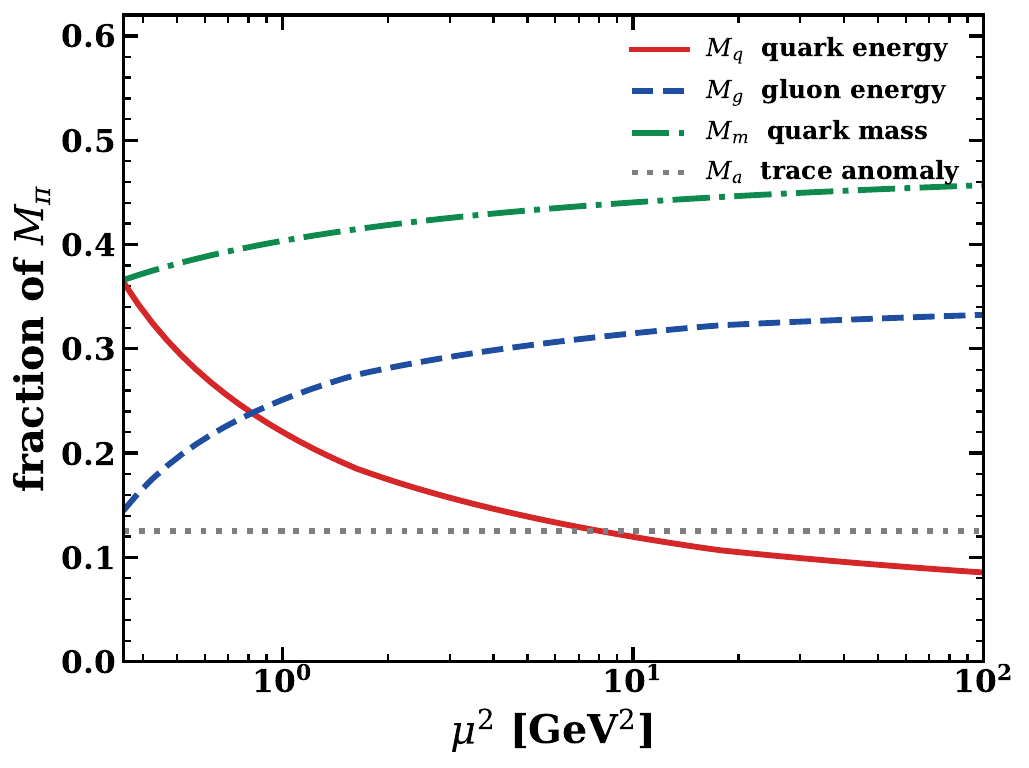}
 \caption{(Color online) Ji mass decomposition against $\mu^2$ with $b=1/2$.}
 \label{fig:T00}
\end{figure}

 The four terms are listed in Table~\ref{tab:massdec}, and in
Fig.~\ref{fig:T00}, we present their scale dependence. At $\mu^2=4\GeV^2$, the quark
kinetic term gives $15\%$ of the pion mass, the gluon kinetic term $30\%$,
the quark mass term $43\%$ and the trace anomaly $12\%$. The gluon kinetic and
anomaly terms together give $42\%$ of the pion mass, although the constituent
gluon of the $\qqbarg$ sector carries only $19\%$ of the momentum at the model
scale. The rest of the gluon contribution is generated by the evolution.

\begin{table}[htbp]\centering
\caption{Ji mass decomposition of Eq.~\eqref{eq:jiterms} at $\mu^2=4$ GeV$^2$ in percent of the pion mass, with $b=1/2$.}
\label{tab:massdec}
\begin{tabular}{lcccc}
\hline\hline
 & $M_q$ & $M_g$ & $M_m$ & $M_a$\\
\hline
Pion & 15\% & 30\% & 43\% & 12\%\\
\hline\hline
\end{tabular}
\end{table}

%==============================================================
\section{Pion-Induced Processes}
\label{sec:pheno}
 \par The pion is unstable, with a lifetime of about $26$~ns, so it cannot be
 used as a target and no pion collider exists. Information on its partonic
 structure therefore comes from secondary pion beams on fixed targets and from
 the pion cloud of the nucleon, and most of these data were taken several
 decades ago. The main source has been the Drell--Yan process,
 $\pi^\pm N\to\mu^+\mu^-X$, measured by NA3~\cite{NA3:1983ejh},
 WA11~\cite{Barate:1979da}, E331~\cite{Anderson:1976ja},
 NA10~\cite{NA10:1985ibr,NA10:1987sqk}, E326~\cite{Greenlee:1985gd},
 E444~\cite{Newman:1979tv}, E537~\cite{Anassontzis:1987hk},
 E615~\cite{E615:1989bda} and, more recently, COMPASS~\cite{COMPASS:2017jbv}.
 At leading order, the dilepton pair is produced by the annihilation
 $\bar q_\pi q_N\to\gamma^*\to\mu^+\mu^-$ of a valence antiquark of the pion
 with a quark of the target, and the gluon enters only at NLO through the
 Compton channel $qg\to\gamma^*q$. The Drell--Yan data therefore mainly
 constrain the valence quark distribution of the pion at large
 $x$~\cite{McGaughey:1999mq,Reimer:2007iy,Chang:2013opa}.

 \par The gluon distribution is probed more directly by $J/\psi$ production,
 $\pi A\to J/\psi X$, which proceeds through gluon fusion $gg\to c\bar c$ and
 quark--antiquark annihilation $q\bar q\to c\bar c$. The $gg$ channel dominates
 at higher beam energies, so these data are sensitive to the pion gluon at
 moderate and large $x$~\cite{Chang:2020rdy}. $J/\psi$ production with pion
 beams was measured by WA39~\cite{Corden:1980rb}, WA11~\cite{McEwen:1982fe},
 NA3~\cite{NA3:1983ltt}, E537~\cite{Tzamarias:1990ij},
 E705~\cite{E705:1992jno} and E672/E706~\cite{E672:1995won}. The most recent
 information comes from leading-neutron production in deep inelastic
 scattering, $ep\to e'nX$, measured at HERA by H1~\cite{H1:2010hym} and
 ZEUS~\cite{ZEUS:2002gig}. In this Sullivan process~\cite{Sullivan:1971kd}, the virtual photon scatters off a $\pi^+$ emitted by the proton, which gives access to the pion structure function $F_2^\pi$ at small $x$ and hence to the pion sea and gluon. The extraction depends on the assumed
 pion flux, which is the main uncertainty of these data. New measurements are
 expected from AMBER at CERN~\cite{Adams:2018pwt} and from leading-neutron
 production at the EIC~\cite{Aguilar:2019teb,Chang:2026tgb} and EicC~\cite{Lu:2025bnm}.
\par In this section, we evolve our model-scale PDFs to the scale of each
measurement and compare the predictions with the pion-induced Drell--Yan,
$J/\psi$ and $\Upsilon$ production and leading-neutron $F_2^\pi$ data.

\subsection{Pion-induced Drell--Yan}
\label{sec:dy}

\par The pion-induced Drell--Yan process, $\pi^-A\to\mu^+\mu^-X$, has been the
main source of information on the valence quark distribution of the pion. The
global analysis have used different parts of these data. The xFitter
analysis~\cite{Novikov:2020snp} and the JAM
analysis~\cite{Barry:2018ort,Barry:2021osv} are based mainly on the E615 and
NA10 measurements, while the statistical model of
Refs.~\cite{Bourrely:2022mjf,Bourrely:2023yzi} also includes the E326 data.
The transverse momentum spectra of E615 and E537 have been used to extract the
pion TMDs~\cite{Vladimirov:2019bfa,Cerutti:2022lmb,Chang:2026rft}. The other
fixed-target measurements are rarely included, although they cover different
beam energies and targets. In this work, we compare our results with all six
available $x_F$ spectra, namely E615~\cite{E615:1989bda},
NA10 at $194$ and $286\GeV$~\cite{NA10:1985ibr},
E326~\cite{Greenlee:1985gd}, E444~\cite{Newman:1979tv} and
E537~\cite{Anassontzis:1987hk}, and with the dilepton mass spectra of E615,
NA3~\cite{NA3:1983ejh}, WA39~\cite{Corden:1980xf}, E326 and E444.

\par We denote the momenta of the pion and the target nucleon by $p_1$ and
$p_2$ and the momenta of the two leptons by $l$ and $l'$. The kinematics of the
process are described by
\begin{equation}
s=(p_1+p_2)^2,\qquad q=l+l',\qquad m^2=q^2,\qquad
Y=\frac12\ln\frac{q^0+q^3}{q^0-q^3},
\label{eq:dykin1}
\end{equation}
where $m$ is the invariant mass and $Y$ the rapidity of the lepton pair in the
hadronic center-of-mass frame, together with
\begin{equation}
x_F=x_1-x_2,\qquad \tau=\frac{m^2}{s},\qquad
z=\frac{m^2}{\hat s}=\frac{\tau}{x_1x_2},\qquad
y=\frac{\frac{x_1}{x_2}e^{-2Y}-z}{(1-z)\big(1+\frac{x_1}{x_2}e^{-2Y}\big)} ,
\label{eq:dykin2}
\end{equation}
where $\hat s=x_1x_2s$ and $x_1$ ($x_2$) is the momentum fraction of the
annihilating parton in the pion (target). They can be written as
\begin{equation}
x_1=\sqrt{\frac{\tau}{z}\,\frac{1-(1-y)(1-z)}{1-y(1-z)}}\;e^{Y},\qquad
x_2=\sqrt{\frac{\tau}{z}\,\frac{1-y(1-z)}{1-(1-y)(1-z)}}\;e^{-Y}.
\label{eq:dyx12}
\end{equation}
At NLO in perturbative QCD, the cross section
reads~\cite{Lan:2019rba,P:2026crg,Anastasiou:2003yy,Anastasiou:2003ds,Becher:2007ty}
\begin{equation}
\frac{m^3\,d^2\sigma}{dm\,dY}=\frac{8\pi\alpha^2}{9}\,\frac{m^2}{s}
\sum_{ij}\int dx_1\,dx_2\;
\widetilde C_{ij}(x_1,x_2,s,m,\mu^2)\,
f_{i/\pi}(x_1,\mu^2)\,f_{j/A}(x_2,\mu^2),
\label{eq:dynlo}
\end{equation}
where $f_{i/\pi}$ are our evolved pion PDFs, $f_{j/A}$ are the target PDFs and
$\widetilde C_{ij}$ are the hard-scattering kernels, expanded in powers of
$\alpha_s(\mu^2)$. We set the factorization and renormalization scales to
$\mu=m$. At leading order, only the annihilation channels $(ij)=(q\bar q)$ and
$(\bar qq)$ contribute. At NLO, the Compton channels $(qg)$, $(gq)$,
$(\bar qg)$ and $(g\bar q)$ enter as well, and the explicit kernels are given
in Refs.~\cite{Anastasiou:2003yy,Lan:2019rba}. The pion gluon therefore enters
Eq.~\eqref{eq:dynlo} at $\mathcal O(\alpha_s)$ only. The measurements are
given in terms of $x_F$ and $\sqrt\tau$. Using
\begin{equation}
x_{1,2}=\frac{\pm x_F+\sqrt{x_F^2+4\tau/z}}{2},\qquad
Y=\frac12\ln\frac{\sqrt{x_F^2+4\tau/z}+x_F}{\sqrt{x_F^2+4\tau/z}-x_F},
\label{eq:dyxF}
\end{equation}
the cross section in these variables becomes
\begin{equation}
\frac{d^2\sigma}{d\sqrt\tau\,dx_F}
=\frac{\sqrt s}{\sqrt{x_F^2+4\tau/z}}\;\frac{d^2\sigma}{dm\,dY},
\label{eq:dyjac}
\end{equation}
and the mass spectra $m^3\,d\sigma/dm$ follow from Eq.~\eqref{eq:dynlo} by
integrating $Y$ over the measured range.

\par All six data sets are cross sections per nucleon, $\sigma_A/A$, measured on
nuclear targets. For each of them, we therefore use the per-nucleon NLO PDF of
its own target nucleus, $f_{j/A}=\frac{Z}{A}f_{j/p}^{A}+\frac{A-Z}{A}f_{j/n}^{A}$,
which accounts for the neutron excess and the nuclear modification of the target. We use the nCTEQ15 tungsten PDFs~\cite{Kovarik:2015cma} for E615,
NA10, E326 and E537, and the nCTEQ15 copper PDFs for E444, which used carbon,
copper and tungsten targets and quoted $1/A$ times the cross section. For the
mass spectra, we use tungsten for E615, WA39 and E326, copper for E444 and the
EPPS16 platinum PDFs~\cite{Eskola:2016oht} for NA3. The
experimental normalization is absorbed into a single factor $K$
per data set. We fit $K$ analytically by minimizing the $\chi^2$ of each
$x_F$ spectrum, so that only the shape of the spectrum tests the pion PDFs.

\par
\begin{figure*}[t]
 \centering
 \includegraphics[width=0.99\textwidth]{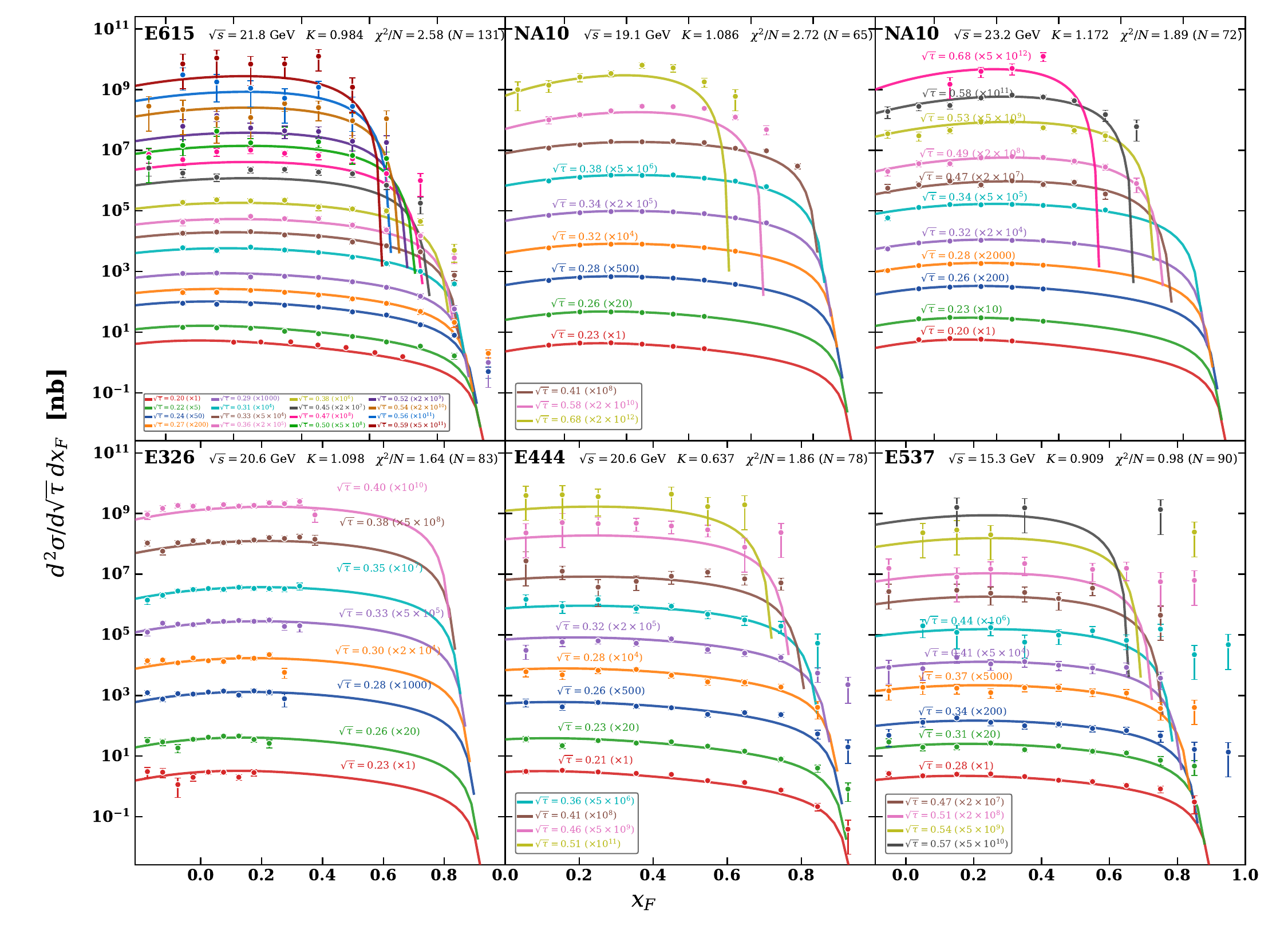}
 \caption{(Color online) Drell--Yan $x_F$ spectra for $\pi^-$ beams from E615~\cite{E615:1989bda}, NA10~\cite{NA10:1985ibr}, E326~\cite{Greenlee:1985gd}, E444~\cite{Newman:1979tv} and E537~\cite{Anassontzis:1987hk}, compared with the model. Each $\sqrt\tau$ bin is scaled as given in the legend.}
 \label{fig:dystack}
\end{figure*}

\begin{figure*}[t]
 \centering
 \includegraphics[width=0.99\textwidth]{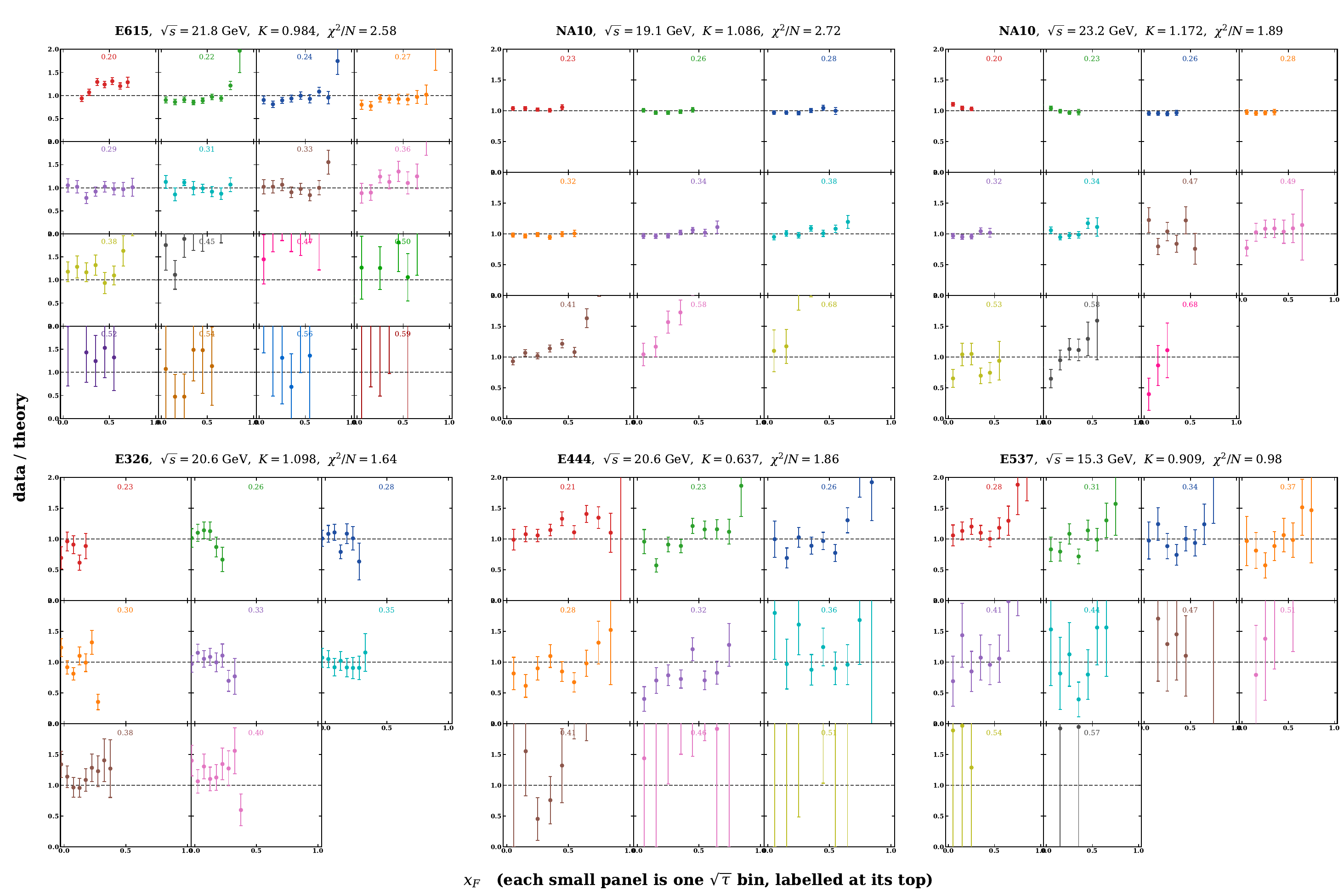}
 \caption{(Color online) Data-to-theory ratio for Fig.~\ref{fig:dystack}, one panel per $\sqrt\tau$ bin.}
 \label{fig:dyratio}
\end{figure*}

 The fitted normalizations and the $\chi^2$ of the $x_F$ spectra are listed
in Table~\ref{tab:dyxf}, and in Fig.~\ref{fig:dystack}, we present the spectra.
The model describes the shape of the E537 spectrum very well, with $\chi^2/N=0.98$, and gives $\chi^2/N=1.64$ for E326, $1.86$ for E444 and $2.58$ for
E615. The ratio of data to theory in Fig.~\ref{fig:dyratio} is close to one for
$\sqrt\tau\lesssim0.35$ over the whole $x_F$ range. At larger $\sqrt\tau$ and large $x_F$, the pion momentum fraction $x_1$ approaches one, the cross section
falls steeply and the data lie above the model, in particular for E615. The two NA10 energies need similar
normalizations, $K=1.09$ at $194\GeV$ and $K=1.17$ at $286\GeV$, with
$\chi^2/N=2.72$ and $1.89$.

\par
\begin{figure*}[t]
 \centering
 \includegraphics[width=0.99\textwidth]{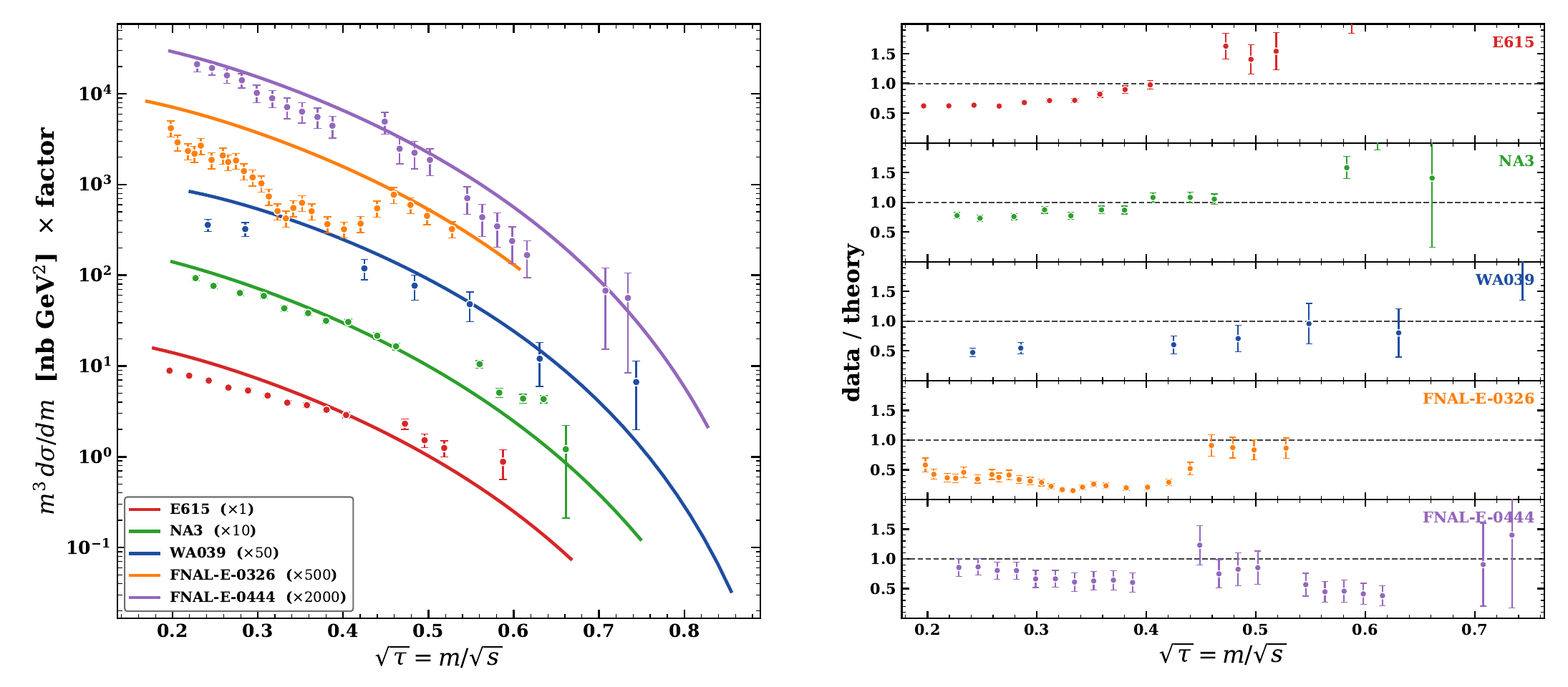}
 \caption{(Color online) Drell--Yan mass spectra $m^3d\sigma/dm$ compared with E615~\cite{E615:1989bda}, NA3~\cite{NA3:1983ejh}, WA39~\cite{Corden:1980xf}, E326~\cite{Greenlee:1985gd} and E444~\cite{Newman:1979tv} (left), and the data-to-theory ratio (right). No normalization is fitted.}
 \label{fig:dymass}
\end{figure*}

 In Fig.~\ref{fig:dymass}, we present the mass spectra $m^3\,d\sigma/dm$.
No normalization is fitted here, so the figure shows the absolute prediction.
For NA3, the ratio of data to theory stays between $0.7$ and $1.1$ up to
$\sqrt\tau\simeq0.45$. For E615, it rises from $0.62$ at small $\sqrt\tau$ to
about one at $\sqrt\tau\simeq0.4$, while the $x_F$ fit gives $K=0.98$. WA39 and E444 lie below the prediction by a factor
between $0.5$ and $0.9$, and E326 lies below it by a factor of two to five.

\begin{table*}[htbp]\centering
\caption{Drell--Yan $x_F$ data sets, with the fitted normalization $K$ and $\chi^2$ for the full input at $\mu_0=0.593$ GeV. Each set uses the per-nucleon nCTEQ15 PDFs~\cite{Kovarik:2015cma} of its own target, tungsten for E615, NA10, E326 and E537 and copper for E444.}
\label{tab:dyxf}
\small
\setlength{\tabcolsep}{4pt}
\begin{tabular}{lccccccc}
\hline\hline
data set & $E_\pi$ [GeV] & $\sqrt s$ [GeV] & $\sqrt\tau$ & $x_F$
 & $N$ & $K$ & $\chi^2$ ($\chi^2/N$)\\
\hline
E615~\cite{E615:1989bda} & 252 & 21.8 & 0.20--0.59 & $-0.05$--0.95
 & 131 & 0.984(11) & 338.0 (2.58)\\
NA10~\cite{NA10:1985ibr} & 194 & 19.1 & 0.23--0.68 & $-0.15$--0.75
 & 65 & 1.086(6) & 176.6 (2.72)\\
NA10~\cite{NA10:1985ibr} & 286 & 23.2 & 0.20--0.68 & $-0.15$--0.65
 & 72 & 1.172(8) & 135.9 (1.89)\\
E326~\cite{Greenlee:1985gd} & 225 & 20.6 & 0.23--0.40 & $-0.17$--0.38
 & 83 & 1.098(21) & 135.9 (1.64)\\
E444~\cite{Newman:1979tv} & 225 & 20.6 & 0.21--0.51 & 0.06--0.93
 & 78 & 0.637(15) & 145.2 (1.86)\\
E537~\cite{Anassontzis:1987hk} & 125 & 15.3 & 0.28--0.57 & $-0.05$--0.95
 & 90 & 0.909(29) & 88.1 (0.98)\\
\hline\hline
\end{tabular}
\end{table*}

\subsection{\texorpdfstring{$J/\psi$ and $\Upsilon$ production}{J/psi and Upsilon production}}
\label{sec:jpsi}

\par $J/\psi$ production in pion-nucleus collisions, $\pi^-A\to J/\psi X$, is
sensitive to the pion gluon, because the gluon-fusion channel $gg\to c\bar c$
gives a large part of the cross section at the energies of the fixed-target
experiments. We describe it with the color evaporation model
(CEM)~\cite{Fritzsch:1977ay,Halzen:1977rs,Gavai:1994gb}, as in
Refs.~\cite{Chang:2020rdy,Lan:2026czc}. In this model, a fixed fraction $F$ of
all $c\bar c$ pairs produced with an invariant mass between the threshold
$2m_c$ and the open-charm threshold forms a $J/\psi$ as
\begin{equation}
 \sigma_{J/\psi}=F\int_{2m_c}^{2m_D}dm_{c\bar c}\;
 \frac{d\sigma_{c\bar c}}{dm_{c\bar c}},
 \label{eq:cem}
\end{equation}
where $m_{c\bar c}$ is the invariant mass of the $c\bar c$ pair. The $x_F$
distribution of the pairs is
\begin{equation}
 \frac{d\sigma_{c\bar c}}{dx_F}=\sum_{ij}
 \int dx_1\,dx_2\;
 f_i^{\pi}(x_1,\mu_F^2)\,f_j^{A}(x_2,\mu_F^2)\;
 \hat\sigma_{ij}\big(\hat s,m_c,\mu_R^2\big)\,
 \delta\big(x_F-x_1+x_2\big),
 \label{eq:cemxf}
\end{equation}
with $\hat s=x_1x_2s$ and $\hat\sigma_{ij}$ the NLO partonic cross sections for
heavy-quark production~\cite{Nason:1987xz,Nason:1989zy,Mangano:1992kq}. At leading order, only the $q\bar q\to c\bar c$ and $gg\to c\bar c$ channels
contribute, and at NLO the $qg$ channel enters as well. We use $m_c=1.5\GeV$,
$m_D=1.87\GeV$, $\mu_F=2m_c$ and $\mu_R=m_c$. We use the nCTEQ15 PDFs of beryllium for E672/E706 and for the beryllium target of E537, of copper and tungsten for the other two E537 targets, and of lithium for E705. WA39 used a hydrogen target, and NA3 quotes its cross section per proton, so for these two experiments we use the CT18 NLO proton PDFs~\cite{Hou:2019efy}. The
normalization $F$ is not predicted by the model and is fitted separately for
each data set. Since Eq.~\eqref{eq:cemxf} is linear in $f_i^\pi$, the
contributions of the $\qqbar$ and $\qqbarg$ sectors add up to the total
exactly once $F$ is fixed. 

\par
\begin{figure*}[t]
 \centering
 \includegraphics[width=0.99\textwidth]{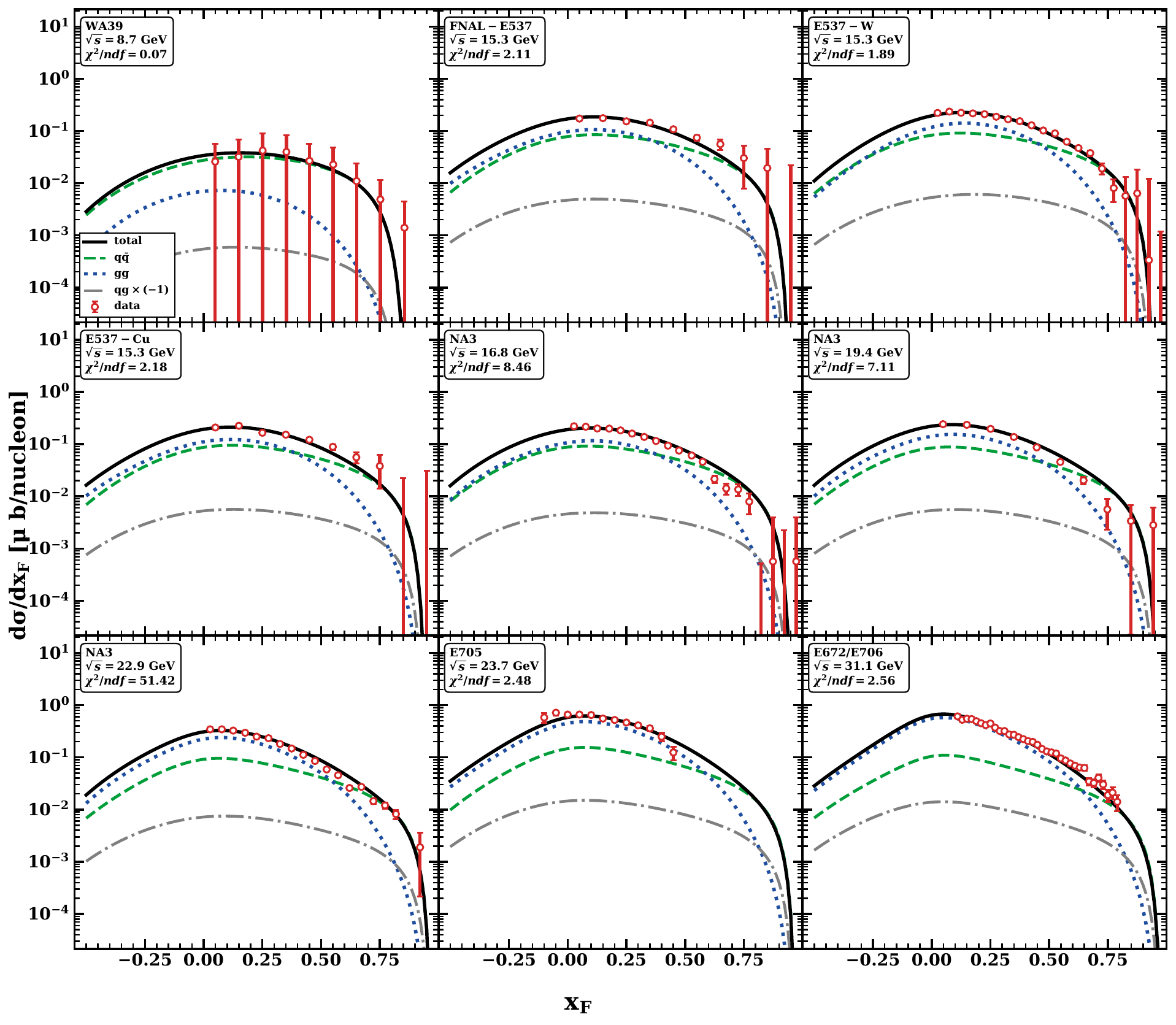}
 \caption{(Color online) $J/\psi$ cross sections $d\sigma/dx_F$ per nucleon compared with WA39~\cite{Corden:1980rb}, E537~\cite{Tzamarias:1990ij}, NA3~\cite{NA3:1983ltt}, E705~\cite{E705:1992jno} and E672/E706~\cite{E672:1995won}, with the $q\bar q$, $gg$ and $qg$ ($\times(-1)$) channels.}
 \label{fig:jpsisectors}
\end{figure*}

 We compare with the $J/\psi$ data of WA39~\cite{Corden:1980rb},
E537~\cite{Tzamarias:1990ij} on beryllium, copper and tungsten,
NA3~\cite{NA3:1983ltt} at three beam energies, E705~\cite{E705:1992jno} and
E672/E706~\cite{E672:1995won}, which cover $8.7\le\sqrt s\le31.1\GeV$ and
contain $145$ data points in total. The combined fit gives
$\chi^2/{\rm ndf}=1339.5/136=9.85$. The results for each data set are given in Table~\ref{tab:jpsifits}. In
Fig.~\ref{fig:jpsisectors}, we present the $x_F$ spectra together with the
contributions of the $q\bar q$, $gg$ and $qg$ channels. The $qg$ channel is
negative at NLO and small, at most $3.5\%$ of the total in magnitude. The
contribution of the $gg$ channel grows with the energy. Integrated over the
measured $x_F$ range, it is $13\%$ for WA39 at $\sqrt s=8.7\GeV$, between $50\%$ and $57\%$ for E537 at $\sqrt s=15.3\GeV$, $59\%$ for NA3 at $\sqrt s=19.4\GeV$ and $81\%$ for E672/E706 at $\sqrt s=31.1\GeV$. The $gg$ and $q\bar q$ channels are therefore
of equal size near $\sqrt s\simeq17\GeV$. At large $x_F$, the $q\bar q$ channel dominates at all energies, because the gluon distribution of the pion falls
faster than the valence distribution at large $x$. The $\qqbarg$ sector gives
between $16\%$ for WA39 and $50\%$ for E705 of the integrated cross section, and up
to $54\%$ locally. Its share is largest where gluon fusion dominates, so
$J/\psi$ production is the process most sensitive to the constituent gluon of
our model.

\par The description is good for WA39, the three E537 targets, E705 and
E672/E706, with $\chi^2/{\rm ndf}$ between $0.07$ and $2.56$. The three NA3
data sets have the smallest quoted uncertainties and give most of the total
$\chi^2$. The NA3 set at $\sqrt s=22.9\GeV$ alone contributes $925.5$ of
$1339.5$. At $\sqrt s=16.8\GeV$, the NA3 data fall below the model for
$0.55\lesssim x_F\lesssim0.75$. The normalization $F$ lies between $0.03$ and
$0.05$. It varies by less than a factor of two between $\sqrt s=8.7$ and
$31.1\GeV$ and across five targets, which is within the accuracy expected for
the color evaporation model.

\par
\begin{figure*}[t]
 \centering
 \includegraphics[width=0.85\textwidth]{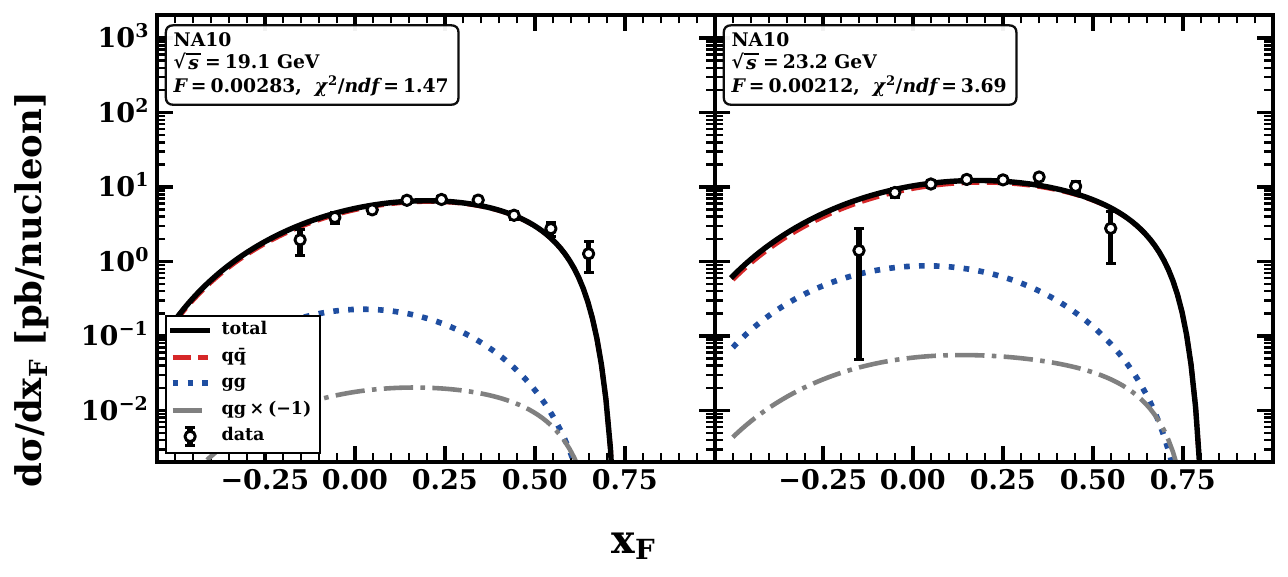}
 \caption{(Color online) $\Upsilon$ cross sections compared with NA10 at $\sqrt s=19.1$ and $23.2$ GeV~\cite{NA10:1985boe,NA10:1986qyc}, with the channels of Fig.~\ref{fig:jpsisectors}.}
 \label{fig:upssub}
\end{figure*}

 We describe $\Upsilon$ production, $\pi^-A\to\Upsilon X$, with the same color evaporation model applied to bottom case. A fixed fraction of the $b\bar b$ pairs with an invariant mass between the threshold $2m_b$ and the open-bottom threshold $2m_B$ forms an $\Upsilon$. We use
$m_b=4.75\GeV$, $m_B=5.28\GeV$, $\mu_F=2m_b$ and $\mu_R=m_b$, and the respective nuclear PDFs for the target, and compare with the NA10 data at
$194$ and $286\GeV$~\cite{NA10:1985boe,NA10:1986qyc}. The combined fit to the
$17$ data points gives $\chi^2/{\rm ndf}=37.5/15=2.50$. The results are listed in
Table~\ref{tab:jpsifits}, and in Fig.~\ref{fig:upssub}, we present the spectra. Because of the larger $b$-quark mass, $\Upsilon$ production probes larger momentum fractions, where the valence quarks dominate. The $q\bar q$ channel gives $98\%$ of the
cross section at $194\GeV$ and $95\%$ at $286\GeV$, and the $gg$ channel only
$3\%$ and $6\%$. The $\qqbarg$ sector gives $7.0\%$ and $12.9\%$ of the
integrated cross section, and at most $37\%$ at the edges of the $x_F$ range.
$\Upsilon$ production is therefore mainly a test of the valence quark
distribution at large $x$, and it complements $J/\psi$ production, which is
dominated by the gluon at the higher energies. The same pion PDFs describe
both processes without any change of parameters.

\begin{table*}[htbp]\centering
\caption{$J/\psi$ and $\Upsilon$ fits per data set, with ${\rm ndf}=N-1$, $m_D=1.87\GeV$ and $m_B=5.28\GeV$. The last three columns give the channel shares in percent.}
\label{tab:jpsifits}
\small
\setlength{\tabcolsep}{4pt}
\begin{tabular}{llccccccc}
\hline\hline
experiment & target & $\sqrt s$ [GeV] & $N$ & $F$ & $\chi^2$ ($\chi^2/{\rm ndf}$)
 & $q\bar q$ [\%] & $qg$ [\%] & $gg$ [\%]\\
\hline
WA39~\cite{Corden:1980rb} & $p$ & 8.7 & 9 & 0.0397(44) & 0.5 (0.07)
 & 88.3 & $-1.6$ & 13.3\\
E537~\cite{Tzamarias:1990ij} & Be & 15.3 & 10 & 0.0492(17) & 19.0 (2.11)
 & 52.9 & $-3.4$ & 50.5\\
E537~\cite{Tzamarias:1990ij} & W & 15.3 & 20 & 0.0529(6) & 35.9 (1.89)
 & 46.9 & $-3.5$ & 56.6\\
E537~\cite{Tzamarias:1990ij} & Cu & 15.3 & 10 & 0.0539(20) & 19.6 (2.18)
 & 51.8 & $-3.4$ & 51.6\\
NA3~\cite{NA3:1983ltt} & $p$ & 16.8 & 20 & 0.0386(7) & 160.7 (8.46)
 & 51.5 & $-3.1$ & 51.6\\
NA3~\cite{NA3:1983ltt} & $p$ & 19.4 & 10 & 0.0317(7) & 64.0 (7.11)
 & 43.8 & $-3.2$ & 59.4\\
NA3~\cite{NA3:1983ltt} & $p$ & 22.9 & 19 & 0.0293(4) & 925.5 (51.42)
 & 35.4 & $-3.2$ & 67.8\\
E705~\cite{E705:1992jno} & Li & 23.7 & 12 & 0.0542(12) & 27.3 (2.48)
 & 26.6 & $-2.8$ & 76.2\\
E672/E706~\cite{E672:1995won} & Be & 31.1 & 35 & 0.0299(5) & 87.1 (2.56)
 & 22.8 & $-3.3$ & 80.5\\
\hline
$J/\psi$ combined & & & 145 & & 1339.5 (9.85) & & & \\
\hline
NA10~\cite{NA10:1985boe} & W & 19.1 & 9 & 0.00283(15) & 11.7 (1.47)
 & 97.6 & $-0.3$ & 2.7\\
NA10~\cite{NA10:1986qyc} & W & 23.2 & 8 & 0.00212(19) & 25.8 (3.69)
 & 94.6 & $-0.5$ & 5.9\\
\hline
$\Upsilon$ combined & & & 17 & & 37.5 (2.50) & & & \\
\hline\hline
\end{tabular}
\end{table*}
\subsection{Pion Structure Function}
\label{sec:f2}
\par The pion structure function $F_2^\pi$ is measured in leading-neutron
production in deep inelastic scattering, $ep\to e'nX$, where the virtual
photon scatters off a $\pi^+$ emitted by the proton in the Sullivan
process~\cite{Sullivan:1971kd,Chang:2026tgb}. In contrast to Drell--Yan, $F_2^\pi$ is
dominated by small $x$, where the sea quarks and the gluon of the pion matter.
At NLO in the $\overline{\rm MS}$ scheme, it reads
\begin{equation}
 F_2^\pi(x,\mu^2)=x\sum_q e_q^2\bigg\{
 \big(q+\bar q\big)\otimes\Big[\delta(1-z)
 +\frac{\alpha_s(\mu^2)}{2\pi}C_q^{(1)}\Big]
 +2\,g\otimes\frac{\alpha_s(\mu^2)}{2\pi}C_g^{(1)}\bigg\},
 \label{eq:f2nlo}
\end{equation}
where the PDFs are taken at $\mu^2=Q^2$, the convolution is
$(f\otimes C)(x)=\int_x^1\frac{dz}{z}f(x/z,\mu^2)\,C(z)$, and the coefficient
functions are
\begin{align}
 C_q^{(1)}(z)&=C_F\left[\frac{1+z^2}{1-z}
 \left(\ln\frac{1-z}{z}-\frac34\right)+\frac14\big(9+5z\big)\right]_{+},
 \label{eq:cq1}\\
 C_g^{(1)}(z)&=T_R\left\{\big[z^2+(1-z)^2\big]\ln\frac{1-z}{z}
 -1+8z(1-z)\right\},
 \label{eq:cg1}
\end{align}
with $C_F=4/3$ and $T_R=1/2$. The gluon enters Eq.~\eqref{eq:f2nlo} directly
through $C_g^{(1)}$, and after evolution it also generates the sea. $F_2^\pi$
is linear in the PDFs, so the contributions of the two Fock sectors add up to
the total.

\par
\begin{figure*}[t]
 \centering
 \includegraphics[width=0.99\textwidth]{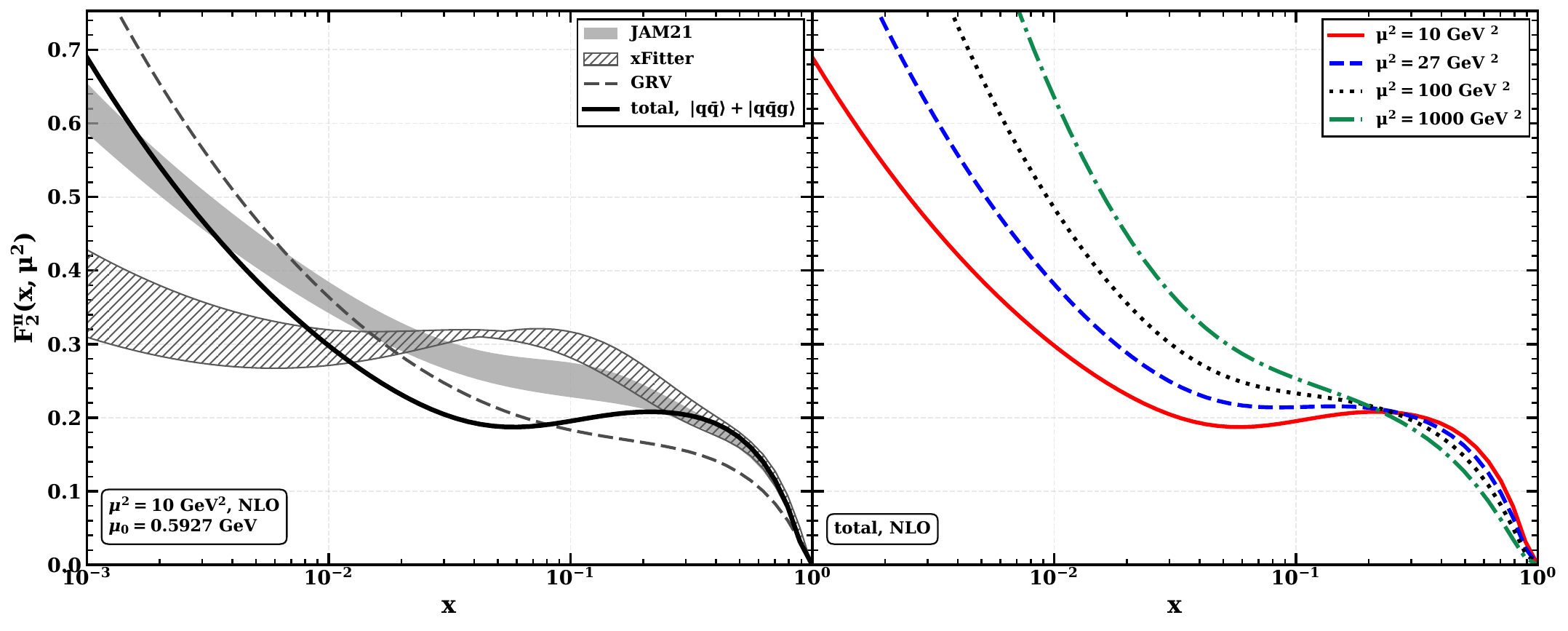}
 \caption{(Color online) $F_2^\pi$ at $\mu^2=10$ GeV$^2$, compared with JAM21~\cite{Barry:2021osv}, xFitter~\cite{Novikov:2020snp} and GRV~\cite{Gluck:1991ey} (left), and at $\mu^2=10$--$1000$ GeV$^2$ (right).}
 \label{fig:f2piQ210}
\end{figure*}

 In Fig.~\ref{fig:f2piQ210}, we present our $F_2^\pi$ at $\mu^2=10\GeV^2$
together with JAM21, xFitter and GRV. Both sectors rise at small $x$ after the evolution, because
the gluon and the sea are radiated from the quarks of both sectors. The
$\qqbarg$ sector gives about $53\%$ of $F_2^\pi$ for $x\lesssim0.1$. At $x=0.1$, we find $F_2^\pi=0.20$, of which $0.09$ comes from the $\qqbar$ sector
and $0.11$ from the $\qqbarg$ sector. Its share falls to $37\%$ at $x=0.3$ and
$18\%$ at $x=0.5$, where the valence quarks dominate. This share is much larger
than in the Drell--Yan mass spectra, where it is between $7$ and $15\%$.
Relative to the JAM21 central curve, the total is $0.82$, $0.72$, $0.71$,
$0.77$ and $1.00$ at $x=0.010$, $0.030$, $0.048$, $0.096$ and $0.31$, while
the valence sector alone gives $0.39$, $0.34$, $0.33$, $0.35$ and $0.64$. The
$\qqbarg$ sector therefore removes more than half of the difference to JAM21 for
$0.03\lesssim x\lesssim0.1$. At smaller $x$, our result rises above xFitter for $x\lesssim0.008$ and above JAM21 for $x\lesssim0.002$, and it lies about $17\%$ below GRV for $0.003\lesssim x\lesssim0.1$.
In the right panel of Fig.~\ref{fig:f2piQ210}, we present the total $F_2^\pi$ at
$\mu^2=10$, $27$, $100$ and $1000\GeV^2$. With increasing $\mu^2$, it rises at
small $x$ and falls at large $x$, and the curves cross near $x\simeq0.25$, as
expected from the DGLAP evolution.

\par
\begin{figure*}[t]
 \centering
 \includegraphics[width=0.99\textwidth]{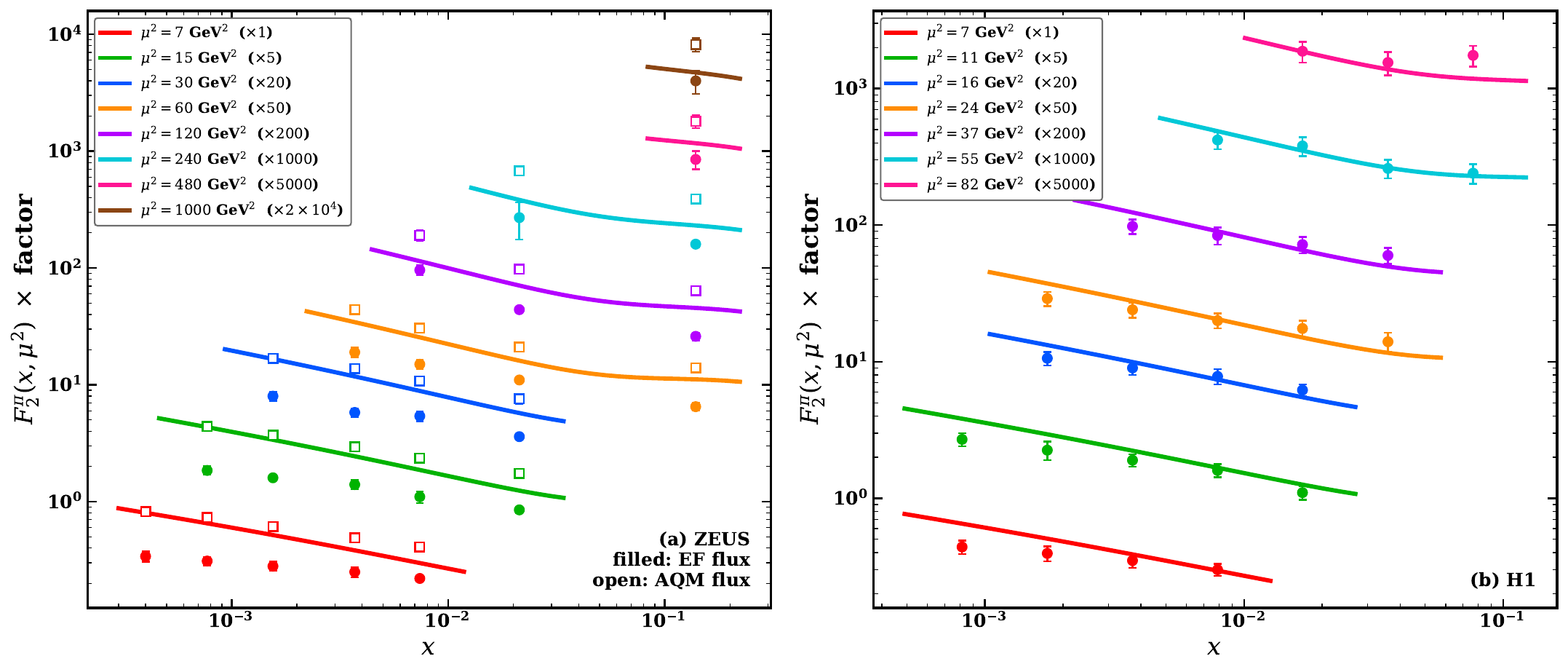}
 \caption{(Color online) $F_2^\pi$ compared with the ZEUS~\cite{ZEUS:2002gig} (left, filled for the effective flux, open for the AQM flux) and H1~\cite{H1:2010hym} (right) leading-neutron data. No normalization is fitted.}
 \label{fig:f2pih1}
\end{figure*}

 In Fig.~\ref{fig:f2pih1}, we compare with the H1~\cite{H1:2010hym} and
ZEUS~\cite{ZEUS:2002gig} leading-neutron data. The experiments extract
$F_2^\pi$ from the measured cross section with an assumed pion flux, and the
normalization of this flux is the main uncertainty of the extraction. ZEUS
gives its results for two flux choices, an effective flux (EF) and the flux
of the additive quark model (AQM), which differ by about a factor of two in
normalization but have the same shape, and we show both. We follow
Ref.~\cite{Chang:2026tgb} and do not fit any normalization, so the curves are
the absolute prediction of the model. The $\chi^2$ values are listed in
Table~\ref{tab:f2h1zeus}. For H1, we obtain $\chi^2/N=2.6$. This value is
dominated by the two lowest bins, $\mu^2=7$ and $11\GeV^2$, where the model
lies above the data, and $\chi^2/N$ falls to $0.4$ and $1.3$ at $\mu^2=55$ and
$82\GeV^2$. For ZEUS, the result depends strongly on the flux. The EF data lie
well below the model, with $\chi^2/N=135.7$, and the AQM data lie much closer
to it, with $\chi^2/N=24.8$. The two ZEUS extractions bracket our result, so
leading-neutron data constrain the shape of $F_2^\pi$ better than its
normalization.

\begin{table}[htbp]\centering
\caption{$\chi^2$ of $F_2^\pi$, against the H1~\cite{H1:2010hym} and ZEUS~\cite{ZEUS:2002gig} data, with no normalization fitted.}
\label{tab:f2h1zeus}
\begin{tabular}{lcccc}
\hline\hline
data set & flux & $N$ & $\chi^2$ & $\chi^2/N$\\
\hline
H1~\cite{H1:2010hym} & --- & 29 & 75.5 & 2.60\\
ZEUS~\cite{ZEUS:2002gig} & effective flux & 25 & 3391.7 & 135.67\\
ZEUS~\cite{ZEUS:2002gig} & additive quark model & 25 & 621.0 & 24.84\\
\hline\hline
\end{tabular}
\end{table}

%==============================================================

\section{Summary and Conclusions}
\label{sec:conclusion}

\par In this work, we have studied the three-dimensional structure of the pion
in the light-cone quark model with a Fock expansion truncated at
$|\pi\rangle=\qqbar+\qqbarg$, so that the wave function contains a dynamical
gluon. The valence sector is described by the BHL wave function with the
Melosh--Wigner spin structure, and the $\qqbarg$ sector is generated from it
by the quark--gluon vertex of light-front QCD. The strong coupling is fixed by
the model scale through Eq.~\eqref{eq:gslock}, and the confinement scale of the
three-body state is the same as that of the valence pair. The quark mass
$m_q=0.304\GeV$ and the scale $\beta=0.463\GeV$ are fixed by the PDG values of
the decay constant and the charge radius of the pion. The masses
$m_g=0.55\GeV$ and $m_f=7.00\GeV$ are taken from the scan of
Appendix~\ref{app:mgmf}, where they are checked against the BLFQ values, and
the model scale $\mu_0=0.593\GeV$ gives $g_s=3.17$. At this scale, the $\qqbar$ sector carries $60\%$ of the
probability and the $\qqbarg$ sector $40\%$, and the constituent gluon carries
$19\%$ of the pion momentum. The number and momentum sum rules are satisfied
without being imposed.

\par From the same wave function, we have calculated the quark and gluon TMDs,
the GPDs and their impact-parameter densities, the PDFs, the electromagnetic
form factor and the mass decomposition. The quarks of the $\qqbarg$ sector carry smaller momentum
fractions and smaller transverse momenta than the valence quarks. After CSS evolution, the quark TMD follows the $b_T$ dependence of lattice QCD in most of
the region where they are compared, while at $\mu^2=25\GeV^2$ it is larger
than the recent Drell--Yan extraction, by up to a factor of about $1.5$. The
evolved quark GPD agrees with lattice QCD for $x\gtrsim0.25$ at small $-t$ and
is close to it for $0.4\lesssim x\lesssim0.8$ at larger $-t$, and the
double-distribution extension satisfies polynomiality. The
charge radius is $0.66\fm$ for the full state and $0.53\fm$ for the valence
sector alone.

\par At $\mu^2=4\GeV^2$, the momentum fractions are
$2\langle x\rangle_v=0.517$, $\langle x\rangle_g=0.398$ and
$\langle x\rangle_S=0.085$. The valence distribution agrees with JAM21,
xFitter and MAP23, and the gluon momentum fraction lies close to JAM21, MAP23
and CoSMAO22. The higher Mellin moments show that the shape of the valence
distribution is nearly the same in all these analysis, while the gluon of the
present work lies between the softer and the harder gluon distributions. In
the Ji decomposition with $b=1/2$ from chiral symmetry, the quark kinetic,
gluon kinetic, quark mass and trace anomaly terms give $15\%$, $30\%$, $43\%$
and $12\%$ of the pion mass at $\mu^2=4\GeV^2$.

\par We have compared the evolved PDFs with pion-induced Drell--Yan, $J/\psi$ and $\Upsilon$ production and the leading-neutron structure function. One normalization is fitted per data set for the Drell--Yan $x_F$ spectra and for quarkonium production, while the Drell--Yan mass spectra and $F_2^\pi$ are absolute predictions. The Drell--Yan $x_F$ spectra are described with $\chi^2/N$ between
$0.98$ and $2.58$ for E537, E326, E444 and E615, and the two NA10 energies
require similar normalizations. Drell--Yan production is
dominated by the valence quarks, and the $\qqbarg$ sector gives between $7$ and
$15\%$ of the mass spectra. $J/\psi$ production is sensitive to the gluon. The
$gg$ channel grows from $13\%$ of the cross section at $\sqrt s=8.7\GeV$ to
$81\%$ at $31.1\GeV$, and the $\qqbarg$ sector gives between $16$ and $50\%$.
$\Upsilon$ production is dominated by the $q\bar q$ channel and tests the
valence quarks at large $x$. For $F_2^\pi$, the $\qqbarg$ sector gives more than half of the structure function at $x\lesssim0.1$ and removes more than half of
the difference between the valence result and JAM21 for
$0.03\lesssim x\lesssim0.1$. The two ZEUS flux assumptions bracket our
prediction, so the leading-neutron data constrain the shape of $F_2^\pi$
better than its normalization.

\par The present work can be extended in several directions. The $T$-odd
distributions, such as the Boer--Mulders function, require the gauge link and
vanish in the present calculation. The next Fock sectors, $|q\bar qgg\rangle$
and $|q\bar qq\bar q\rangle$, together with the instantaneous interactions of
light-front QCD, would give a sea at the model scale.
The upcoming measurements at AMBER, the EIC and the EicC will test the gluon
and sea distributions of the pion obtained here.

\section*{Acknowledgments}

S.~P. acknowledges Prof.~Wen-Chen Chang for valuable discussions. S.~P. also thanks Prof.~Craig D.~Roberts for suggesting a modification to the BHL model that improves its endpoint behavior. S.~P. acknowledges Mr.~Ashutosh Dwibedi for helpful discussions during this work. S.~P. is grateful to Dr.~Valentine Moos for numerous discussions on pion TMDs and for providing the extracted pion TMD data. S.~P. also acknowledges Dr.~Lorenzo Rossi for providing the pion TMD extraction from the MAP Collaboration. S.~P. thanks Mr.~Abhishek K.~P. for his assistance with the TMD evolution code and for valuable discussions that helped clarify its implementation. S.~P. also thanks Prof.~C.~Lorcé for his valuable comments during Pacific Spin 2026 and Dr.~Jai D.~More for fruitful discussions during the International Conference on Frontiers in High Energy Physics (ICFHEP) at IIT Bhilai. S.~P. acknowledges the Institute of Physics, Academia Sinica, and particularly Prof.~Wen-Chen Chang, for providing access to an upgraded NVIDIA GeForce RTX 5080 GPU for high-performance computing. Finally, S.~P. acknowledges the use of ChatGPT for assistance with writing and editing the manuscript.

\section{Appendix}

\subsection{Choice of the gluon and vertex masses}
\label{app:mgmf}
Adding the $\qqbarg$ sector brings in two new parameters, the gluon mass $m_g$
and the vertex mass $m_f$. The coupling is not a free parameter, since $g_s$
follows from the model scale through Eq.~\eqref{eq:gslock}, and $m_q$ and $\beta$
are fixed by the decay constant and the charge radius of the pion, as described in
Sec.~\ref{sec:lfqm}. The model scale, however, depends on $m_g$ and $m_f$. For every pair $(m_g,m_f)$, the E615 fit of Sec.~\ref{sec:lfqm} is therefore repeated,
and $\mu_0$, $\alpha_s(\mu_0)$ and $g_s$ change from one pair to the next.

We scan $m_g$ from $0.10$ to $1.00$ GeV in steps of $0.05$ GeV and $m_f$ from
$0.5$ to $8.0$ GeV in steps of $0.5$ GeV. Since $m_f$ enters the quark--gluon
vertex linearly, the norm of the $\qqbarg$ amplitude and its quark and gluon
densities are quadratic in $m_f$ at fixed $m_g$, and three values of $m_f$ for
each $m_g$ are enough to build the full grid. For each pair, we determine $\mu_0$
from the minimum of the E615 $\chi^2/N$, compute the probability $P_{\qqbarg}$
and the gluon momentum fraction $\langle x\rangle_g(\mu_0)$, and evolve the gluon
to $\mu^2=4~\GeV^2$, where it is compared with the JAM21
gluon~\cite{Barry:2021osv} for $0.01\le x\le0.8$ with the JAM21 $68\%$ band as
the uncertainty.

\begin{figure*}[htbp]
\centering
\includegraphics[width=0.99\textwidth]{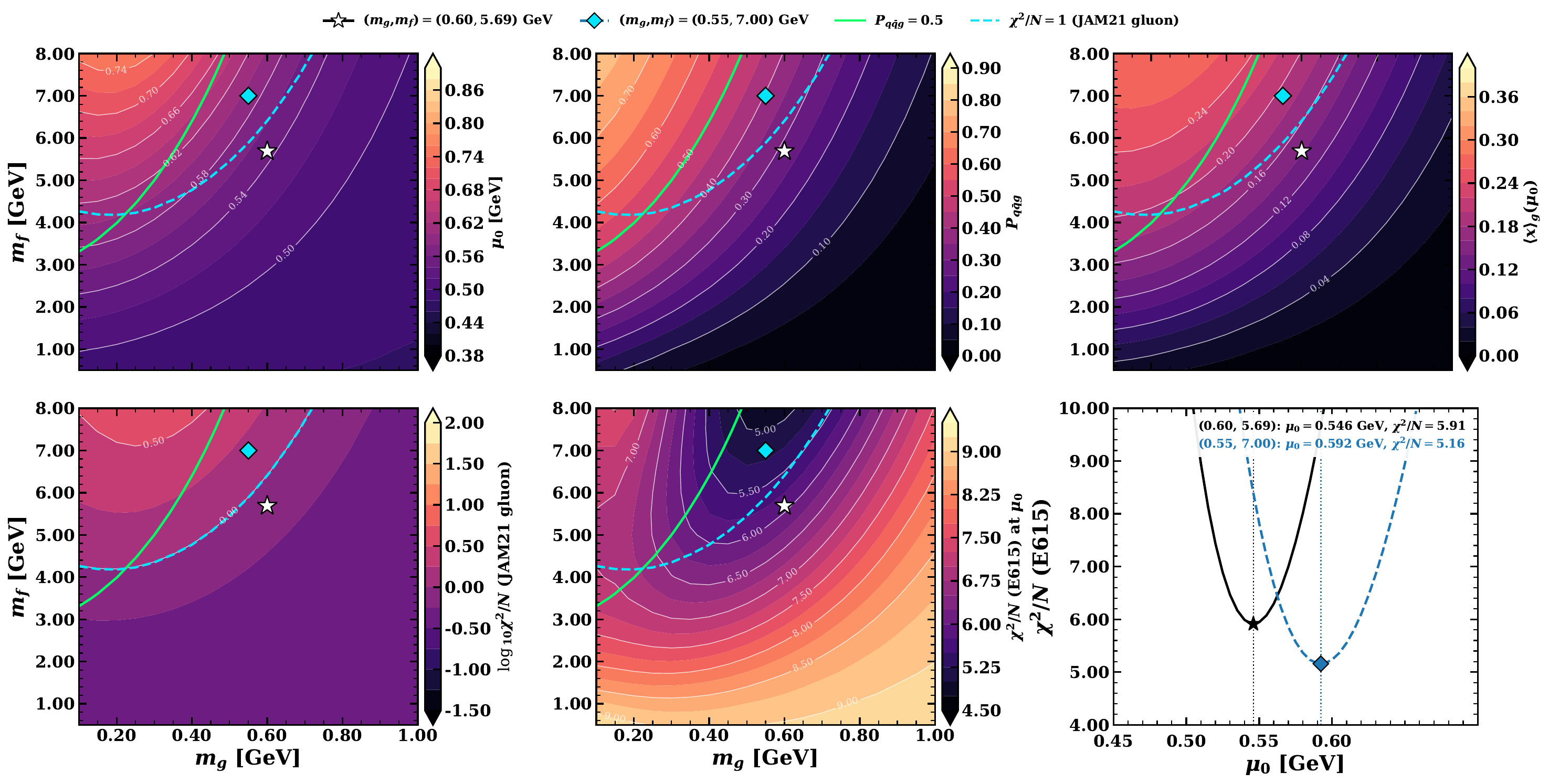}
\caption{Dependence of the model on $m_g$ and $m_f$. Upper row, from left to
right, the model scale $\mu_0$, the probability $P_{\qqbarg}$, and the gluon
momentum fraction $\langle x\rangle_g(\mu_0)$. Lower row, from left to right,
$\log_{10}$ of the $\chi^2/N$ of the evolved gluon with respect to JAM21 at
$\mu^2=4~\GeV^2$, the E615 $\chi^2/N$ at the fitted $\mu_0$, and the E615
$\chi^2/N$ as a function of $\mu_0$ for our pair and the BLFQ pair. The solid line
marks $P_{\qqbarg}=0.5$ and the dashed line $\chi^2/N=1$ with respect to JAM21.}
\label{fig:mgmf}
\end{figure*}

In Fig.~\ref{fig:mgmf}, we present the results of the scan. The $\qqbarg$ sector
grows toward a light gluon and a large vertex mass, and the model scale grows
with it, from $0.48$ to $0.76$ GeV. After evolution to $\mu^2=4~\GeV^2$, the gluon momentum fraction lies between $0.39$ and $0.40$ over the whole grid, so
the evolved gluon keeps little memory of the input gluon and the JAM21 gluon
alone does not fix $m_g$ and $m_f$. The E615 data are more sensitive. Their
$\chi^2/N$ varies between $4.8$ and $9.2$ across the grid. It falls as the
$\qqbarg$ sector grows up to $P_{\qqbarg}\simeq0.4$--$0.5$, because the quark of
this sector is softer and brings the large-$x$ fall-off of the evolved valence
distribution closer to the data, and it rises again for larger $P_{\qqbarg}$.

In the present work, we take $m_g=0.55$ GeV and $m_f=7.00$ GeV, together with
$m_q=0.304$ GeV and $\beta=\beta_3=0.463$ GeV. This pair lies in the region
where the E615 $\chi^2/N$ is smallest, and it keeps $P_{\qqbarg}$ below $0.5$, so
that the $\qqbar$ state remains the leading Fock component of the truncated
pion. The lower right panel of Fig.~\ref{fig:mgmf} shows that its E615
$\chi^2/N$ has its minimum, $5.16$, at $\mu_0=0.592$ GeV. The main fit gives
$\mu_0=0.593$ GeV, as listed in Table~\ref{tab:params}. At this scale, $P_{\qqbarg}=0.40$, and the gluon carries $19\%$ of the pion momentum.

\begin{figure}[htbp]
\centering
\includegraphics[width=0.6\textwidth]{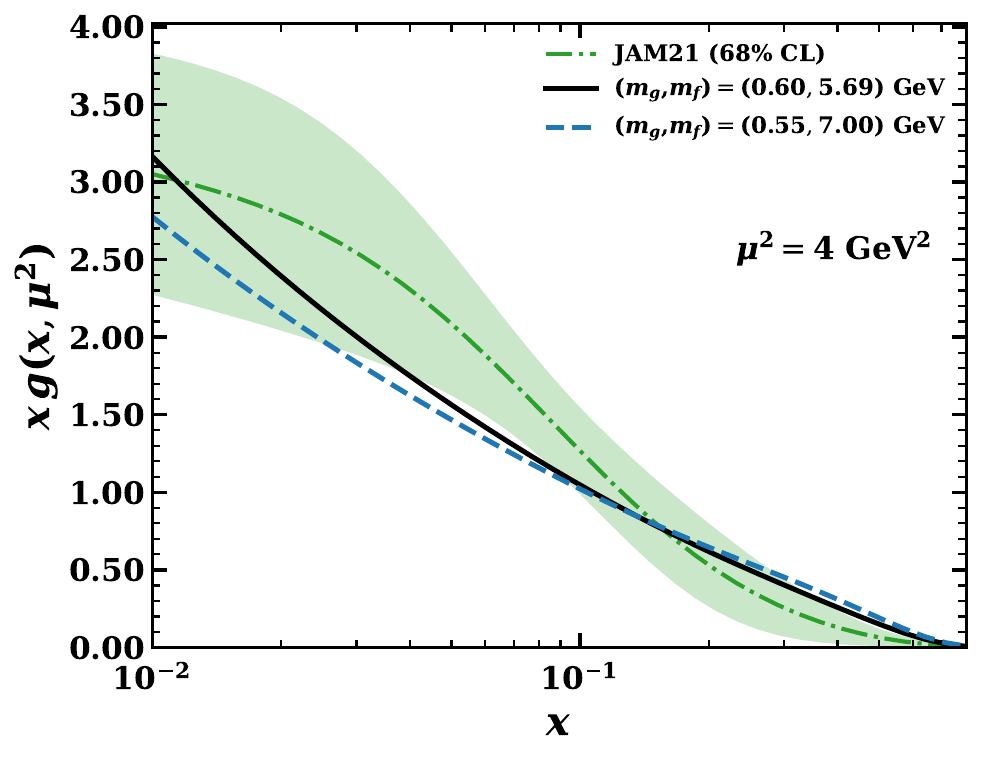}
\caption{Gluon distribution $xg(x,\mu^2)$ at $\mu^2=4~\GeV^2$ for our pair and the BLFQ pair
of $(m_g,m_f)$ of Fig.~\ref{fig:mgmf}, compared with
JAM21~\cite{Barry:2021osv}.}
\label{fig:mgmf_jam}
\end{figure}

We have checked this choice against the BLFQ values $m_g=0.60$ GeV and
$m_f=5.69$ GeV~\cite{Lan:2021wok,Lan:2024ais,Lan:2025qio}, with the same $m_q$
and $\beta$. They give $\mu_0=0.546$ GeV, an E615 $\chi^2/N$ of $5.91$,
$P_{\qqbarg}=0.27$ and $\langle x\rangle_g(\mu_0)=0.14$. Our pair therefore
describes the E615 data better and carries a larger $\qqbarg$ sector. In
Fig.~\ref{fig:mgmf_jam}, we present the evolved gluon distributions of the two
pairs at $\mu^2=4~\GeV^2$ together with JAM21.

\subsection{Light-front conventions and notation}
\label{app:conv}

Light-cone components are $a^{\pm}=a^{0}\pm a^{3}$, so that
\begin{equation}
a\cdot b=\tfrac12\big(a^{+}b^{-}+a^{-}b^{+}\big)
-\bm{a}_\perp\!\cdot\!\bm{b}_\perp ,
\qquad
g^{+-}=g^{-+}=2,\qquad g^{11}=g^{22}=-1 ,
\label{eq:metric}
\end{equation}
and $\gamma^{\pm}=\gamma^{0}\pm\gamma^{3}$. On mass shell
$a^{-}=(\bm{a}_\perp^2+m^2)/a^{+}$, so the minus component is not independent,
and $x$ and $\kperp$ form a complete set of variables for a parton.

A parton $i$ inside a hadron of momentum
$P=(P^{+},P^{-},\bm{P}_\perp)$ carries
\begin{equation}
p_i=\big(p_i^{+},\,p_i^{-},\,\bm{p}_{\perp i}\big),
\qquad
x_i=\frac{p_i^{+}}{P^{+}},
\qquad
\bm{p}_{\perp i}=\bm{k}_{\perp i}+x_i\bm{P}_\perp ,
\label{eq:intrinsic}
\end{equation}
where $\bm{k}_{\perp i}$ is the intrinsic transverse momentum, measured with
respect to the hadron, subject to
\begin{equation}
\sum_i x_i=1,\qquad \sum_i \bm{k}_{\perp i}=\bm{0} .
\label{eq:conserve}
\end{equation}
The distinction in Eq.~\eqref{eq:intrinsic} between the physical momentum
$\bm{p}_{\perp i}$ and the intrinsic momentum $\bm{k}_{\perp i}$ matters. The
light-front wave functions depend on $\bm{k}_{\perp i}$, while the Dirac
spinors depend on $\bm{p}_{\perp i}$, and away from the forward limit the two
differ. This is the origin of the $(1-x)\Dperp/2$ shift of
Eq.~\eqref{eq:shifts} in Sec.~\ref{sec:gpd}.

The $n$-particle phase-space measure used throughout is
\begin{equation}
\int[\mathrm{d}n]\;\equiv\;
\prod_{i=1}^{n}\int\frac{\mathrm{d}x_i\,\mathrm{d}^2\bm{k}_{\perp i}}{16\pi^3}
\;16\pi^3\,\delta\Big(1-\sum_i x_i\Big)\,
\delta^{(2)}\Big(\sum_i\bm{k}_{\perp i}\Big),
\label{eq:measure}
\end{equation}
which is the Lepage--Brodsky normalization used in Eqs.~\eqref{eq:valstate}
and~\eqref{eq:psi3}. Wave functions in the
$\mathrm{d}^3p/[(2\pi)^3 2p^{+}]$ convention are converted by
$\Psi_{\rm LB}=\psi/\sqrt{\prod_i x_i}$.

Throughout, $\Delta=P'-P$ is the momentum transfer, the skewness is
$\xi=-\Delta^{+}/(2\bar P^{+})$ with $\bar P=(P+P')/2$, and every result of
this work is at $\xi=0$, where
\begin{equation}
t=\Delta^{2}=-\Dperp^{2}\equiv-Q^{2} .
\label{eq:tdef}
\end{equation}
We write $Q^{2}$ when the observable is the electromagnetic form factor and
$-t$ otherwise, but they are the same variable.

A single frame is used throughout. The GPDs of Sec.~\ref{sec:gpd} are
evaluated in the symmetric frame, $\bm{P}_\perp=-\Dperp/2$ and
$\bm{P}'_\perp=+\Dperp/2$, so that $\bar{\bm P}_\perp=\bm 0$. The initial
(final) intrinsic momenta are denoted $\bm k'_{\perp i}$ ($\bm k''_{\perp i}$),
the struck parton and the spectators shift as in Eq.~\eqref{eq:shifts}, and the
overlaps are integrated over the average momenta $\bm k_{\perp i}$.

The Dirac matrices are in the Dirac representation, with
$\beta_{\rm LB}=\gamma^{0}$, $\alpha^{i}=\gamma^{0}\gamma^{i}$, and the light-front
spinors of Eq.~\eqref{eq:spinors} are built on the two-component objects
$X_\uparrow=\tfrac1{\sqrt2}(1,0,1,0)^{T}$ and
$X_\downarrow=\tfrac1{\sqrt2}(0,1,0,-1)^{T}$. They satisfy
\begin{equation}
\bar u_{{s'}}(p')\gamma^{+}u_{{s}}(p)=2\sqrt{p'^{+}p^{+}}\,
\delta_{{ss'}},
\qquad
\sum_{{s}} u_{{s}}(p)\bar u_{{s}}(p)=\slashed p+m ,
\label{eq:spinorid}
\end{equation}
the first of which makes the $\gamma^{+}$ correlator a pure overlap,
Eqs.~\eqref{eq:tmdqov} and~\eqref{eq:gpdqov}. The gluon polarization vectors are those of
Eq.~\eqref{eq:pol}, transverse and physical in the light-cone gauge $A^{+}=0$,
with
\begin{equation}
\sum_{\lambda=\pm1}\epsilon^{i}_{\lambda}\epsilon^{*j}_{\lambda}
=\delta^{ij},
\qquad
\epsilon_\lambda\!\cdot\!k=0 ,
\label{eq:polsum}
\end{equation}
so that the two physical polarizations, and no others, propagate.

Finally, $k^{R,L}=k^{1}\pm ik^{2}$ is used throughout for the transverse
raising and lowering combinations, and $\bm{b}_\perp$ always denotes the
impact parameter conjugate to $\Dperp$. It is different from the variable
$b_T$ conjugate to $\kperp$ that appears in the Collins--Soper--Sterman
evolution of the TMDs.

\subsection{TMD evolution}
\label{app:sec:tmdevol}

The TMDs above are the model-scale objects, defined at $\mu_0$ with the gauge
link set to unity. Comparison with any measurement requires the
Collins--Soper--Sterman (CSS) evolution of the transverse-momentum
dependence~\cite{Collins:1984kg}, which is most naturally carried out in the
conjugate transverse-separation space,
\begin{equation}
 \tilde f_1(x,b_T;\mu)
 = \int\!\mathrm{d}^2\kperp\;e^{-i\kperp\cdot\bm{b}_T}
 f_1(x,k_\perp^2;\mu)
 = 2\pi\!\int_0^\infty\!\!\mathrm{d}k_\perp\,k_\perp\,
 J_0(k_\perp b_T)\,f_1(x,k_\perp^2;\mu),
 \label{eq:bspaceapp}
\end{equation}
where $b_T$ is conjugate to the parton transverse momentum and differs from
the impact parameter $b_\perp$ of the GPDs. In $b_T$ space, the
evolution is multiplicative,
\begin{equation}
 \tilde f_1(x,b_T;\mu)
 = \tilde f_1\big(x,b_T;\mu_0\big)\,
 \exp\!\Big[-S_{\rm pert}(b_*,\mu)\Big]\,
 \exp\!\Big[-S_{\rm NP}(b_T,\mu)\Big] .
 \label{eq:cssapp}
\end{equation}
The perturbative Sudakov factor $S_{\rm pert}$ is evaluated using
\begin{equation}
 b_*(b_T)=\frac{b_T}{\sqrt{1+b_T^2/b_{\rm max}^2}},
 \qquad
 \mu_b=\frac{C_0}{b_*},\qquad C_0=2e^{-\gamma_E}\simeq1.1229,
 \label{eq:bstarapp}
\end{equation}
so that $b_*\to b_T$ at small $b_T$, $b_*\to b_{\rm max}$ at large $b_T$, and
$\mu_b$ never reaches the Landau pole. Writing
$a_s(\mu)\equiv\alpha_s(\mu)/4\pi$, the Sudakov exponent implemented here is
\begin{equation}
 S_{\rm pert}(b_*,\mu)=
 \int_{\mu_b}^{\mu}\frac{\mathrm{d}\bar\mu}{\bar\mu}
 \left[\Gamma_{\rm cusp}\big(a_s(\bar\mu)\big)\,
 \ln\frac{\mu^{2}}{\bar\mu^{2}}
 -2\,\gamma_F\big(a_s(\bar\mu)\big)\right],
 \label{eq:sudakov}
\end{equation}
which is the standard CSS form
$\int\mathrm{d}\bar\mu^{2}/\bar\mu^{2}\,[A\ln(\mu^{2}/\bar\mu^{2})+B]$ with
$A=\Gamma_{\rm cusp}/2$ and $B=-\gamma_F$, i.e., $A^{(1)}=C_F/2$ and $B^{(1)}=-\tfrac34 C_F$ in the $\alpha_s/\pi$ normalization. It is the Sudakov factor of a single TMD. A Drell--Yan cross section carries it squared, once for each hadron, which gives the familiar $A^{(1)}=C_F$ and $B^{(1)}=-\tfrac32 C_F$. The integral is evaluated using a ten-point
Gauss--Legendre rule in $\ln\bar\mu$ and is set to zero whenever
$\mu\le\mu_b$.

Both anomalous dimensions are carried to $a_s^3$,
\begin{equation}
 \Gamma_{\rm cusp}=\sum_{n=0}^{2}\Gamma_n\,a_s^{\,n+1},
 \qquad
 \gamma_F=\sum_{n=0}^{2}\gamma_n\,a_s^{\,n+1},
 \label{eq:andimexp}
\end{equation}
with the cusp coefficients
\begin{align}
 \Gamma_0&=4C_F,
 \label{eq:gam0}\\
 \Gamma_1&=C_FC_A\Big(\tfrac{268}{9}-\tfrac{4\pi^{2}}{3}\Big)
 -\tfrac{40}{9}\,C_F n_f,
 \label{eq:gam1}\\
 \Gamma_2&=\tfrac12\Big[
 -\tfrac{32}{27}C_Fn_f^{2}
 +C_AC_Fn_f\Big(-\tfrac{1672}{27}+\tfrac{160\pi^{2}}{27}
 -\tfrac{224\zeta_3}{3}\Big)
 \nonumber\\
 &\qquad
 +C_A^{2}C_F\Big(\tfrac{980}{3}-\tfrac{1072\pi^{2}}{27}
 +\tfrac{88\pi^{4}}{45}+\tfrac{176\zeta_3}{3}\Big)
 +C_F^{2}n_f\Big(-\tfrac{220}{3}+64\zeta_3\Big)\Big],
 \label{eq:gam2}
\end{align}
and the TMD anomalous dimension
\begin{align}
 \gamma_0&=3C_F,
 \label{eq:gf0}\\
 \gamma_1&=\tfrac12\Big[
 C_F^{2}\big(3-4\pi^{2}+48\zeta_3\big)
 +C_FC_A\Big(\tfrac{961}{27}+\tfrac{11\pi^{2}}{3}-52\zeta_3\Big)
 +C_Fn_f\Big(-\tfrac{130}{27}-\tfrac{2\pi^{2}}{3}\Big)\Big],
 \label{eq:gf1}\\
 \gamma_2&=\tfrac12\Big[
 C_F^{2}n_f\Big(-\tfrac{2953}{27}+\tfrac{26\pi^{2}}{9}
 +\tfrac{28\pi^{4}}{27}-\tfrac{512\zeta_3}{9}\Big)
 +C_Fn_f^{2}\Big(-\tfrac{4834}{729}+\tfrac{20\pi^{2}}{27}
 +\tfrac{16\zeta_3}{27}\Big)
 \nonumber\\
 &\qquad
 +C_F^{3}\Big(29+6\pi^{2}+\tfrac{16\pi^{4}}{5}+136\zeta_3
 -\tfrac{32\pi^{2}\zeta_3}{3}-480\zeta_5\Big)
 \nonumber\\
 &\qquad
 +C_A^{2}C_F\Big(\tfrac{139345}{1458}+\tfrac{7163\pi^{2}}{243}
 +\tfrac{83\pi^{4}}{45}-\tfrac{7052\zeta_3}{9}
 +\tfrac{88\pi^{2}\zeta_3}{9}+272\zeta_5\Big)
 \nonumber\\
 &\qquad
 +C_AC_Fn_f\Big(\tfrac{17318}{729}-\tfrac{2594\pi^{2}}{243}
 -\tfrac{22\pi^{4}}{45}+\tfrac{1928\zeta_3}{27}\Big)
 \nonumber\\
 &\qquad
 +C_AC_F^{2}\Big(\tfrac{151}{2}-\tfrac{410\pi^{2}}{9}
 -\tfrac{494\pi^{4}}{135}+\tfrac{1688\zeta_3}{3}
 +\tfrac{16\pi^{2}\zeta_3}{3}+240\zeta_5\Big)\Big],
 \label{eq:gf2}
\end{align}
with $C_F=4/3$, $C_A=3$, $T_R=1/2$, $n_f=5$, $\zeta_3=1.2020569$ and
$\zeta_5=1.0369278$.

At small $b_T$, the TMD is matched onto the collinear PDF by the operator
product expansion, which we take at $\mathcal{O}(\alpha_s)$,
\begin{equation}
 f^{\rm coll}_{q}(x,\mu_b)
 =f_{q}(x,\mu_b)
 +a_s(\mu_b)\Big[
 \big(C^{(1)}_{qq}\otimes f_{q}\big)(x,\mu_b)
 +\delta^{(1)}_{qq}f_{q}(x,\mu_b)
 +\big(C^{(1)}_{qg}\otimes f_{g}\big)(x,\mu_b)\Big],
 \label{eq:ope}
\end{equation}
where $(C\otimes f)(x)=\int_x^1(\mathrm{d}z/z)\,C(z)\,f(x/z)$ and
\begin{equation}
 C^{(1)}_{qq}(z)=2C_F(1-z),\qquad
 \delta^{(1)}_{qq}=-\frac{C_F\pi^{2}}{6},\qquad
 C^{(1)}_{qg}(z)=4T_R\,z(1-z).
 \label{eq:wilson}
\end{equation}
The convolution is evaluated on a Gauss--Legendre grid in $\ln(1/z)$.

The non-perturbative evolution is the remaining rapidity-dependent piece.
It is governed by the Collins--Soper kernel, whose non-perturbative part we
take in the standard quadratic form
\begin{equation}
 g_K(b_T)=-\frac{g_2^{2}}{2}\,b_T^{2},
 \qquad
 S_{\rm NP}(b_T,\mu)=-\frac{1}{2}g_K(b_T)\,\ln\frac{\mu^{2}}{Q_0^{2}}
 =\frac{g_2^{2}}{4}\,b_T^{2}\,\ln\frac{\mu^{2}}{Q_0^{2}} .
 \label{eq:snp}
\end{equation}
Writing the parameter squared enforces $g_K\le0$, the sign required for the
large-$b_T$ suppression, so $g_2$ here is not the $g_2$ of the BLNY
parametrization, and the two should not be compared numerically. The pair
$(g_2,b_{\rm max})$ is fixed jointly, and neither may be varied alone.

Collecting Eqs.~\eqref{eq:bstarapp}--\eqref{eq:snp}, the evaluated TMD is
\begin{equation}
 \tilde f_1(x,b_T;\mu)
 = f^{\rm coll}_{q}(x,\mu_b)\;
 e^{-S_{\rm pert}(b_*,\mu)}\;
 \hat f_{\rm NP}(x,b_T)\;
 e^{-S_{\rm NP}(b_T,\mu)},
 \label{eq:cssfull}
\end{equation}
where the model enters through
\begin{equation}
 \hat f_{\rm NP}(x,b_T)
 =\frac{\displaystyle\int_0^\infty\!\mathrm{d}k_\perp\,k_\perp\,
 J_0(k_\perp b_T)\,f_1(x,k_\perp^{2};\mu_0)}
 {\displaystyle\int_0^\infty\!\mathrm{d}k_\perp\,k_\perp\,
 f_1(x,k_\perp^{2};\mu_0)} ,
 \label{eq:fnp}
\end{equation}
normalized to $\hat f_{\rm NP}(x,0)=1$. The numerical values of every
parameter entering Eqs.~\eqref{eq:bstarapp}--\eqref{eq:fnp} are collected in
Table~\ref{tab:evolpar}.

\begin{table}[htbp]
\centering
\caption{Parameters of the TMD evolution, Eqs.~\eqref{eq:bstarapp}--\eqref{eq:fnp}.}
\label{tab:evolpar}
\begin{tabular}{lll}
\hline\hline
quantity & value & role \\
\hline
$\mu_0$ & $0.593\GeV$ & model scale, where $\hat f_{\rm NP}$ is defined\\
$\mu$ & $2\GeV$ or $5\GeV$ & final scale (lattice; extractions)\\
$b_{\rm max}$ & $1.5\GeV^{-1}$ & $b_*$ saturation, Eq.~\eqref{eq:bstarapp}\\
$C_0$ & $2e^{-\gamma_E}=1.1229$ & matching-scale constant\\
$g_2$ & $0.248\GeV$ & non-perturbative CS kernel, Eq.~\eqref{eq:snp}\\
$Q_0$ & $1\GeV$ & reference scale of $g_2$\\
$n_f$ & $5$ & active flavors in $\Gamma_{\rm cusp}$, $\gamma_F$\\
OPE order & $\mathcal{O}(\alpha_s)$ & Eq.~\eqref{eq:ope}\\
$\Gamma_{\rm cusp},\gamma_F$ & to $a_s^{3}$ & Eqs.~\eqref{eq:gam0}--\eqref{eq:gf2}\\
$\alpha_s$ & from the collinear set & two-loop, thresholds at $m_c$, $m_b$\\
\hline\hline
\end{tabular}
\end{table}
\bibliographystyle{JHEP}
\bibliography{ref}

\end{document}